\documentclass[fontsize=10pt]{article}
\usepackage[hyphens]{url}
\usepackage[a4paper, total={6in, 8in}]{geometry}
\usepackage{graphicx}
\usepackage[sort]{natbib}
\usepackage[font=small,labelfont=bf]{caption}

\usepackage{mathtools}
\usepackage{amsthm}
\usepackage{amsmath}
\usepackage{amsfonts}
\usepackage{amssymb}
\usepackage{booktabs}
\usepackage{float}
\usepackage[ruled]{algorithm2e}

\usepackage[hidelinks,breaklinks]{hyperref}
\usepackage{cleveref}

\usepackage{authblk}

\usepackage{import}
\usepackage{preamble}

\Crefname{theorem}{Theorem}{Theorems}
\newtheorem{proposition}{Proposition}
\Crefname{proposition}{Proposition}{Propositions}
\newtheorem{corollary}[proposition]{Corollary}
\newtheorem{lemma}[proposition]{Lemma}
\Crefname{lemma}{Lemma}{Lemmata}
\newtheorem{example}[proposition]{Example}

\Crefname{rem}{Remark}{Remarks}

\theoremstyle{definition}
\newtheorem{definition}[proposition]{Definition}
\Crefname{definition}{Definition}{Definitions}

\title{From Independence of Clones to Composition Consistency: \\ A Hierarchy of Barriers to Strategic Nomination}

\author[1,2]{Ratip Emin Berker}
\author[3]{S\'ilvia Casacuberta}
\author[3]{Isaac Robinson}
\author[4]{Christopher Ong}
\author[1,2,3]{Vincent Conitzer}
\author[5]{Edith Elkind}

\affil[1]{Carnegie Mellon University}
\affil[2]{Foundations of Cooperative AI Lab (FOCAL)}
\affil[3]{University of Oxford}
\affil[4]{Harvard University}
\affil[5]{Northwestern University}

\begin{document}

\maketitle

\begin{abstract}
    We study two axioms for social choice functions that capture the impact of similar candidates: independence of clones (IoC) and composition consistency (CC). We clarify the relationship between these axioms by observing that CC is strictly more demanding than IoC, and investigate whether common voting rules that are known to be independent of clones (such as STV, Ranked Pairs, Schulze, and Split Cycle) are composition-consistent. While for most of these rules the answer is negative, we identify a variant of Ranked Pairs that satisfies CC. Further, we show how to efficiently modify any (neutral) social choice function so that it satisfies CC, while maintaining its other desirable properties. Our transformation relies on the hierarchical representation of clone structures via PQ-trees. We extend our analysis to social preference functions. Finally, we interpret IoC and CC as measures of robustness against strategic manipulation by candidates, with IoC corresponding to strategy-proofness and CC corresponding to obvious strategy-proofness.
\end{abstract}

\section{Introduction}\label{sec:intro}

On November 6th, 1934, Oregonians took to the polls to elect their 28th governor. Earlier, in a contested Republican primary, Senator Joe E.~Dunne had narrowly defeated Peter Zimmerman, who then decided to run as an independent.
In the subsequent general election, each candidate received the following share of votes~\citep{Needham21:1934}:
\begin{center}
    \begin{tabular}{ |c|c|c|c| } 
    \hline
    Charles Martin & Peter Zimmerman & Joe Dunne \\
    \hline
    116,677 & 95,519 & 86,923 \\ 

    \hline
    \end{tabular}
\end{center}
The Democratic candidate (Martin) won, even though the two Republicans collectively won nearly $60\%$ of the vote. This example motivates the following question: \textit{How can we ensure that similar candidates in an election do not `spoil' the election, preventing each other from winning?}

Naturally, some winner determination rules, or \emph{social choice functions (SCFs)}, are more resilient
to this ``spoilage'' effect than others. 
The field of social choice offers a rich variety of SCFs and formulates various desirable criteria
(axioms) for them. In particular, \citet{Tideman87:Independence} introduces the concept of a {\em clone set}, \emph{i.e.}, a group of candidates (clones) that are ranked consecutively in all voter's rankings, and puts forward an axiom called \emph{independence of clones (IoC)}, which asks 
that if a candidate is an election winner, this should remain the case even if we add\footnote{For example, under veto/anti-plurality, a non-IoC SCF that picks the candidate(s) ranked bottom by the least number of voters, introducing clones can {\em help} the cloned candidate, as the clones split the last places in votes. \label{footnote:veto}} or remove clones of her opponents. In the context of political elections, IoC means that a political party need not be strategic about the number of party representatives participating in an election, as long as it does not care which of its candidates wins. Conversely, if a rule fails IoC, adding/deleting clones may be a viable strategy to change the outcome (Tideman himself recalls winning a grade school election after nominating his opponent's best friend). Moreover, \citet{Elkind11:Cloning} show that the algorithmic problem of purposeful cloning (\emph{i.e.}, to change the outcome of an election) is easy for many common SCFs. Thus, the appeal of SCFs satisfying IoC goes beyond the theoretical.

In settings with more abstract candidates, it is even easier to introduce clones. For example, when candidates consist of drafts of a text---\emph{e.g.}, we are voting over drafts of the guiding principles of our organization---it is straightforward to introduce a near-duplicate of an existing draft.  When candidates are AI systems---\emph{e.g.}, we are ranking LLMs, as done for example on Chatbot Arena~\citep{Chiang24:Chatbot}, to determine the best one---one can introduce a second version of a model, one that is fine-tuned only slightly differently (as has already been pointed out by~\citet{Conitzer24:Position}).  Without IoC, such clones can critically affect the outcome. 

On the other hand, IoC may not be \emph{enough} to dissuade strategic cloning. While IoC dictates that the cloning of a candidate should not change whether \emph{one} of the clones wins, it does not say anything about \emph{which} clone should win, even though one of them can be significantly preferred to the others by the voters. As such, even with an IoC rule, cloning can have a significant impact on the result by changing which candidate among the clones/political party wins.

Moreover, a rule being IoC does not reveal how \emph{obviously} robust it is against strategic nomination. As demonstrated by \citet{Li17:Obviously} in the context of obvious strategy-proofness, the benefits of a property of a mechanism might only be materialized if the agents actually \emph{believe} that the property indeed holds. Even when using an IoC rule, it is not clear that the average voter or candidate can be easily convinced of this property---resulting, for example, in a candidate unnecessarily dropping out of the race, either out of fear of hurting their party, or of being blamed by their voters for doing so.

These drawbacks of IoC might be one potential explanation for why major parties in the United States still hold internal primaries to pick a single nominee for the handful of elections that pick a winner using single transferable vote~\citep{Richie23:Case}, which is in fact IoC. This happens even though consolidating party support behind a single candidate does not improve the chances of the party winning the election, and they could provide the voters with a wider range of choices just by letting \emph{all} of their willing candidates run in the general election.

To this end, we turn to the \emph{stronger} axiom of \emph{composition consistency (CC)}, introduced by \citet{Laffond96:Composition}, which dictates not just which clone sets win, but which clones among those sets win too. CC, as we will argue, also exposes the \emph{obviousness} of a rule's robustness against strategic nomination. When introducing CC, \citeauthor{Laffond96:Composition} were seemingly unaware of the IoC definition by \citet{Tideman87:Independence}, despite using equivalent (but differently-named) concepts. This, among other factors, has led the literature on IoC and CC to progress relatively independently, with few papers identifying them as comparable axioms. By studying these axioms in a unified framework, we hope to help dispel this ambiguity.

\subsection{Our Contributions}
\begin{itemize}[leftmargin=*]
    \item We clarify the relationship between IoC and CC---which has historically been ambiguous (see \Cref{sec:related})---by formally showing that CC is strictly more demanding (\Cref{prop:cctoioc}). 
    \item We provide (to the best of our knowledge) the first ever analysis of whether SCFs that are known to be IoC (\emph{e.g.}, Ranked Pairs, Beatpath/Schulze Method, and Split Cycle) also satisfy the stronger property of CC, thereby also establishing where each rule falls in our hierarchy of barriers to strategic nomination (\Cref{sec:ioc_rules}). While for most of these rules the answer is negative, we identify a variant of Ranked Pairs that satisfies CC. We connect our results to the literature on \emph{tournament solutions (TSs)}, and show that while not every CC TS is a CC SCF, a certain extension of Uncovered Set is.
    \item We introduce an efficient algorithm that modifies \emph{any} (neutral) SCF into a new rule satisfying CC, while preserving various desirable properties, \emph{e.g.}, Condorcet/Smith consistency, among others (\Cref{sec:cc_transform}). Our transformation relies on the observation that clone sets (which can be nested and overlapping) can be represented in PQ-trees~\citep{Elkind10:Clone}, allowing us to recursively zoom into the ``best'' clone set. We show that the resulting rule is poly-time computable if the starting SCF is, and fixed-parameter tractable in terms of the properties of the PQ-tree for all other SCFs (additionally proving the fixed-parameter tractability of all SCFs that are CC to begin with).
    \item We provide the first extension of CC to social preference functions (SPF) and prove that many of our characterization results generalize (\Cref{sec:spf}). Nevertheless, we give a negative result showing that no anonymous SPF can be CC, and discuss ways in which this can be circumvented.
    \item We formalize the connection of IoC/CC to strategic behavior by candidates via the model of \emph{strategic candidacy} \citep{Dutta01:Strategic} (\Cref{sec:oioc}). We show that if the candidates' preferences over each other are dictated by their clone structure, IoC rules ensure running in the election is a dominant strategy, hence achieving a stronger version of \emph{candidate stability}. However, IoC is not enough for \emph{obvious} strategy-proofness, which we show can be achieved by CC rules using our PQ-tree algorithm. 
\end{itemize} 
\subsection{Related Work}\label{sec:related}

\paragraph{Independence of Clones and Composition Consistency.} In this section, we give a conceptual overview of the literature of IoC and CC, and point out potential origins for some inconsistencies. We formalize these claims in \Cref{appsec:bg}. For definitions of all concepts sufficient for our results, see \Cref{sec:bg}.  

In order to identify candidates ``close'' to one another (at least according to voters), \citet{Tideman87:Independence} introduces the notion of \emph{clone sets} and the property of \emph{independence of clones (IoC)} for SCFs that are robust to changes in these sets. Seemingly unaware of \citeauthor{Tideman87:Independence}'s work, \citet{Laffond96:Composition} tackle a similar question by introducing the concept of \emph{components} of candidates and the property of \emph{composition consistency (CC)}. Importantly, \citeauthor{Laffond96:Composition} provide two separate definitions for components, one for a \emph{tournament} (a single asymmetric binary relationship on candidates) and one for a \emph{preference profile} (individual preferences of voters over candidates), and thus two different definitions of CC for \emph{tournament solutions} and \emph{SCFs}, which respectively map tournaments and profiles to winners. While \citeauthor{Laffond96:Composition}'s components for profiles is \emph{equivalent} to \citeauthor{Tideman87:Independence}'s clone sets, later work has described the former as ``more liberal''/``weaker'' \citep{Conitzer24:Position,Holliday24:Simple}, potentially due to focusing on components of tournaments instead (See our \Cref{ex:clones_v_components}).

The relationship between IoC and CC has been similarly unclear, despite the latter being strictly more demanding for SCFs (\Cref{prop:cctoioc}). Potentially due to SCFs taking a relatively small space in the work of \citet{Laffond96:Composition}, subsequent papers have primarily studied CC tournament solutions~\citep{Laslier97:Tournament,Brandt11:Fixed,Brandt18:Extending}, even describing components/CC to be ``analogue'' of clones/IoC for tournaments \citep{Elkind11:Cloning,Dellis13:Multiple,Karpov22:Symmetric}. Other works, while identifying a link between IoC and CC, have not been precise on their relationship~\citep{Brandt09:Some,Öztürk20:Consistency,Koray07:Self,Laslier12:Loser,Lederer24:Bivariate,Saitoh22:Characterization,Elkind17:What,Camps12:Continuous,Laslier16:Strategic}, describing them as ``similar'' notions \citep{Laslier97:Tournament,Heitzig10:Some} or ``related'' \citep{Brandt11:Fixed}.

There are papers that come much closer in identifying CC as a stronger axiom than IoC: working in a more general setting where voters' preferences are neither required to be asymmetric (ties are allowed)  nor transitive, \citet{Laslier00:Aggregation} introduces the notion of \textit{cloning consistency}, which he explains is weaker than CC and is the ``same idea'' as \citeauthor{Tideman87:Independence}'s IoC in voting theory. However, there are significant differences between \citeauthor{Laslier00:Aggregation}'s definition and IoC: first, his ``clone set'' definition requires every voter being \textit{indifferent} between any two alternatives in the set (as opposed to having the same relationship to all other candidates), and his cloning consistency dictates that if one clone wins, then so must every other member of the same clone set. Another property for tournament solutions named \emph{weak composition consistency} (this time in fact analogous to IoC) is discussed by \citet{Brandt18:Extending,Kruger18:Permutation}, and \citet{Laslier97:Tournament}, although none of them point out the connection to IoC. Perhaps the work that does the most justice to the relationship between CC and IoC is  by \citet{Brandl16:Consistent}, who explicitly state that \citeauthor{Tideman87:Independence}'s IoC (which they refer to as cloning consistency) is weaker than \citeauthor{Laffond96:Composition}'s CC. Since they work with the more general model of probabilistic social choice functions (PSCF), their observation that CC is stronger than IoC applies to our setting (although the adaptation is not obvious); we formalize this in \Cref{prop:cctoioc}. The PSCFs analyzed by \citeauthor{Brandl16:Consistent} are all non-deterministic; hence, to the best of our knowledge, no previous work has studied whether IoC SCFs also satisfy CC, which we do in \Cref{sec:ioc_rules}.

\paragraph{Strategic Cloning.}
Cloning and voting rules that are IoC are also of interest from a game theory perspective due to their connection to strategic manipulation in elections \citep{Sornat21:Fine,Elkind10:Clone,Delemazure24:Generalizing,Caragiannis10:Socially}. While most of (computational) social choice research treats the set of candidates as fixed, cloning inevitably goes beyond this assumption. To study manipulative behavior in settings with varying candidates, \citet{Dutta01:Strategic,Dutta02:Successsive} have initiated the study of \emph{strategic candidacy}, where candidates too have preferences over each other and may choose whether to run in the election. They define an SCF as \emph{candidate stable} if no candidate can benefit from not running given that all others are running, and show that this property is failed by every non-dictatorial SCF satisfying unanimity. Subsequent work has analyzed the computational aspects of strategic candidacy in extensions of this model, such as when candidates incur a small cost for running \citep{Obraztsova15:Strategic}, when candidates can decide to rejoin the election after dropping out 
(albeit with possibly less support) 
\citep{Polukarov15:Convergence}, or when both voters and candidates are behaving strategically \citep{Brill15:Strategic}. Many of these papers allude to, but do not formally define, a connection between strategic candidacy and cloning. We do so in \Cref{sec:oioc}, where we use the model of strategic candidacy to analyze the strength of robustness of IoC/CC rules against spoilage by clones.

Recently, \citet{Holliday23:Split} have introduced novel robustness criteria they call \emph{immunity to spoilers} and \emph{immunity to stealers}, which study the impact of adding candidates that are not necessarily clones (but must fulfill some other conditions). These criteria are incomparable in strength to IoC/CC, and while we focus exclusively on the impact of similar candidates (\emph{i.e.}, clones), we believe the methods we develop may be of future interest for studying different types of spoilers.
    
\section{Preliminaries}\label{sec:bg}
\paragraph{Profiles and clones.}\label{subsec:clones} We consider a set of \textit{candidates} $\cand$ with $|\cand|=m$ and a set of \textit{voters} $\voter = \{1,\ldots, n\}$. A {\em ranking} over $\cand$ is an asymmetric, transitive, and complete binary relation $\succ$ on $\cand$. Let $\mathcal{L}(\cand)$ denote the set of all rankings over $\cand$; $a \succ_r b$ indicates that $a$ is ranked above $b$ in a ranking $r$. Each voter $i \in \voter$ has a ranking $\sigma_i \in \mathcal{L}(\cand)$; we collect the rankings of all voters in a \textit{preference profile} $\profile \in \mathcal{L}(\cand)^n$. The next definition helps identify sets of similar candidates (according to voters). 

\begin{definition}[{\citealt[\oldS I]{Tideman87:Independence}; \citealt[Def. 4]{Laffond96:Composition}}]\label{def:clones}
    Given a preference profile $\profile$ over candidates $\cand$, a nonempty subset of candidates $\clone \subseteq \cand$ is a \emph{set of clones} with respect to $\profile$ if for each $a, b\in K$ and each $c\in \cand\setminus K$, no voter ranks $c$ between $a$ and $b$.
\end{definition}

All preference profiles admit two types of \textit{trivial} clone sets:\footnote{\citet{Tideman87:Independence} in fact excludes trivial clone sets. We use the definition from the work of \citet{Elkind10:Clone}.} (1) the entire candidate set $\cand$, and (2) for each $a\in \cand$, the singleton $\{a\}$. We call all other clone sets \textit{non-trivial}. For example, for the profile in \Cref{fig:eg_prof}, the only non-trivial clone set is $\{b,c\}$.

\begin{figure}[t]
    \centering
    \begin{tabular}{|c|c|c|c|}
        \hline
        6 voters & 5 voters & 2 voters & 2 voters \\ \hline
        \cgreen $b$ & \cred $d$  & \cyellow $a$ & \cyellow $a$ \\ \hline
       \cblue $c$ & \cblue $c$ & \cred $d$  & \cred $d$  \\ \hline
        \cyellow $a$ &\cgreen $b$ & \cgreen $b$ & \cblue $c$ \\ \hline
        \cred $d$ & \cyellow $a$& \cblue $c$ & \cgreen $b$ \\
        \hline
    \end{tabular}
    \caption{A preference profile. Columns show rankings, with the bottom row ranked last. The first row shows the number of copies of each ranking (\emph{e.g.}, leftmost column indicates 6 voters rank $b\succ c\succ a\succ d$).}\label{fig:eg_prof}
\end{figure} 

\paragraph{Social choice functions and axioms.}\label{subsec:axioms}
A \textit{social choice function (SCF)} is a mapping $f$ that, given a set of candidates $A$ and a profile $\profile$ over $A$,
outputs a nonempty subset of $A$; the candidates in $f(\profile)$ are the {\em winners} in $\profile$ under $f$. An SCF $f$ is {\em decisive} on $\profile$ if $|f(\profile)|=1$. \Cref{tab:scfs} contains the descriptions of the SCFs that we consider.\footnote{While we describe some SCFs by their winner determination \textit{procedures}, the SCFs themselves are the \textit{functions} that  output the respective winners, and these functions may be computed by other---possibly more efficient---algorithms.} 

We list some desirable properties (\emph{axioms}) for SCFs. For example, an SCF is \emph{neutral} (resp., \emph{anonymous}) if its output is robust to relabeling the candidates (resp., voters); for a formal definition, see {\citet[Def. 2.4, 2.5]{Brandt16:Handbook}}. Further, an SCF $f$ satisfies Smith (resp., Condorcet) consistency if $f(\profile) \subseteq Sm(\profile)$ for all $\profile$ (resp., for all $\profile$ with $|Sm(\profile)|=1$), where $Sm$ is defined in Table~\ref{tab:scfs}.

Next, we will present two axioms that both aim to capture the idea of robustness against strategic nomination. In what follows, we write $\profile \setminus \cand'$ to denote the profile obtained by removing the elements of $\cand' \subset \cand$ from each voter's ranking in $\profile$ while preserving the order of all other candidates.

\begin{definition}[\citealt{Zavist89:Complete}]\label{def:ioc}
An SCF $f$ is \emph{independent of clones (IoC)} if for each profile $\profile$ over $A$ and each non-trivial clone set $\clone \subset A$ with respect to $\profile$,

(1) for all $a \in \clone$, 
\[
    \clone \cap f(\profile) \neq \emptyset \Leftrightarrow \left(\clone \setminus \{a\}\right) \cap f(\profile \setminus \{a\}) \neq \emptyset;  
\]
\indent (2) for all $a \in \clone$ and all $b \in A \setminus \clone$,
\[
    b \in f(\profile) \Leftrightarrow b \in f(\profile \setminus \{a\}).
\]
\end{definition}

Intuitively, IoC dictates that deleting one of the clones in a non-trivial clone set $K$ must not alter the winning status of $K$ as a whole, or of any candidate not in $K$. In particular, it implies that one cannot substantively change the election outcome by nominating new copy-cat candidates. 
 \begin{figure}[t]
    \centering
    \begin{tabular}{|c|c|c|c|}
        \hline
    3 voters &  2 voters & 4 voters & 3 voters \\
    \hline
    \cyellow$a_1$ & \cred$a_2$ & \cgreen$b$ & \cblue$c$ \\ \hline
    \cred$a_2$ & \cyellow$a_1$ & \cblue$c$ & \cred$a_2$ \\ \hline
    \cgreen$b$ & \cgreen$b$ & \cred$a_2$ & \cyellow$a_1$ \\ \hline
    \cblue$c$ & \cblue$c$ & \cyellow$a_1$ & \cgreen$b$ \\
    \hline
    \end{tabular} $\xRightarrow[\text{remove }a_2]{}$ \begin{tabular}{|c|c|c|}
        \hline
    5 voters & 4 voters & 3 voters \\  
    \hline
    \cyellow$a_1$ & \cgreen$b$ & \cblue$c$ \\ \hline
      \cgreen$b$ & \cblue$c$ & \cyellow$a_1$ \\ \hline
     \cblue$c$  &  \cyellow$a_1$ & \cgreen$b$ \\ 
    \hline
    \end{tabular} \caption{(Left) Example profile $\profile$. (Right)  $\profile\setminus \{a_2\}$.}\label{fig:bg_eg}
\end{figure}

\begin{example}\label{eg:ioc}
    In the profile $\profile$ in Fig.~\ref{fig:bg_eg} (left), $K = \{a_1,a_2\}$ is a clone set.
    Plurality Voting (PV) outputs $b$ as the unique winner.
    However, $\mathit{PV}(\profile \setminus \{a_2\}) = \{a_1\}$, since with $a_2$ gone, $a_1$ now has 5 voters ranking it first (\Cref{fig:bg_eg}, right). This violates both conditions~(1) and~(2) from Definition~\ref{def:ioc}. 

    In contrast, $\STV$ eliminates
    $a_2$ (whose votes then transfer to $a_1$), then $c$ and finally $b$, 
    so that $a_1$ is elected; moreover, it produces the same result on $\profile\setminus\{a_2\}$.
    This is in line with the fact that $\STV$ is IoC \citep{Tideman87:Independence}. 
\end{example}

To formulate the second axiom that deals with strategic nomination, we must first introduce a few additional concepts.
  \begin{definition}\label{def:clone_grouping}
Given a preference profile $\profile$ over candidates $\cand$, a set of sets $\decomp=\{\clone_1,\clone_2,\ldots,\clone_\ell\}$, where $\clone_i \subseteq A$ for all $i\in [\ell]$, is a \emph{(clone) decomposition} with respect to $\profile$ if
\begin{enumerate}
    \item $\decomp$ is a partition of $\cand$ into pairwise disjoint subsets, and
    \item Each $\clone_i$ is a non-empty clone set with respect to $\profile$.
\end{enumerate}
\end{definition}
Every profile has at least two decompositions: the \textit{null} decomposition $\decomp_{null}=\{A\}$ and the \textit{trivial} decomposition $\decomp_{triv}=\{ \{a\}\}_{a \in A}$. Given a decomposition $\decomp$ with respect to $\profile$, for each $i \in N$ let $\sigma^\decomp_i$ be voter $i$'s ranking over the sets in $\decomp$; this is well-defined, since each clone set forms an interval in $\sigma_i$. The profile $\profile^\decomp = \{\sigma^\decomp_i\}_{i \in N}$ over $\decomp$ is called the \textit{summary} of $\profile$ with respect to the decomposition $\decomp$. For each $K \in \decomp$, we write $\profile|_{K}$ to denote the restriction of $\profile$ to $K$, so that $\profile|_{K} \equiv \profile \setminus (A \setminus K)$. 

\begin{definition}\label{def:gloc} 
    The \emph{composition product} function of an SCF $f$ is a function $\gloc_f$ that takes as input a profile $\profile$ and a clone decomposition $\decomp$ with respect to $\profile$ and outputs $\gloc_f(\profile, \decomp) \equiv \bigcup_{K \in f\left(\profile^\decomp\right)  }f(\profile|_\clone)$.
\end{definition}

Intuitively, $\gloc_f$ first runs the input SCF $f$ on the summary (as specified by $\decomp$), collapsing each clone set into a meta-candidate $\clone_i$. It then ``unpacks'' the clones of each winning clone set, and runs $f$ once again on each.
\begin{example}\label{eg:GLOC}
    For the profile $\profile$ from Figure~\ref{fig:bg_eg} (left), it holds that $\decomp=\{K_a,K_b,K_c\}$ with $K_a=\{a_1,a_2\}$, $K_b=\{b\}$, $K_c=\{c\}$ is a valid clone decomposition with respect to $\profile$. Figure~\ref{fig:bg_eg_2} shows $\profile^\decomp$ and $\profile|_{\clone_a}$. We have $\STV(\profile^\decomp)=\{\clone_a\}$ and $\STV(K_a)=\{a_2\}$, implying $\gloc_{\STV}(\profile, \decomp) =\{a_2\}$.
\end{example}

Together, \Cref{eg:ioc,eg:GLOC} imply that $\STV(\profile) \neq \gloc_{\STV}(\profile, \decomp)$ for this $\profile$ and $\decomp$; \emph{i.e.}, that $\STV$ does not respect this clone decomposition---even though the winners in $\STV$ and $\gloc_\STV$ are from the same clone set. We now state the composition consistency axiom, which precisely requires a rule to respect all possible clone decompositions.

\begin{definition}[{\citealt[Def. 11]{Laffond96:Composition}}]\label{def:oioc}
      A neutral\footnote{\citet{Laffond96:Composition} define composition consistency (CC) only for neutral SCFs; this is without loss of generality, as they treat $\profile^\decomp$ as a profile over candidates $\{1,2,\ldots, |\decomp|\}$, in which case CC automatically implies neutrality by the trivial decomposition. \citet{Brandl16:Consistent} instead use a definition where $\profile^\decomp$ is simply $\profile$ with all but one candidate removed from each $K_i$; nevertheless, they show that in this model too CC implies neutrality (their Lemma 1). We explicitly state this as a prerequisite for simplicity.} SCF $f$ is \emph{composition-consistent (CC)} if for all preference profiles $\profile$ and all clone decompositions $\decomp$ with respect to $\profile$, we have $f(\profile) =\gloc_f(\profile, \decomp)$.
\end{definition}

\begin{figure}[t]
\centering

 \begin{tabular}{|c|c|c|}
        \hline
    5 voters & 4 voters & 3 voters \\
    \hline
    \corange$\clone_a$  & \cgreen$\clone_b$ & \cblue$\clone_c$ \\ \hline
    \cgreen$\clone_b$ & \cblue$\clone_c$ & \corange$\clone_a$ \\ \hline
   \cblue$\clone_c$ & \corange$\clone_a$ & \cgreen$\clone_b$ \\
    \hline
    \end{tabular} \quad \quad
  \begin{tabular}{|c|c|}
        \hline
    3 voters & 9 voters \\
    \hline
  \cyellow{$a_1$} & \cred{$a_2$} \\
  \hline
  \cred{$a_2$} & \cyellow{$a_1$} 
  \\
  \hline
    \end{tabular}

    \captionsetup{width=.8\linewidth}
    \caption{(Left) $\profile^\decomp$, where clone sets from $\profile$ in \Cref{fig:bg_eg} are condensed into candidates $\clone_a$, $\clone_b$, and $\clone_c$. (Right) $\profile|_{\clone_a}$, where $\profile$ is limited to members of $\clone_a$.}
    \label{fig:bg_eg_2}
\end{figure}

CC dictates that an SCF should choose the ``best'' candidates from the ``best'' clone sets. In contrast, IoC is much more permissive when it comes to choosing a candidate from a best clone set. Indeed, in \Cref{prop:cctoioc} we show that CC implies IoC. On the other hand, \Cref{eg:ioc,eg:GLOC} demonstrate that the converse is false: they show that $\STV$, which is IoC, is not CC. Later, we analyze other IoC rules to determine whether they are CC (\Cref{sec:ioc_rules}).  

\paragraph{Social choice functions considered in this paper.}\label{subsec:rules}
Prior work has shown each SCF in \Cref{tab:scfs} (with the exception of $\mathit{PV}$) to be IoC (see \citet{Holliday23:Split} for an overview). For some, winner determination may require tie-breaking (\emph{e.g.}, under $\STV$, candidates may tie for the lowest plurality score). We define the output of such SCFs as the set of candidates that win for \textit{some} tie-breaking rule, also called \emph{parallel-universes tiebreaking}~\citep{Conitzer09:Preference}. Crucially, this variant of $\RP$ is \textit{not} IoC~\citep{Zavist89:Complete}, a nuance we will address in detail in \Cref{subsec:rp}. Lastly, $\Sm, \Sz, \UCg$ are SCF extensions of \emph{tournament solutions} that are known to either fail or satisfy CC. However, whether they satisfy CC as SCFs is a more subtle issue, which we address in \Cref{sec:majoritarian}.

\begin{table*}[h!]
    \begin{threeparttable}
        \centering
        \begin{tabular}{|m{0.1548\linewidth}|m{0.0497\linewidth}|m{0.7095\linewidth}|}
            \hline  \centering \emph{Name of SCF} & \centering  $f$ &  \emph{Description of the SCF's output on input profile $\profile$} \\
            \hline\hline  \cpink \centering Plurality &\cpink \centering $\mathit{PV}$ & \cpink Outputs the candidate(s) ranked first by the most number of voters.\\
            \hline \centering \cellcolor{SeaGreen!20} Single~Transferable Vote &\cellcolor{SeaGreen!20} \centering $\STV$ & \cellcolor{SeaGreen!20} At each round, the candidate ranked top by the fewest voters is eliminated. Eventually a single candidate remains, becoming the winner.  \\\hline \centering  \cpink  \text{Ranked Pairs} \citep{Tideman87:Independence,Zavist89:Complete}& \cpink \centering $\RP$ & \cpink    Given a profile $\profile$ over candidates $\cand =\{a_i\}_{i \in [m]}$, construct the \textit{majority matrix} $M$, whose $ij$ entry is the number of voters who rank $a_i$ ahead of $a_j$ minus those who rank $a_j$ ahead of $a_i$. Construct a digraph over $\cand$ by adding edges for each $M[ij]\geq 0$ in non-increasing order, skipping those that result in a cycle. The winner is the source node.\\
            \hline \centering \cellcolor{SeaGreen!20} Beatpath (Schulze Method) \citep{Schulze10:New} & \cellcolor{SeaGreen!20} \centering  $\BP$  & \cellcolor{SeaGreen!20} Construct $M$ as in $\RP$, and the corresponding weighted digraph over $\cand$ without skipping edges\tnote{$\dagger$}. Let $S[i,j]$ be the width (min. weight edge) of the widest path from $a_i$ to $a_j$, computed, \emph{e.g.}, with the Floyd--Warshall algorithm. Then $a_i$ is a winner iff $S[i, j] \geq S[j,i]$ for all $j \in [m]$.\\
            \hline \cpink  \centering \citet{Schwartz86:Logic} set (GOCHA)&\cpink \centering $\Sz$ &   \cpink Given $\profile$, we say that $B \subseteq A$ is \emph{undominated} if no $a \in A \setminus B$ pairwise defeats (preferred to by a strict majority of voters) any $b \in B$. The winners are the union of minimal (by inclusion) undominated sets.\\
            \hline \centering \cellcolor{SeaGreen!20} \citet{Smith73:Aggregation} set (GETCHA) &\cellcolor{SeaGreen!20} \centering $\Sm$ & \cellcolor{SeaGreen!20} Outputs the smallest set of candidates who all pairwise defeat every candidate outside the set.\\ \hline \centering   \cpink Alternative-Smith~\citep{Tideman06:Collective}&  \cpink \centering $\AS$ &  \cpink (1) Eliminate all candidates not in $\Sm(\profile)$. (2) In the remaining profile, eliminate the candidate ranked top by the fewest voters. Repeat (1)-(2) until a single candidate remains; this is the winner.\\
            \hline \cellcolor{SeaGreen!20} \centering  Split Cycle \citep{Holliday23:Split}& \cellcolor{SeaGreen!20} \centering $\SC$ & \cellcolor{SeaGreen!20}Construct $M$ as in $\RP$ and the corresponding weighted digraph $G$ over $\cand$ without skipping edges. For each \textit{simple cycle} (cycles visiting each vertex at most once) in $G$, label the edge(s) with the smallest weight in that cycle. Discard all labeled edges (at once) to get $G'$. The winners are the candidates with no incoming edge in $G'$. \\
            \hline \centering  \cpink Uncovered Set \citep{Gillies59:Solutions} &  \cpink \centering $\UCg$ & \cpink Given $\profile$ and $a,b \in \cand$, we say that $a$ \emph{left-covers} $b$ if any $c \in \cand$ that pairwise defeats $a$ also pairwise defeats $b$. The winners are all $a \in \cand$ such that there is no $b \in \cand$ that left-covers \emph{and} pairwise defeats $a$.
            \\\hline
        \end{tabular}
        \caption{SCFs considered in this paper. Second column indicates our notation for the SCF as a function.}
        \label{tab:scfs}
        \begin{tablenotes}
            \footnotesize
                \item[$\dagger$]More generally, $\BP$ can be defined with various choices for edge weights \citep{Schulze10:New}. We use the ``margin'' variant; our results easily generalize to others.
        \end{tablenotes}
    \end{threeparttable}
\end{table*}
\section{Analysis of IoC Social Choice Functions}\label{sec:ioc_rules}
In this section, we analyze whether the IoC rules in \Cref{tab:scfs} satisfy CC. The answer turns out to be positive for $\RP$ with a specific tie-breaking rule by \citet{Zavist89:Complete} and $\UCg$, but negative for all other SCFs. We first formalize the CC to IoC relationship (\emph{cf.} \citealt{Brandl16:Consistent}).
\begin{restatable}{proposition}{cctoioc}\label{prop:cctoioc}
    If a given SCF is CC, then it is also IoC.
\end{restatable}

All omitted proofs are in the appendix. We show that the converse of Proposition~\ref{prop:cctoioc} is not true.

\begin{restatable}{theorem}{ccfails}\label{thm:cc_fails}
    $\STV$, $\BP$, $\AS$, and $\SC$ all fail CC.
\end{restatable}
\begin{proof}
Since CC requires $f(\profile)=\gloc_f(\profile, \decomp)$ for \textit{all} profiles $\profile$ and clone decompositions $\decomp$, a single counterexample is sufficient to show that a rule fails CC. The statement for $\STV$ follows from \Cref{eg:ioc,eg:GLOC}. For the $\profile$ and $\decomp$ from these examples, $\AS(\profile)=\STV(\profile)$ and $\gloc_{\AS}(\profile, \decomp)=\gloc_{\STV}(\profile, \decomp)$; hence they also show $\AS$ is not CC. For $\BP$ and $\SC$, we use the profile from Fig.~\ref{fig:eg_prof} (say, $\profile'$), with $\decomp'=\{\{a\},\{b,c\}, \{d\}\}$. We have $\BP(\profile')=\SC(\profile')=\{b,c\}$, whereas $\gloc_{\BP}(\profile',\decomp)=\gloc_{\SC}(\profile',\decomp')=\{b\}$, so both $\BP$ and $\SC$ fail CC;
see \Cref{appsec:extended_ioc} for detailed calculations.
\end{proof}

\subsection{Ranked Pairs}\label{subsec:rp}
Our definition of SCFs deals with ties by returning all candidates that win via \textit{some} tie-breaking method. In particular, for $\RP$ the majority matrix $M$ may contain ties, so we need a \textit{tie-breaking order over unordered pairs} to decide the order of adding edges to the digraph. \citet{Tideman87:Independence} originally defined Ranked Pairs as returning all candidates that win for some tie-breaking order (we refer to this rule as $\RP$); later, \citet{Zavist89:Complete} showed that this rule is \emph{not} IoC. 

\begin{proposition}[\citealt{Zavist89:Complete}]\label{prop:rpfailioc}
        $\RP$ fails IoC.
\end{proposition}

By Proposition~\ref{prop:cctoioc}, this also implies $RP$ fails CC. \citet{Zavist89:Complete} propose breaking ties based on the vote of a fixed voter $i \in \voter$, which makes $\RP$ satisfy IoC.  
Specifically, they use $\sigma_i$ to construct a \emph{tie-breaking ranking} $\Sigma_i$ over unordered pairs in $A$ as follows: (1) order the elements within each pair according to $\sigma_i$; (2) rank the pairs according to $\sigma_i$'s ranking of their first elements; (3) rank pairs with the same first element according to $\sigma_i$'s ranking of the second elements. 
\begin{example}
    For $\cand=\{a,b,c\}$ and $\sigma_i: a\succ b \succ c $ we get
      $\Sigma_i: \{a,a\}\succ\{a,b\}\succ\{a,c\} \succ\{b,b\}\succ \{b,c\}\succ\{c,c\}.$
\end{example}

Using $\Sigma_i$, we can construct a complete \emph{priority order} $\mathcal{L}$ over ordered pairs: pairs are ordered (in non-increasing order) according to $M$, with ties broken by $\Sigma_i$ (in the special case where $M[a,b]=0$, we rank $(a,b) \succ_{\mathcal{L}} (b,a)$ if and only if $a \succ_i b$). Then, \textit{the Ranked Pairs method using voter $i$ as a tie-breaker} (which we will call $\RP_i$) adds edges from $M$ to a digraph according to $\mathcal{L}$, skipping those that create a cycle. \citet{Zavist89:Complete} show that $\RP_i$ is IoC. We now strengthen this result.

\begin{restatable}{theorem}{rpcc}\label{thm:rp}
    $\RP_i$ is CC for any fixed $i \in N$. 
\end{restatable}

\begin{proof}[Proof sketch] The proof relies on an equivalence between the topological orders of the final $RP$ graphs and \emph{stacks} over $\cand$, which are rankings $r$ where $a\succ_r b$ implies there is a path in $M$ from $a$ to $b$ consistent with the ranking $r$ and with each link at least as strong as $M[b,a]$ \citep{Zavist89:Complete}. We extend this equivalence to the specific case of stacks with respect to a priority order $\mathcal{L}$, and show that this definition is satisfied by the $\RP_i$ ranking, its summary using any $\decomp $, and its restriction to any clone set. This allows us to establish an agreement between $\RP_i$ and $\gloc_{\RP_i}$, proving $\RP_i$ is CC. The full proof can be found in \Cref{appsec:rpi}.
\end{proof}

Moreover, $\RP_i$ is poly-time computable, whereas the outputs of $\RP$ are \NP-hard to compute~\citep{Brill12:Price}. However, for any fixed $i$ this rule breaks anonymity, \emph{i.e.}, it fails to treat all voters equally.  \citet{Holliday23:Split} suggest (based on personal communication with Tideman) returning 
$\RP_N(\profile) \equiv \bigcup_{i \in N} \RP_i(\profile)$, \emph{i.e.}, 
declaring an $a \in \cand$ to be a winner if and only if $a \in RP_i(\profile)$ for \emph{some} $i \in N$.  This modification recovers anonymity while preserving IoC and tractability, but we show that it loses CC.
\begin{proposition}\label{prop:rpn}
    $\RP_N$ is IoC, but not CC.
\end{proposition}
\begin{proof}
    IoC holds since $RP_i$ is IoC for all $i \in N$. For failure of CC, consider $\profile$ with $n=2$, $\cand=\{a,b,c\}$, and rankings $a \succ_1 b \succ_1 c$ and  $c \succ_2 b \succ_2 a$. We have $RP_N(\profile)=\{a,c\}$. Using decomposition $\decomp=\{K, \{c\}\}$ with $K=\{a,b\}$, we get $RP_N(\profile^\decomp)=\{K, \{c\}\}$ and $RP_N(\profile|_K)=\{a,b\}$. Hence, $\gloc_{RP_N}(\profile, \decomp)= \{a,b,c\} \neq RP_N(\profile)$.
\end{proof}

Thus, Ranked Pairs without tie-breaking ($\RP$) is neither IoC, CC, nor tractable. Using a voter to break ties ($\RP_i$), we get all three, but lose anonymity. Recovering anonymity by taking a union over all voters ($\RP_N$) keeps IoC and tractability, but loses CC.\footnote{For probabilistic SCFs (PSCFs), a tempting approach is to pick an $i\in N$ uniformly at random and return $\RP_{i}(\profile)$. The counterexample from Prop.~\ref{prop:rpn} also shows that this variant fails the CC definition for PSCFs given by \citet{Brandl16:Consistent}.}  

\usetikzlibrary{backgrounds}

\begin{figure}[t]
    \centering
\begin{tikzpicture}[on background layer]

    \definecolor{set1}{RGB}{255, 204, 204}
    \definecolor{set2}{RGB}{204, 255, 204}
    \definecolor{set3}{RGB}{204, 204, 255}

    \fill[set1] (-4, -1.7) rectangle (4, 1.2);
    \draw[black] (-4, -1.7) rectangle (4, 1.2);
    \node at (0,0.95) {\textbf{\textit{Satisfies neither CC nor IoC}}};
    \node at (0,0.6) {
        \begin{tabular}{l l l}
            $\bullet RP$ &  $\bullet PV$ & $\bullet \UCf^\dagger$
        \end{tabular}
    };

    \fill[set2] (-3.5, -1.6) rectangle (3.5, 0.3);
    \draw[black] (-3.5, -1.6) rectangle (3.5, 0.3);
    \node at (0,0.05) {\textbf{\textit{Satisfies IoC}}};
    \node at (0,-0.35) { 
        \begin{tabular}{@{\hskip 0.2cm} l @{\hskip 0.2cm} l @{\hskip 0.2cm} l @{\hskip 0.2cm} l @{\hskip 0.2cm} l @{\hskip 0.2cm} l @{\hskip 0.2cm} l @{\hskip 0.2cm} l @{\hskip 0.2cm} l}
            $\bullet STV$ & $\bullet AS$ & $\bullet BP$ & $\bullet RP_N$ & $\bullet SC$ & $\bullet Sz^\dagger$ & $\bullet Sm^\dagger$
        \end{tabular}
    };

    \fill[set3] (-2, -1.5) rectangle (2, -0.6); 
    \draw[black] (-2, -1.5) rectangle (2, -0.6);
    \node at (0,-0.85) {\textbf{\textit{Satisfies CC}}}; 
    \node at (0,-1.2) { 
        \begin{tabular}{l l}
             $\bullet RP_i$ & $\bullet \UCg^\dagger$
        \end{tabular}
    };
    
\end{tikzpicture}
\caption{Behavior of SCFs from \Cref{tab:scfs} w.r.t~IoC/CC. $\dagger$ indicates majoritarian SCFs.}\label{fig:diagram}
\end{figure}

\subsection{Aside: Majoritarian SCFs}\label{sec:majoritarian}

While \emph{tournaments} (complete and asymmetric binary relationships over $\cand$) are not the main focus of this paper, it is worth briefly discussing how our results in this section relate to prior results on CC \emph{tournament solutions} (TSs), which map tournaments to sets of winners. As noted in \Cref{sec:related}, \citet{Laffond96:Composition}  introduce two separate definitions of components, in tournaments and in profiles (see \Cref{appdef:component_tour,appdef:component_prof} in our \Cref{appsec:bg}), and thus two separate definitions of CC for SCFs and for TSs. Subsequent work has primarily focused on the latter, showing TSs such as uncovered set, the minimal covering set, and the Banks set are CC~\citep{Laffond96:Composition,Laslier97:Tournament}. 

Of course, if $|\voter|$ is odd, the pairwise defeats in $\profile$ define a tournament, so any TS can be thought of as an SCF that maps $\profile$ to the winners of this induced tournament. However, for a TS to be well-defined as an SCF (without assuming odd $|N|$), it must be \emph{extended} to cases where the pairwise defeat relationship may contain ties (equivalently, to incomplete tournaments). Such induced SCFs are called \emph{majoritarian}. For example, $\UCg$ in \Cref{tab:scfs} is an extension of the TS \emph{uncovered set}. Another extension of the same TS follows from the work of \citet{Fishburn77:Condorcet}, and is defined as follows (recall that we say $a$ \emph{left-covers} $b$ if any $c$ that pairwise defeats $a$ also pairwise defeats $b$):
\begin{align*}
    \UCf(\profile)=\{a\in \cand: \nexists b \in A\text{ such that }b\text{ left covers }a\text{ but }a\text{ does not left-cover }b\}.
\end{align*}
It can be checked that $\UCf(\profile)=\UCg(\profile)$ whenever pairwise defeats have no ties, \emph{i.e.}, they are extensions of the same TS. Crucially, even though uncovered set is CC as a TS, $\UCf$ is not even IoC~\citep{Holliday23:Split}! This demonstrates that \textbf{a TS being CC is not sufficient for its SCF extension to be CC}. As we show next, $\UCg$ (\Cref{tab:scfs}) in fact maintains the CC property.
\begin{restatable}{proposition}{ucg}\label{prop:ucg}
    $\UCg$ is CC.
\end{restatable}

The disparity between $\UCf$ and $\UCg$ motivates future work in investigating whether other TSs known be CC can be extended into SCFs while maintaining CC. Existing negative results for TSs, on the other hand, readily generalize to any of their extensions. This is because \citet{Laffond96:Composition} show that for any tournament and a decomposition $\decomp$ into its (tournament) components, there exists some preference profile (that induces this tournament) for which $\decomp$ is once again a valid decomposition (their Prop. 1); this can be used to show that the CC definitions for TSs and their SCF interpretations coincide under the odd $|N|$ assumption \cite[Prop. 2]{Laffond96:Composition}. Since $\Sm$ and $\Sz$ (\Cref{tab:scfs}) are both SCF extensions of the TS \emph{top cycle}, which is not CC, we get:
\begin{proposition}[{Consequence of \citet[Props. 2, 5]{Laffond96:Composition}}] $\Sm$ and $\Sz$ are not CC.
    
\end{proposition}
\noindent For a summary of our results from \Cref{sec:ioc_rules}, see \Cref{fig:diagram}.

\section{CC Transformation}
\label{sec:cc_transform}

As shown in \Cref{sec:ioc_rules}, almost all IoC SCFs considered, except for $\RP_i$ and $\UCg$, fail CC. Out of these two, $\RP_i$ violates anonymity, and $\UCg$ selects a unique winner only if there is a Condorcet winner ($|\Sm(\profile)|=1$) \citep{Holliday23:Split}, implying it is not decisive for any $\profile$ without this property. As both anonymity and decisiveness are fundamental properties for SCFs (for ensuring the result is fair and conclusive, respectively), having CC rules that also satisfy them would be desirable, especially considering the strong guarantees of CC against strategic behavior, which we will further discuss in \Cref{sec:oioc}. To this end, we show that any neutral SCF can be efficiently modified to satisfy CC, while preserving its desirable properties, including anonymity and decisiveness, among others.

\subsection{Background: Clone Structures and PQ-Trees}\label{subsec:pqtrees}
For a profile $\profile$, \citet{Elkind10:Clone} define the \emph{clone structure} $\family(\profile) \subseteq \mathcal{P}(\cand)$ as the family of \textit{all} clone sets with respect to $\profile$. For example, for $\profile$ from Fig.~\ref{fig:eg_prof}, $\family(\profile)=\{\{a\},\{b\}, \{c\}, \{d\}, \{b,c\}, \{a,b,c,d\}\}$. They identify two types of \emph{irreducible} clone structures: a \emph{maximal} clone structure (also called a \emph{string of sausages}) and a \emph{minimal} clone structure (also called a \emph{fat sausage}). A string of sausages arises when each ranking in  $\profile$ is either a fixed linear order (say, $\sigma_1 : a_1 \succ a_2 \succ \cdots \succ a_m$) or its reversal. In this case, $\family(\profile) = \{ \{a_k\}_{i\le k\le j}: i\le j\}$, \emph{i.e.}, all intervals in $\sigma_1$. The \emph{majority ranking} of the string of sausages is $\sigma_1$ or its reverse, depending on which one appears more frequently in $\profile$. A fat sausage occurs when $\family(\profile) = \{A\} \cup \{\{a_i\}\}_{i \in [m]}$, \emph{i.e.}, the structure only has the trivial clone sets. 

Our CC transformation uses \emph{PQ-trees}: a data structure first defined by \citet{Booth76:Testing} and later used by \citet{Elkind10:Clone} to represent clone sets.
Here, we present the definitions required for our construction; for the full treatment, see \citet{Elkind10:Clone}
(and our \Cref{subsec:extended_pqtrees}).

A {\em PQ-tree} $T$ over $\cand$ is an ordered tree whose leaves correspond to the elements of $\cand$. To represent a clone structure $\family(\profile)$ as a PQ-tree, we iteratively identify irreducible subfamilies of $\family(\profile)$, and collapse them into a single meta-candidate. 
If the subfamily corresponds to a fat sausage, we group its members under an internal node of type P, denoted as a $\odot$-product of its children.
On the other hand, if the subfamily corresponds to a string of sausages, we group its members under an internal node of type Q, denoted as a $\oplus$-product of its children. 
In rankings compatible with $\family(\profile)$, the children of a P-node can be permuted arbitrarily; the order of the children of a Q-node must follow its majority ranking or its reversal. Crucially, the order of collapsing is not important, as the irreducible subfamilies of a clone structure are non-overlapping \cite[Prop. 4.2.]{Elkind10:Clone}. 

\begin{example}\label{ex:pq-tree}
    Let $\profile$ be a profile on $A=\{a, b, c, d\}$ with two rankings: $a\succ b \succ c \succ d$ and $d \succ c \succ a \succ b$.
    Then, $\family(\profile) = \{\{a\}, \{b\}, \{c\}, \{d\}, \{a,b\}, \{c, d\}, \{a, b, c\}, A\}$.
    Collapsing the irreducible subfamily $K_1 = \{a, b\}$, the updated $\family(\profile)$  is $\{\{c\}, \{d\}, \{K_1\}, \{c, d\}, \{K_1, c\}, \{K_1, c, d\}\}$.
    With size two, $K_1$ is both a string of sausages and a fat sausage; by convention we treat it as a fat sausage (\emph{i.e.}, of type \emph{P}).
    The updated $\family(\profile)$ is a string of sausages itself, so the algorithm terminates by picking the root of the tree as a type \emph{Q} node. The resulting PQ-tree is illustrated in \Cref{fig:tree} (left).
    
    \begin{figure}[t]
        \centering
        \begin{tikzpicture}
            [
              grow=down,
              sibling distance=4em,
              level distance=2.5em,
              edge from parent/.style={draw, -latex},
              every node/.style={draw, rectangle, rounded corners, align=center, scale=0.8}
            ]
            \node {$(a \odot b) \oplus c \oplus d$}
              child { node {$a \odot b$}
                child { node {$a$} }
                child { node {$b$} }
              }
              child { node {$c$} }
              child { node {$d$} };
        \end{tikzpicture} \quad \quad \quad \quad
        \begin{tikzpicture}
            [
              grow=down,
              sibling distance=4em,
              level distance=2.5em,
              edge from parent/.style={draw, -latex},
              every node/.style={draw, rectangle, rounded corners, align=center, scale=0.8}
            ]
            \node {$b\odot (a_1 \odot a_2) \odot c$}
              child { node {$b$} }
              child { node {$a_1 \odot a_2$}
                child { node {$a_1$} }
                child { node {$a_2$} }
              }
              child { node {$c$} };
        \end{tikzpicture}
        \caption{(Left) The PQ-tree representing $\family(\profile)$ from  \Cref{ex:pq-tree} . (Right) The PQ-tree of $\profile$ from \Cref{fig:bg_eg}.}\label{fig:tree}
    \end{figure}
\end{example}

We now formulate two useful properties of PQ-trees, as observed by 
\citet{Cornaz13:Kemeny}.

\begin{restatable}[\citealt{Cornaz13:Kemeny}]{lemma}{pqpoly} \label{lemma:pq-poly}
    PQ-trees can be constructed in $O(|\voter|\cdot|\cand|^3)=O(nm^3)$ time.
\end{restatable}

\begin{restatable}[{\citealt[Prop. 4]{Cornaz13:Kemeny}}]{lemma}{pqclones} \label{lemma:pq_clones}
    Given $\profile$ and its PQ-tree $T$, a set of candidates $K \subseteq A$ is a clone set if and only if it satisfies one of the following: (1) $K$ exactly corresponds to the leaves of a subtree in $T$ (where leaves are also subtrees of size 1), or (2) $K$ exactly corresponds to the leaves of a set of subtrees the roots of which are adjacent descendants of a Q-node in $T$. 
\end{restatable}

\citet{Cornaz13:Kemeny} use PQ-trees to prove fixed-parameter tractability of computing a \emph{Kemeny ranking} of a profile, which obeys a special case of CC (see \Cref{sec:spf} for CC properties of social preference functions, which return aggregate rankings over $\cand$ rather than subsets). Similarly, \citet{Brandt11:Fixed} show that any CC tournament solution is fixed-parameter tractable with respect to the properties of an analogous construct for tournaments (decomposition trees) by running the TS recursively on the nodes of the tree.\footnote{A similar idea of breaking the computation into subsets of candidates is used by \citet{Conitzer06:Computing} for computing Slater rankings of a profile. The eligible subsets in this case (which the author simply refers to as sets of \emph{similar candidates}) exactly correspond to the components of the tournament induced by the profile (see our \Cref{appdef:component_tour}).} Our key observation is that even if we start with an SCF that does \emph{not} satisfy CC (or any weaker version thereof), running it on the PQ-tree enables us to define a \emph{new} SCF that is in fact CC, while maintaining many desirable properties of the original SCF. In the following section, we introduce this transformation and formalize its properties.

\subsection{CC-Transformed SCFs}\label{subsec:transform}
We now present Algorithm~\ref{alg:cc-transform}, which is based on a similar algorithm by \citet{Brandt11:Fixed} for efficiently implementing CC tournament solutions. Given $T=PQ(\profile)$ (the PQ-tree for a profile $\profile$), we refer to its nodes by the subset of candidates in their subtrees. For $\node \subseteq \cand$, is\_p\_node($\node,T$) returns True (resp., False) if the node corresponding to $B$ in $T$ is a P- (resp., Q-) node, raising an error if no such node exists. Let decomp($\node,T$) return the clone decomposition $\decomp$ corresponding to node $\node$, where each $K\in \mathcal{K}$ is a child node of $\node$ (these are clone sets by Lemma~\ref{lemma:pq_clones}).
For example, if $T$ is the tree from Fig.~\ref{fig:tree}(left),
decomp($A,T)=\{\{a,b\}, \{c\}, \{d\}\}$ and  decomp($\{a,b\},T)=\{\{a\}, \{b\}\}$.  For a Q-node $\node$, let $\node_i(\node,T)$ be the $i$-th child of $\node$ according to its majority ranking $\sigma_1$. 

\begin{definition}\label{def:cc-transform}
    Given an SCF $f$, the \emph{CC-transform} of $f$ is an SCF $f^{CC}$ that, on input profile $\boldsymbol{\sigma}$, outputs the candidates consistent with the output of Algorithm~\ref{alg:cc-transform} on input $f$ and $\boldsymbol{\sigma}$. 
\end{definition}

Intuitively, $f^{CC}$ recursively runs $f$ on the PQ-tree of $\profile$, starting at the root. At every P-node $B$, $f^{CC}$ runs $f$ on the summary induced by that node ($\profile^{\text{decomp($\node,T$)}}$), and continues with the winner children. At every Q-node $\node$, it runs $f$ on the summary of the node \emph{restricted} to its first two child nodes ($B_1(\node,T)$ and $B_2(\node,T)$). If the winner is $B_1(\node,T)$ (resp., $B_2(\node,T)$), it continues with the first (resp., last) child node of $\node$, \emph{i.e.}, $B_1(\node, T)$ (resp, $B_{|\text{decomp($\node,T$)}|}(\node, T)$); if both are winners, then $f$ continues with \emph{all} the children of $\node$. The intuition for this is that for any Q-node $\node$ of $T$, the pairwise relationship between $B_i(\node,T)$ and $B_j(\node,T)$ is the same for all $i<j$, so if $B_1(\profile, \decomp)$ defeats $B_2(\profile, \decomp)$ according to $f$ (in a pairwise comparison), it will also defeat $B_j(\profile, \decomp)$ for any $j>1$ by the neutrality of $f$. If $B_2(\profile, \decomp)$ defeats $B_1(\profile, \decomp)$ according to $f$, on the other hand, then $B_j(\profile, \decomp)$ will defeat $B_i(\profile, \decomp)$ for any $j>i$ by the neutrality of $f$, naturally leading us to the last child node, $B_{|\text{decomp($\node,T$)}|}(\node, T)$. Lastly, if both $B_1(\profile, \decomp)$ and $B_2(\profile, \decomp)$ are winners, this implies $f$ cannot choose between any pair of child nodes of $\node$, which is why we continue with all child nodes.

\begin{algorithm}[t]
    \caption{CC transformation for SCF}\label{alg:cc-transform}
    \SetAlgoNoLine
    \KwIn{SCF $f$, preference profile $\boldsymbol{\sigma}$ over candidates $\cand$}
    \KwOut{Winner candidates $W \subseteq \cand$}
    $W =\emptyset$  \tcp*{Winner list, initialized as empty}
    $T = PQ(\profile)$\tcp*{Constructs the PQ-tree for $\profile$}
    $\queue = (\cand)$ \tcp*{Queue of nodes, starting with root node}
    \While{$|\queue| \neq 0$}{
        $\node =$ Dequeue($\queue$)\;
        \eIf(\tcp*[f]{$\node$ is a leaf node}){$|\node|=1$}{
            $W = W \cup \node$ \tcp*{Add $B$ to list of winners}}{
            $\decomp =$decomp$(\node,T)$\ \tcp*{Each $B' \in \decomp$ is a child node of $\node$}
            \uIf(\tcp*[f]{$\node$ is a P-node}){is\_p\_node$(\node,T)$}{
                \lFor(\tcp*[f]{Run $f$ on summary, enqueue winners}){$K \in f(\profile^\mathcal{K})$}{Enqueue($\queue,K$)}
            }
            \Else(\tcp*[f]{$\node$ is a Q-node}){
                $W'=f(\profile^\mathcal{K})|_{\{\node_1(\node,T),\node_2(\node,T)\}}$ \tcp*{Run $f$ on the first two child nodes}
                \uIf{$W'=\{\node_1(\node,T)\}$}{Enqueue($\queue,\node_1(\node,T)$)\tcp*[f]{Enqueue the first child of $B$}}
                \uElseIf{$W'=\{\node_2(\node,T)\}$}{Enqueue($\queue,\node_{|\decomp|}(\node,T)$)\tcp*[f]{Enqueue the last child of $B$}}
                \Else(\tcp*[f]{$W'=\{\node_1(\node,T),\node_2(\node,T)\}$}){
                    \lFor(\tcp*[f]{Enqueue \emph{all} children of $B$}){$K \in \decomp$}{Enqueue($\queue,K$)}
                }

            }

            }
        }
\end{algorithm}

We will shortly show that $f^{CC}$ satisfies CC, even if $f$ fails it. Of course, a useless transformation like $f^{CC'}(\profile)=\cand$ for all $\profile$ would also achieve this. As such, we want to show that $f^{CC}$ \textit{preserves} some of $f$'s desirable properties. It is straightforward to see that anonymity and neutrality are preserved, as \Cref{alg:cc-transform} is robust to relabeling of candidates and/or voters. Further, as we will show, Condorcet and Smith consistency, as well as decisiveness, are among the preserved properties.

Unfortunately, $f^{CC}$ does not preserve monotonicity, independence of Smith-dominated alternatives (ISDA), or participation. This is since changing an existing vote or adding a new candidate/voter can alter the clone structure of $\profile$, and thus its PQ-tree. We introduce relaxations of these axioms that require robustness against changes that \textit{respect} the clone structure. We first define the relaxation of monotonicity. 
\begin{restatable}{definition}{camono}\label{def:weak-mono}
    An SCF $f$ satisfies \emph{clone-aware monotonicity (monotonicity$\ca$)} if $a \in f(\profile)$ implies $a \in f(\profile')$ whenever (1) $\family(\profile)=\family(\profile')$ and (2) for all $i \in N$ and $b,c \in A \setminus\{a\}$, we have $a \succ_{\sigma_i} b \Rightarrow a \succ_{\sigma'_i} b$ and $b \succ_{\sigma_i} c \Rightarrow b \succ_{\sigma'_i} c$.
\end{restatable}
The only difference between Definition~\ref{def:weak-mono} and the usual definition of monotonicity (\emph{i.e.}, promoting a winner in some votes while keeping all else constant should not cause them to lose) is the requirement that $\profile$ and $\profile'$ have the same clone structure. ISDA$\ca$ and participation$\ca$ are defined analogously; see \Cref{appsec:ca-axioms} for formal definitions and examples showing why we need these relaxations. These new axioms implicitly assume that clone structures are \textit{inherent}, based on candidates' location in some shared perceptual space (in line with the original interpretation by \citet{Tideman87:Independence}), so any ``realistic'' change to $\profile$ will not alter its clone sets.

Lastly, in order to analyze the computational complexity of $f^{CC}$, we introduce the \emph{decomposition degree} of a tree, which we adapt from the definition of the decomposition degree of a tournament introduced by \citet{Brandt11:Fixed} in their study of CC tournament solutions. 
Following their fixed-parameter tractability framework, we will state the runtime of \Cref{alg:cc-transform} in terms of a parameter $\delta$ (which corresponds to the decomposition degree of a PQ-tree, formalized below in \Cref{def:delta}) and the running time of the input SCF $f$. 

\begin{definition}\label{def:delta}
    Given a PQ-tree $T$ for a profile $\profile$, let $\mathcal{P}$ denote the set of P-nodes in $T$. 
    The \emph{decomposition degree} $\delta(T)$ of the PQ-tree $T$ is defined as
    \begin{align*}
        \delta(T)= 
        \begin{cases}
            \max_{\node \in \mathcal{P}} |\text{decomp}(\node,T)| & \mathrm{ if } \, \mathcal{P} \neq \emptyset, \\ 
            2 & \mathrm{otherwise}.
        \end{cases}
    \end{align*}
\end{definition}
Intuitively, $\delta(T)$ is the maximum number of candidates with which \Cref{alg:cc-transform} will run $f$, \emph{e.g.}, in the two PQ-trees from \Cref{fig:tree}, $\delta(T)=3$. We now present our main result on CC-transforms.

\begin{restatable}{theorem}{cctransform} \label{thm:cc_transform}
    For any neutral SCF $f$, $f^{CC}$ satisfies:
    \begin{enumerate}[label={(\arabic*)}]
        \item If $\boldsymbol{\sigma}$ has no non-trivial clone sets, $f^{CC}(\boldsymbol{\sigma})=f(\boldsymbol{\sigma})$.
        \item $f^{CC}$ is composition-consistent.
        \item If $f$ is composition-consistent, then $f^{CC}=f$, \emph{i.e.}, they agree for all $\profile$.
        \item If $f$ satisfies any of \{anonymity, Condorcet consistency, Smith consistency, decisiveness (on all $\profile$), monotonicity$\ca$, ISDA$\ca$, participation$\ca$\}, then $f^{CC}$ satisfies this property as well.
        \item Let $g(n, m)$ be an upper bound on the runtime of an algorithm that computes $f$ on profiles with $n$ voters and $m$ candidates; then, $f^{CC}(\profile)$ can be computed in time $O(nm^3) + m \cdot g(n, \delta(PQ(\profile)))$.
    \end{enumerate}
\end{restatable}

\noindent Taken together, (2) and (3) immediately imply that our CC transformation is idempotent.
\begin{corollary}
    For any neutral SCF $f$, we have $(f^{CC})^{CC}=f^{CC}$, \emph{i.e.}, they agree for all $\profile$.
\end{corollary}

\noindent Further, (5) from \Cref{thm:cc_transform} implies that our CC-transform preserves efficient computability. 
\begin{corollary}
    If $f$ is polynomial-time computable, then so is $f^{CC}$.
\end{corollary}

Even if $f$ is not polynomial-time computable, (5) in \Cref{thm:cc_transform} gives us a running time that depends on the decomposition degree $\delta(T)$ of the PQ-tree. Therefore, we have shown a stronger result: namely, we obtain fixed-parameter tractability for $f^{CC}$ (in terms of $\delta(T)$) for all (neutral) SCFs $f$ with runtime that is polynomial in $n$. 
For example, this includes SCFs that are \NP-hard to compute when the number of candidates $m$ is arbitrarily large, but is polynomial-time computable for constant $m$. Indeed, as mentioned in \Cref{subsec:rp}, a well-known SCF that falls in this category is (anonymous) $RP$ \citep{Brill12:Price}.
Moreover, by (3) in \Cref{thm:cc_transform}, fixed-parameter tractability holds not just for the CC-transform $f^{CC}$ of any (neutral) SCF $f$, but also for all SCFs that are CC to begin with. 

Despite the above theoretical guarantees of \Cref{alg:cc-transform} regarding tractability and axiomatic properties, one can still question how useful our CC-transform is \emph{in practice}, as it does not modify the SCF unless actual clone sets exist (by (1) of \Cref{thm:cc_transform}). While clone sets are unlikely in political elections (where the number of candidates is reasonably bounded anyway), they may easily occur in settings where candidates or voters are not human. For example, if the candidates are AI outputs---\emph{e.g.}, for reinforcement learning from human feedback (RLHF)---it is relatively easy to introduce minorly tweaked versions of the same output into the evaluation process (see \Cref{sec:disc} for a discussion of why our results may be highly relevant for RLHF). Further, voters too could be not human~\citep{Xu24:Aggregating}. For example, if voters are benchmarks against which we are testing AI models (and we are supposed to choose a model by aggregating the ranking resulting from each benchmark) \citep{Lanctot25:Evaluating}, variants of the same model are likely to have similar performance on each benchmark. More classically, meta-search engines aggregate results from various ranking algorithms, each of which plays the role of a voter \citep{Dwork01:Rank}, and cloned webpages are likely to be ranked together by each algorithm. The guarantees of \Cref{thm:cc_transform} can be even more critical in settings such as these, as (a) the cost of cloning can be arbitrarily low, making it all the more important that the SCF used cannot be manipulated by such clones, and (b) the number of candidates can be very large due to such cloning, making tractability a significant concern.

Before ending our discussion of CC-transforms, it is worth comparing $f^{CC}$ to two similar notions from prior literature. First, in addition to CC, \citet{Laffond96:Composition} defined the \textit{CC hull} of an SCF $f$: the smallest (by inclusion) CC solution containing $f$. However, the CC hull does not necessarily preserve Condorcet consistency and achieves CC by adding candidates to the returned set, which sacrifices decisiveness. Second, in an unpublished preprint, \citet{Heitzig02:Social} introduces a similar recurrent CC transformation for SCFs. However, his transformation does not specify the order in which clone sets need to be collapsed and requires the original SCF to satisfy additional axioms, including Condorcet consistency and anonymity.

\section{Social Preference Functions}\label{sec:spf}

We now turn to \emph{social preference functions (SPFs)}, which, given input $\profile$, return a set of rankings of $A$, rather than a subset of candidates. Indeed, SPFs may be more useful than SCFs in certain settings, such as the meta-search engine example in the previous section. We will first present the definition of IoC for SPFs as introduced by \citet{Freeman14:Axiomatic}.

For a ranking $r$ over $A$ and a non-empty $K\subseteq A$, let $r\neg_{K\to z}$ be the ranking obtained from $r$ by replacing the highest-ranked element of $K$ with a new candidate $z$ and removing all other candidates in $K$.
 For example, if $r= (a\succ b\succ c\succ d$) and $K=\{b,d\}$, then $r \neg_{\{b,d\}\to z} = (a \succ  z \succ c)$. For a set of rankings $R$, let $R\neg_{K\to z} = \{r \neg_{K\to z}: r \in R\}$. 

\begin{restatable}{definition}{iocspf}[{\citealt[Def. 4]{Freeman14:Axiomatic}]\label{def:ioc_spf}}
An SPF $F$ is \emph{independent of clones (IoC)} if for all $\boldsymbol{\sigma}$, each non-trivial clone set $\clone$, and $a \in \clone$, we have $F(\boldsymbol{\sigma}) \neg_{\clone \to z}= F(\boldsymbol{\sigma} \setminus \{a\}) \neg_{(\clone \setminus \{a\})\to z}$.
\end{restatable}

Much like its SCF precursor, the IoC criterion for SPFs focuses on the performance of \emph{some} clone in $K$, which is not necessarily the clone that would have ranked the highest if the same SPF was applied to members of $K$ alone. Once again, we would like to strengten this property.

To the best of our knowledge, CC has not been studied for SPFs in prior work. Hence, we now introduce a natural extension of \Cref{def:oioc}. Given rankings $r$ and $r'$ over different sets, where $a$ appears in $r$, let $r(a \rightarrow r')$ be $r$ with $a$ replaced by $r'$, in order. For example, if $r= (a\succ b\succ c$) and  $r'= (d\succ e)$, then $r(b \rightarrow r')= a\succ d\succ e\succ  c$. For sets of rankings $R, R'$, and $R''$, we write $R(a\rightarrow R')= \{r(a\rightarrow r'): r \in R, r'\in R'\}$ and $R(a\rightarrow R', b\rightarrow R'')=R(a\rightarrow R')(b \rightarrow R'').$

\begin{restatable}{definition}{ccspf}[CC for SPFs] \label{def:cc_spf} 
A neutral SPF $F$ is \emph{composition-consistent (CC)} if for all $\boldsymbol{\sigma}$ and all clone decompositions $\decomp$, we have $F(\boldsymbol{\sigma}) = F(\boldsymbol{\sigma}^\decomp)(K \rightarrow F(\boldsymbol{\sigma}|_\clone)\text{ for }\clone \in \decomp)$.    
\end{restatable}

If $F$ is CC, clone sets must appear as intervals in $F(\boldsymbol{\sigma})$ in the order(s) specified by $F(\profile^\decomp)$, and the order(s) within each clone set $K$ is specified by $F(\profile|_\clone)$. Next, we show that the definition of IoC for SPFs by \citet{Freeman14:Axiomatic} and our novel definition of CC for SPFs are consistent with the ones for SCFs.
\begin{restatable}{proposition}{spfscf}\label{prop:SPFtoSCF}
    Let $f$ be the SCF that corresponds to SPF $F$, \emph{i.e.}, $f(\boldsymbol{\sigma})=\{\text{top}(r): r\in F(\boldsymbol{\sigma})\}$. If $F$ is IoC, then $f$ is IoC. If $F$ is CC, then $f$ is CC.
\end{restatable}

It is straightforward to see that the reverse of \Cref{prop:SPFtoSCF} is false: given an SCF $f$ that is CC/IoC, we can always construct an SPF that picks the top ranked candidate according to $f$, and then orders the remaining candidates according to some arbitrary order (\emph{e.g.}, by their plurality scores). Intuitively, such an SPF cannot be expected to obey \emph{any} reasonable definition of IoC/CC for SPFs. Further, there are also more ``natural'' counterexamples, as we show in \Cref{appsec:nested}.

Further, we show that the hierarchy between CC and IoC (\Cref{prop:cctoioc}) extends to SPFs.

\begin{restatable}{proposition}{spfccioc}\label{prop:cctoiocspf}
    If a given SPF is CC, then it is also IoC.
\end{restatable}

We can interpret each version of $\RP$ as an SPF outputting the topological sorting(s) of the final graph(s); \citet{Schulze10:New} shows that $\BP$, too, admits an interpretation as an SPF. Finally, we can view $\STV$ as an SPF outputting candidates in reverse order of elimination; see \Cref{appsec:spfdefs} for formal definitions. Our results generalize to each of these SPFs.

\begin{restatable}{theorem}{spftaxonomy}\label{thm:spftaxonomy}
    Each of $\{\STV, \BP, \RP, \RP_i,\RP_N\}$ satisfies IoC/CC (for all $\profile$) if and only if its SPF version does. 
\end{restatable}

While the above results are intuitive, they rely on a careful definition of CC/IoC for SPFs. For example, \citet[Appendix A]{Boehmer23:Rank} provide an alternative definition of IoC for SPFs where the bottom-ranked clone is replaced in \Cref{def:ioc_spf} rather than the top-ranked one. They show that under this alternative definition, an SPF that iteratively adds the veto winner (see our Footnote~\ref{footnote:veto}) to the ranking and deletes it from the profile would be IoC (with bottom replacement), whereas $\STV$ would not. Hence, both our Prop.~\ref{prop:SPFtoSCF} and Thm.~\ref{thm:spftaxonomy} would fail under this alternative IoC definition.

\Cref{thm:spftaxonomy} gives us a single CC SPF: $\RP_i$, which (like its SCF counterpart), fails anonymity. We next give a (to us, surprising) negative result that this weakness is inevitable.

\begin{restatable}{theorem}{anonspf}\label{thm:anonspf}
    No anonymous SPF can be CC.
\end{restatable}

\Cref{thm:anonspf} has strong implications. First, it shows that to design CC SPFs, a non-anonymous tie-breaker (such as the one by \citet{Zavist89:Complete} for $RP_i$) is not only sufficient but necessary. Indeed, in \Cref{appsec:nested}, we design a novel CC SPF that uses a similar tie-breaking rule, inspired by a nested version of $\STV$ introduced by \citet{Freeman14:Axiomatic}. Second, it shows that in settings where anonymity is a must and ties are likely to occur, CC is too demanding for SPFs, and motivates studying its relaxations. For example, as mentioned in \Cref{subsec:pqtrees}, the Kemeny SPF (which returns the ranking(s) with the minimum total Kendall-Tau distance to voters' rankings) obeys a weaker form of CC, which requires $F(\profile)$ and $F(\boldsymbol{\sigma}^\decomp)(K \rightarrow F(\boldsymbol{\sigma}|_\clone)\text{ for }\clone \in \decomp)$ to have a nonempty intersection under certain decompositions (where the summary is single-peaked or single-crossing), rather than being equal for all decompositions~\citep{Cornaz13:Kemeny}. Since Kemeny is not IoC as an SCF~\citep{Tideman87:Independence}, this relaxation is incomparable with IoC for SPFs by our \Cref{prop:SPFtoSCF}. 

We also observe that the impossibility in \Cref{thm:anonspf} can be circumvented if $|\voter|$ is assumed to be odd. In this regime, for a neutral SPF $F$, we can define $F^{CC}$ analogously to \Cref{def:cc-transform} to obtain CC (by forcing Q-nodes to output their majority ranking). We leave formalizing this transformation and identifying which SPF-specific axioms (\emph{e.g.}, independence of the last-ranked alternatives) are preserved by $F^{CC}$ for future work.

\section{Obvious Independence of Clones}\label{sec:oioc}
 We now investigate what IoC and CC tell us about \emph{strategic behavior} under an election using an IoC/CC SCF. Unlike \citet{Elkind11:Cloning}, who study strategic cloning in a model where each clone set has a central ``manipulator'' that single-handedly decides how many clones run in the election, we will be looking at the strategies of individual candidates, who can personally decide to run in the election or drop out. Our motivation for this is that, as we have seen in \Cref{sec:cc_transform}, a single candidate can be a member of multiple non-trivial (potentially nested) clone sets of different size, and its preference over these sets (and thus over other candidates in them) may vary. In order to formalize this intuition, we now define a clone-based metric over the candidate set $\cand$. Given $\profile$, for each  $a,b \in \cand$, define $d_{\profile}(a,b)=  |K| -1$, where $K$ is the smallest clone set containing both $a$ and $b$.

\begin{restatable}{proposition}{clonemetric}\label{prop:metric}
    For any $\profile$, $d_{\profile}$ is a metric over $\cand$. 
\end{restatable}
 
Now, given $\profile$ and an SCF $f$, consider a normal-form game (NFG) $\Gamma^f_{\profile}$ where the players are the candidates, and each of them has two actions: run ($R$) and drop out ($D$). For simplicity, we assume $f$ is decisive; \emph{i.e.}, it outputs a single candidate. If exactly $S \subseteq \cand$ play $R$, the utility of any $a \in \cand$ is a (strictly) decreasing function of $d_{\profile}(a, f(\profile|_S))$. This follows from the assumption that clones represent proximity in some space (\emph{e.g.}, for political elections, this could be the ideological landscape): the closer the winner is to a candidate, the happier that candidate is.  If \emph{all} candidates pick $D$, then the election has no winner, which we assume gives everyone the worst utility. 

An action is a \emph{dominant strategy} of player $a$ if it brings (weakly) higher utility than any other action, no matter how $a$'s opponents play. A (pure) strategy profile  $\boldsymbol{s}=(s_a)_{a \in A}$ specifies an action $s_a \in \{R,D\}$ for each player $a \in A$. We say $\boldsymbol{s}$ is a \emph{pure-strategy Nash equilibrium (PNE)} if no player $a\in A$ can strictly increase her utility by unilaterally changing her action~\citep{Nash50:Non}. 

The setting of $\Gamma^f_{\profile}$ is a restriction of the more general \emph{strategic candidacy} model introduced by \citet{Dutta01:Strategic}, where candidates also have a preference over each other, and accordingly choose to run or not. Since $d_{\profile}(a,b)=0$ if and only if $a=b$ by \Cref{prop:metric}, our setting fulfills the condition of \emph{self-supporting} preferences (\emph{i.e.}, all candidates like themselves the best), which is taken as a natural domain restriction by \citeauthor{Dutta01:Strategic}~and much of the subsequent work on strategic candidacy. An SCF is called \emph{candidate stable} if for all profiles, the action profile where \emph{all} candidates are running is a PNE. For example, in our setting with  $\Gamma^f_{\profile}$, $PV$ is not candidate stable.

\begin{example}
    Consider $\Gamma^{PV}_{\profile}$, where $\profile$ is the profile from \Cref{fig:bg_eg} (left). In the strategy profile in which all candidates play $R$ (say $\boldsymbol{s}^R$), $b$ wins. However, if $a_2$ deviates to $D$, then $a_1$ wins. Since $d_{\profile}(a_1,a_2)=1<3=d_{\profile}(b,a_2)$, this deviation increases the utility of $a_2$, proving $\boldsymbol{s}^R$ is not a PNE.
\end{example}

Crucially, \citet{Dutta01:Strategic} show that in the general setting with no restrictions on the preferences of voters or of candidates (except the latter being self-supporting), the \emph{only} SCF that is both unanimous (a candidate is picked if all voters rank her first) and candidate-stable is \emph{dictatorship} (\emph{i.e.}, a single voter decides the outcome). As we show next, this is not the case in our restricted setting.

\begin{restatable}{proposition}{iocds}\label{prop:ioc_ds}
    If $f$ is IoC, then $R$ is a dominant strategy in $\Gamma^f_{\profile}$ for all candidates. 
\end{restatable}

As such, in our setting, IoC rules not only achieve candidate stability, but strengthen it, as all candidates running is a \emph{dominant-strategy Nash equilibrium}. \Cref{prop:ioc_ds} is in line with prior results showing how restricting the rankings of voters (\emph{e.g.}, to profiles with a Condorcet winner~\citep{Dutta01:Strategic}) or the rankings of both the voters and candidates (\emph{e.g.}, to single-peaked preferences~\citep{Samejima07-Strategic}) can circumvent the impossibility result by \citet{Dutta01:Strategic} in the general setting. In our setting, we require that all preferences are consistent with \emph{some} inherent clone structure, but since there are infinitely many such structures, this effectively puts no restriction on the preferences of the voters, but only on those of the candidates. Further, \Cref{prop:ioc_ds} formalizes the interpretation of IoC as ``strategy-proofness for candidates,'' in the sense that any candidate who is willing to run will not drop out due to fear of hurting  like-minded candidates (\emph{e.g.}, her political party).\footnote{This is not the only advantage of an IoC/CC rule; \emph{e.g.}, candidates also cannot make their opponents' clone set lose by nominating more clones in this set. In our model, we focus on the choice of whether to run.}

However, as demonstrated by \citet{Li17:Obviously}, the benefits of a property of a mechanism might only be materialized if the agents actually \emph{believe} that the property does indeed hold. If a candidate needs to go through complicated ``what if'' steps in order to believe that an SCF is indeed IoC, she might still drop out of the race (even though running is her dominant strategy), either out of fear of hurting her party, or of being blamed by the voters for doing so.

In order to characterize the ``obviousness'' of IoC, we turn to \emph{obviously dominant strategies}~\citep{Li17:Obviously}, which inherently deal with extensive-form games (EFG), in which players take actions in turns. Intuitively, an EFG can be represented by a rooted tree; at each node, the player associated with that node takes an action, each leading to a child node. The nodes belonging to each player are partitioned into \emph{information sets}; a player cannot tell any two nodes in the same information set apart. Below, we introduce the definition of obviously dominant strategies informally for games where each player acts once; the full definition (along with that of EFGs) is in \Cref{appsec:efg-ods}.

\begin{definition}[{\citealt[Informal]{Li17:Obviously}}]
     An action $s$ is \emph{obviously dominant} for player $a$ if for any other action $s'$, starting from the point in the game when $a$ must take an action, the best possible outcome from $s'$ is no better than the worst possible outcome from $s$.
\end{definition}

Here, the ``best'' and ``worst'' outcomes are defined over the actions of candidates that act along with or after $a$. For example, interpreting $\Gamma^f_{\profile}$ from above as an EFG where all candidates act simultaneously, running ($R$) may \emph{not} be an obviously dominant strategy, even if $f$ is IoC, as the next example shows.

\begin{figure}[t]
    \tikzset{
        every path/.style={-},
        every node/.style={draw},
    }
    \forestset{
    subgame/.style={regular polygon,
    regular polygon sides=3,anchor=north, inner sep=1pt},
    }
    \centering
        \begin{forest}
            [\scriptsize{$c$},cgs,name=c1,s sep=14pt,l sep=10pt
                [\scriptsize{$b$},bgs,name=b1,el={1}{R}{},s sep=14pt,l sep=10pt
                    [\scriptsize{$a_2$},a2gs,name=a21,el={3}{R}{},s sep=14pt,l sep=10pt
                        [\scriptsize{$a_1$},a1gs,name=a11,el={4}{R}{},s sep=14pt,l sep=10pt
                            [\text{$a_1$},terminal,el={2}{R}{},yshift=-3.3pt]
                            [\text{$a_2$},terminal,el={2}{D}{},yshift=-3.3pt]
                        ]
                        [\scriptsize{$a_1$},a1gs,name=a12,el={4}{D}{},s sep=14pt,l sep=10pt
                            [\text{$a_1$},terminal,el={2}{R}{},yshift=-3.3pt]
                            [\text{$b$},terminal,el={2}{D}{},yshift=-3.3pt]
                        ]
                    ]
                    [\scriptsize{$a_2$},a2gs,name=a22,el={3}{D}{},s sep=14pt,l sep=10pt
                        [\scriptsize{$a_1$},a1gs,name=a13,el={4}{R}{},s sep=14pt,l sep=10pt
                            [\text{$c$},terminal,el={2}{R}{},yshift=-3.3pt]
                            [\text{$c$},terminal,el={2}{D}{},yshift=-3.3pt]
                        ]
                        [\scriptsize{$a_1$},a1gs,name=a14,el={4}{D}{},s sep=14pt,l sep=10pt
                            [\text{$c$},terminal,el={2}{R}{},yshift=-3.3pt]
                            [\text{$c$},terminal,el={2}{D}{},yshift=-3.3pt]
                        ]
                    ]
                ]
                [\scriptsize{$b$},bgs,name=b2,el={1}{D}{},s sep=14pt,l sep=10pt
                    [\scriptsize{$a_2$},a2gs,name=a23,el={3}{R}{},s sep=14pt,l sep=10pt
                        [\scriptsize{$a_1$},a1gs,name=a15,el={4}{R}{},s sep=14pt,l sep=10pt
                            [\text{$a_2$},terminal,el={2}{R}{},yshift=-3.3pt]
                            [\text{$a_2$},terminal,el={2}{D}{},yshift=-3.3pt]
                        ]
                        [\scriptsize{$a_1$},a1gs,name=a16,el={4}{D}{},s sep=14pt,l sep=10pt
                            [\text{$a_1$},terminal,el={2}{R}{},yshift=-3.3pt]
                            [\text{$b$},terminal,el={2}{D}{},yshift=-3.3pt]
                        ]
                    ]
                    [\scriptsize{$a_2$},a2gs,name=a24,el={3}{D}{},s sep=14pt,l sep=10pt
                        [\scriptsize{$a_1$},a1gs,name=a17,el={4}{R}{},s sep=14pt,l sep=10pt
                            [\text{$a_2$},terminal,el={2}{R}{},yshift=-3.3pt]
                            [\text{$a_2$},terminal,el={2}{D}{},yshift=-3.3pt]
                        ]
                        [\scriptsize{$a_1$},a1gs,name=a18,el={4}{D}{},s sep=14pt,l sep=10pt
                            [\text{$a_1$},terminal,el={2}{R}{},yshift=-3.3pt]
                            [\text{$\emptyset$},terminal,el={2}{D}{},yshift=-3.3pt]
                        ]
                    ]
                ]
            ]
            \draw[infosetb] (b1)--(b2);
            \draw[infoseta2] (a21)--(a22)--(a23)--(a24);
            \draw[infoseta1] (a11)--(a12)--(a13)--(a14)--(a15)--(a16)--(a17)--(a18);
        \end{forest}
    \caption{EFG representation of $\Gamma_{\profile}^{\STV}$ for $\profile$ from Fig.~\ref{fig:bg_eg}. Terminals show the winner under that action profile. Information sets are joined by dotted lines. For $a_1$, the worst outcome of running is  $c$ winning, and the best outcome of dropping out is $a_2$ winning, so running is not an obviously dominant strategy for $a_1$.
    \label{fig:nfgstv}}
\end{figure}
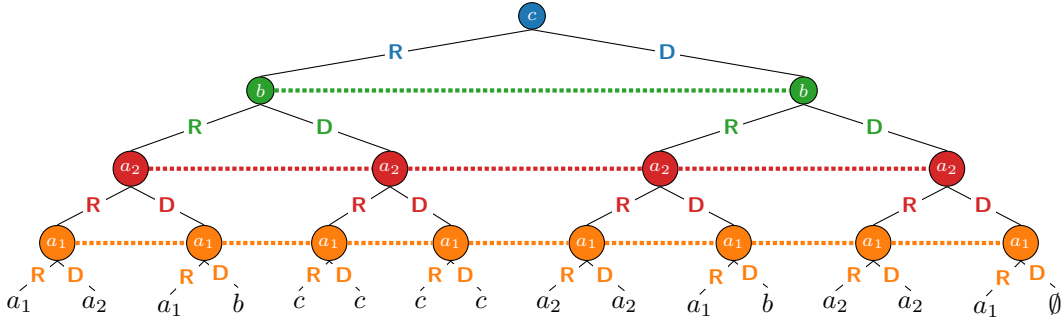

\begin{example}

    Consider $\Gamma^{\STV}_{\profile}$, where $\profile$ is from Fig.~\ref{fig:bg_eg}. From the perspective of $a_1$, the \emph{worst} outcome of running ($R$) is if $a_2$ and $c$ also play $R$, but $b$ plays $D$, making $c$ the winner. The \emph{best} outcome of $a_1$ dropping out ($D$) is if everyone else plays $R$, making $a_2$ win. Since $d_{\profile}(a_1,c)=3>1=d_{\profile}(a_1,a_2)$, $R$ is \emph{not} an obviously-dominant strategy for $a_1$. A tree representation of $\Gamma^{\STV}_{\profile}$ is in \Cref{fig:nfgstv}.
\end{example}

What if we had used a composition-consistent SCF $f$ instead? Recall that by (3) of \Cref{thm:cc_transform}, $f$ can be implemented using \Cref{alg:cc-transform}, \emph{i.e.}, by running it on the PQ-tree of the input profile. The key observation is that when running \Cref{alg:cc-transform}, we can postpone asking a candidate if she is running or not until we reach the internal node that is the immediate ancestor of that candidate. More formally, consider then an alternate EFG $\Lambda^f_{\profile}$ where the actions and utilities are the same as  $\Gamma^f_{\profile}$, but the winner is determined by running \Cref{alg:cc-transform} on inputs $f$ and $\profile$, with the following process after each node $\node$ is dequeued from $\queue$:
\begin{itemize}
    \item If $\node$ is a P-node, then all the children of $\node$ that are actual candidates (\emph{i.e.}, leaf nodes) are asked (simultaneously) to pick $R$ or $D$. Given $S'$ are the child nodes that chose $D$, $f$ is run on $\profile^{\text{decomp}(\node,T)}\setminus S'$ to decide which branch to follow. If the winner is a leaf, the game is over. 
    \item If $\node$ is a Q-node, say $W'=f(\profile^\mathcal{K})|_{\{\node_1(\node,T),\node_2(\node,T)\}}$. If $W'=\node_1(\node,T)$ and $\node_1(\node,T)$ is an internal node, then it is enqueued. Otherwise, the (single) candidate corresponding to $\node_1(\node,T)$ is asked to pick $R$, in which case it is the winner, or $D$, in which case the process is repeated with $\node_2(\node,T), \node_3(\node,T),\ldots$ until either an internal node or a candidate that plays $R$ is encountered. If $W'=\node_2(\node,T)$, on the other hand, the identical process is followed, except starting from $\node_\ell(\node,T)$, where $\ell=|\text{decomp}(\node,T)|$, and moving backwards to $\node_{\ell-1}(\node,T),\node_{\ell-2}(\node,T),\ldots$
    
    \item In either case, if all the children of $\node$ are leaf nodes and all play $D$, then the algorithm moves back to the parent node of $\node$, and repeats the computation there (without re-asking all the leaf nodes) with $\node$ also dropped out of the summary.
\end{itemize}

\begin{figure}[t]
    \tikzset{
        every path/.style={-},
        every node/.style={draw},
    }
    \forestset{
    subgame/.style={regular polygon,
    regular polygon sides=3,anchor=north, inner sep=1pt},
    }
    \centering
        \begin{forest}
            [\scriptsize{$c$},cgs,name=c1,s sep=14pt,l sep=10pt
                [\scriptsize{$b$},bgs,name=b1,el={1}{R}{},s sep=14pt,l sep=10pt
                    [\scriptsize{$a_2$},a2gs,name=a21,el={3}{R}{},s sep=14pt,l sep=10pt
                        [\scriptsize{$a_1$},a1gs,name=a11,el={4}{R}{},s sep=14pt,l sep=10pt
                            [\text{$a_2$},terminal,el={2}{R}{},yshift=-3.3pt]
                            [\text{$a_2$},terminal,el={2}{D}{},yshift=-3.3pt]
                        ]
                        [\scriptsize{$a_1$},a1gs,name=a12,el={4}{D}{},s sep=14pt,l sep=10pt
                            [\text{$a_1$},terminal,el={2}{R}{},yshift=-3.3pt]
                            [\text{$b$},terminal,el={2}{D}{},yshift=-3.3pt]
                        ]
                    ]
                        [\text{$c$},terminal,el={3}{D}{},yshift=-3.3pt]
                ]
                [\scriptsize{$b$},bgs,name=b2,el={1}{D}{},s sep=14pt,l sep=10pt
                    [\scriptsize{$a_2$},a2gs,name=a23,el={3}{R}{},s sep=14pt,l sep=10pt
                        [\scriptsize{$a_1$},a1gs,name=a15,el={4}{R}{},s sep=14pt,l sep=10pt
                            [\text{$a_2$},terminal,el={2}{R}{},yshift=-3.3pt]
                            [\text{$a_2$},terminal,el={2}{D}{},yshift=-3.3pt]
                        ]
                        [\scriptsize{$a_1$},a1gs,name=a16,el={4}{D}{},s sep=14pt,l sep=10pt
                            [\text{$a_1$},terminal,el={2}{R}{},yshift=-3.3pt]
                            [\text{$b$},terminal,el={2}{D}{},yshift=-3.3pt]
                        ]
                    ]
                    [\scriptsize{$a_2$},a2gs,name=a24,el={3}{D}{},s sep=14pt,l sep=10pt
                        [\scriptsize{$a_1$},a1gs,name=a17,el={4}{R}{},s sep=14pt,l sep=10pt
                            [\text{$a_2$},terminal,el={2}{R}{},yshift=-3.3pt]
                            [\text{$a_2$},terminal,el={2}{D}{},yshift=-3.3pt]
                        ]
                        [\scriptsize{$a_1$},a1gs,name=a18,el={4}{D}{},s sep=14pt,l sep=10pt
                            [\text{$a_1$},terminal,el={2}{R}{},yshift=-3.3pt]
                            [\text{$\emptyset$},terminal,el={2}{D}{},yshift=-3.3pt]
                        ]
                    ]
                ]
            ]
            \draw[infosetb] (b1)--(b2);
            \draw[infoseta2] (a21)--(a23)--(a24);
            \draw[infoseta1] (a11)--(a12)--(a15)--(a16)--(a17)--(a18);
        \end{forest}
    \caption{$\Lambda^{STV^{CC}}_{\profile}$, for $\profile$ from Fig.~\ref{fig:bg_eg}, the PQ-tree of which is in Fig.~\ref{fig:tree} (right). For $a_1$, best outcome of not running is $a_2$ winning, which is no better than the worst outcome of running, which is also $a_2$ winning. Therefore, running is an obviously dominant strategy for $a_1$. A similar analysis applies for all other candidates.
    \label{fig:efgpvcc}}
\end{figure}
Intuitively, $\Lambda_{\profile}^f$ asks each candidate whether she is running only when this decision becomes relevant. Just like in $\Gamma^f_{\profile}$, each player in $\Lambda^f_{\profile}$ has a single information set, since she is not aware of the actions of the players that are acting before or simultaneously with her; she only knows her parent node is reached. The winner in $\Lambda_{\profile}^f$ is precisely the winner of applying $f^{CC}$ directly to $\profile \setminus S'$, where $S'$ are the players that picked $D$.\footnote{There is a slight caveat here: the leaf nodes that are children of internal nodes that never got visited did not get to play $R$ or $D$ in  $\Lambda^f_{\profile}$. Any choice these candidates could have made does not change the result of $f$ as long as \emph{at least one} candidate of each non-trivial clone set were to pick $R$. This is in line with the assumption made by \citet{Elkind11:Cloning} that at least one clone of each clone set will be in the election. Indeed, this is not a far-fetched assumption: in practice, it is the leadership of a political party that decides to participate in an election, before individual members of the party make up their mind about whether to run.} If $f$ is CC to begin with, this exactly corresponds to $f(\profile \setminus S')$ by \Cref{thm:cc_transform}(3). \Cref{fig:efgpvcc} shows the game tree for $\Lambda_{\profile}^f$ for $\profile$ from \Cref{fig:bg_eg_2} and $f=STV^{CC}$.

Crucially, this implementation of $f^{CC}$ allows us to strengthen Prop.~\ref{prop:ioc_ds}, achieving obviousness.
\begin{restatable}{theorem}{ccods}\label{thm:cc_ods}
    For any neutral $f$, $R$ is an obviously-dominant strategy in $\Lambda^f_{\profile}$ for all candidates.
\end{restatable}
The proof relies on the observation that in  $\Lambda^f_{\profile}$, when a candidate is asked to decide between $R$ and $D$, \Cref{alg:cc-transform} has already reached her parent node, which is the smallest non-trivial clone set containing her. Thus, the best case of $D$ and worst case of $R$ are both one of her second-favorite group of candidates (after herself) winning, achieving obvious strategy-proofness (OSP).

\Cref{thm:cc_ods} has strong practical implications: since any CC rule can be implemented with \Cref{alg:cc-transform}, the decision of a candidate to drop out of an election can be postponed until \emph{after} she learns whether her smallest clone set has won. Hence, using a CC rule, the election result will not change if we replace the candidates' names on the ballots with party names, and hold in-party primaries for the winners afterwards. In contrast, with rules that are just IoC, the results within the party vary based on whether internal primaries are held (\Cref{eg:GLOC}). Without primaries, CC rules can also derive clone sets \emph{a posteriori} from the votes, rather than simply assuming a party to be a clone set.

Importantly, obviousness is also relevant for contexts where candidates (or, for settings with abstract candidates, their deployers) are perfectly capable of reasoning about an SCF and its properties, but they worry about manipulation of the SCF by the entity implementing it (also called \emph{agenda control}~\citep{Lang13:New}). As shown by \citet{Li17:Obviously}, choice rules that are OSP-implementable offer a significant advantage in these settings, as they are exactly those that are supported by \emph{bilateral commitments} (partial commitments by the planner such that, if violated, those violations can be observed by the agents themselves without communicating among each other). In the context of $\Lambda^f_{\profile}$, instead of committing to using a specific SCF, the planner can commit to each candidate he interacts with that, if she decides to run, the winner will be some member of her smallest non-trivial clone set. This (1) is enough to convince the candidate to run and (2) ensures that a violation of this commitment can be observed by the candidate (by looking at the outcome of the election).

In general, the connection we establish between IoC and CC yields new and natural interpretations of the two properties: we can view CC as a way of exposing the \emph{obviousness} of IoC.\footnote{In fact, in prior work, we have introduced a definition for \emph{obvious independence of clones} that is identical to composition consistency, without realizing the connection at the time~\citep{Berker22:Obvious}.}

\section{Conclusion and Future Work}\label{sec:disc}

There are several ways in which our analysis of strategic behavior can be further improved. First, while CC/IoC rules ensure that candidates, by running, do not hurt their clone sets when voter preferences are fixed, there may yet be practical reasons for consolidating support behind a single candidate, \emph{e.g.}, a fixed party campaign budget. Thus, it is worth considering extensions of strategic candidacy where running may be costly~\citep{Obraztsova15:Strategic}. Second, our analysis inherently assumes that clone sets possess some structural affinity, such as ideological closeness. This need not be the case: two extremist candidates on opposite sides of the spectrum might be ranked at the bottom by all voters, making them a clone set. In such cases, other metrics of similarity, such as proximity in the \emph{societal axis} of \emph{single-peaked elections}, may be more appropriate. As such, a natural future avenue of research is defining more general notions of clones, and analyzing whether our results extend.

Another exciting direction is to study the role that IoC and CC can play in the context of AI alignment. Methods such as reinforcement learning from human feedback (RLHF) require aggregating data representing diverse human opinions, for which social choice methods are well-suited. It is relatively easy to copy AI model responses, or even entire models, and perform small tweaks to them (\emph{e.g.}, via fine-tuning). Such tweaks are likely to not outperform other significantly better models, hence forming a clone set. \citet{Xu24:RLHF} demonstrate that using non-IoC aggregation rules for RLHF can result in egregious behavior,\footnote{While the~\citet{Xu24:RLHF} present these undesirable outcomes as a failure to meet the independence of irrelevant alternatives property by~\citet{Luce59:Individual}, the demonstrated pathology would also be prevented if the aggregation rule being used was IoC.} a result that is especially concerning as standard RLHF approaches implicitly use Borda Count~\citep{Siththaranjan24:Distributional}, which fails IoC. Thus, as pointed out by~\citet{Conitzer24:Position}, IoC (and thereby CC) becomes highly relevant for social choice for AI models. In line with this agenda, \citet{Procaccia25:Clone} have recently studied how existing RLHF algorithms can be modified to increase their robustness against clones. A natural strengthening of this goal for future work is developing RLHF approaches that implicitly use CC rules.

\section*{Acknowledgements}
We are grateful to Ariel Procaccia for his valuable contributions to the conceptualization of this project and helpful feedback. We also thank Felix Brandt and Jobst Heitzig for useful discussions. R.E.B. and V.C. thank the Cooperative AI Foundation, Polaris Ventures (formerly the Center for Emerging Risk Research) and Jaan Tallinn's donor-advised fund at Founders Pledge for financial support. R.E.B. is also supported by the Cooperative AI PhD Fellowship. 

\bibliography{refs}

\appendix

\section{Further Background}\label{appsec:bg}
In this section, we give a more extended analysis of the concepts and definitions introduced by \citet{Tideman87:Independence} and by \citet{Laffond96:Composition}, and how they have been interpreted and used in subsequent literature. To give a more complete picture, this section repeats some of the information given in \Cref{sec:related} (\nameref{sec:related}) and \Cref{sec:bg} (\nameref{sec:bg}) of the main body of the paper.

\begin{figure}[b]
\centering \begin{tabular}{|c|c|c|}
  \hline
  Voter 1 & Voter 2 & Voter 3\\ \hline 
  \cellcolor{blue!25} $b$ & \cellcolor{yellow!25}  $a$ &\cellcolor{yellow!25} $a$ \\
  \hline
\cellcolor{green!25} $c$ &  \cellcolor{red!25} $d$ &\cellcolor{blue!25} $b$ \\
  \hline
      \cellcolor{yellow!25} $a$ & \cellcolor{green!25} $c$ & \cellcolor{green!25} $c$  \\
  \hline
     \cellcolor{red!25} $d$ & \cellcolor{blue!25} $b$ &\cellcolor{red!25} $d$ \\
  \hline
\end{tabular}
\caption{A preference profile. Columns show rankings.}\label{fig:appeg_prof}
\end{figure} 

\subsection{Preference profiles and clones}\label{appsubsec:clones}
We consider a finite set of \textit{voters} $\voter = \{1,\ldots, n\}$ and a finite set of \textit{candidates} $\cand$ with $|\cand|=m$. A {\em ranking} over $\cand$
is an asymmetric, transitive, and complete binary relation $\succ$ on $\cand$; we denote the set of all rankings over $\cand$ by $\mathcal{L}(\cand)$.
Each voter $i \in \voter$ has a \textit{ranking} $\sigma_i \in \mathcal{L}(\cand)$; we 
write $a \succ_i b$ to indicate that $i$ ranks $a$
above $b$, and collect the rankings of all voters in
a \textit{preference profile} $\profile \in \mathcal{L}(\cand)^n$. Given $\profile$, how can we identify candidates that are ``close'' to one another, at least from the point of view of the voters?  \citet{Tideman87:Independence} addressed this question:
\begin{definition}[{\citealt[\S I]{Tideman87:Independence}}]\label{appdef:clones}
Given a preference profile $\profile$ over candidates $\cand$, a nonempty subset of candidates, $\clone \subseteq \cand$, is a \emph{set of clones} with respect to $\profile$ if no voter ranks any candidate outside of $K$ between any two elements of $\clone$.
\end{definition}
Note that all profiles ways have two types of \textit{trivial} clone sets:\footnote{\citet{Tideman87:Independence} in fact excludes trivial clone sets. We use the definition followed by \citet{Elkind10:Clone}} (1) the entire candidate set $\cand$, and (2) for each $a\in \cand$, the singleton $\{a\}$. We call all other clone sets \textit{non-trivial}. For example, in the preference profile in Figure \ref{fig:appeg_prof}, the only non-trivial clone set is $\{b,c\}$.

Seemingly unaware of Tideman's definition, \citet{Laffond96:Composition} tackled a similar question of identifying similar candidates. First, they focused on the context of a \textit{tournament} $T$, which is a complete asymmetric binary relation on the candidates $\cand$ (\emph{i.e.}, a ranking with the transitivity condition relaxed). For a tournament, they defined:
\begin{definition}[{\citealt[Def. 1]{Laffond96:Composition}}]\label{appdef:component_tour}
Given tournament a $T$ over candidates $\cand$, a nonempty $\comp \subseteq \cand$ is a \emph{component} of $T$ if for all $y,y' \in \comp$ and all $x\in \cand \setminus \comp$: $y\succ_T x \Leftrightarrow y'\succ_{T} x$. 
\end{definition}
Of course, any preference profile $\profile$ with odd number of voters (to avoid ties) can be interpreted as a tournament $T_{\profile}$ over the same set of candidates, where the binary relation is given by the pairwise defeats of $\profile$; \emph{i.e.}, $\forall a,b \in \cand$: 
\begin{align*}
 a \succ_{T_{\profile}} b \Leftrightarrow |\{i \in N: a \succ_{\sigma_i} b\}|-|\{i \in N: b \succ_{\sigma_i} a\}|>0.
\end{align*}
Considering the similarities between Definitions \ref{def:clones} and \ref{appdef:component_tour} (they both group up candidates that have an identical relationship to other candidates), one might expect them to respect this transformation; that is, for $\clone$ to be a clone set of $\profile$ if and only if it is a component of $T_{\profile}$. However, this is not the case, as demonstrate next.

\begin{example} \label{ex:clones_v_components}
    Consider again the profile in \Cref{fig:appeg_prof}. Here, $\{a,c\}$ is a component in $T_{\profile}$ (they both defeat $d$ and both lose to $b$), but they are not a set of clones in $\profile$, since Voter 3 ranks $d$ between them. The intuition behind this is that when interpreting a preference profile as a tournament, we lose information about the preferences of individual voters. Indeed, it is easy to see that the implication holds in one direction: if $K$ is a set of clones of $\profile$, then it is a component of $T_{\profile}$. 
\end{example}

However, \citet{Laffond96:Composition} introduce a separate definition for components for a preference profile or, more accurately, to the more general notion of a \textit{tournament profiles} $(T_i)_{i \in \profile}$, where voters are allowed to submit tournaments (\emph{i.e.}, votes need not be transitive):

\begin{definition}[{\citealt[Def. 4]{Laffond96:Composition}}]\label{appdef:component_prof}
Given tournament profile $\boldsymbol{T}=(T_i)_{i \in N}$, a nonempty $C \subseteq A$ is a \emph{component} of $\boldsymbol{T}$ if $C$ is a component of $T_i$ for all $i\in N$. 
\end{definition}

One can see from Definition \ref{appdef:component_tour} that if a tournament is transitive (\emph{i.e.}, a ranking), then $\comp$ is a component of $T$ if and only if it appears as an interval in that ranking. As such, given a tournament profile $\boldsymbol{T}=\{T_{i}\}_{i \in N}$ where all voters submit transitive tournaments, $\comp$ is a component of $\boldsymbol{T}$ if and only if it appears as an interval in the vote of every voter. As such, \citeauthor{Laffond96:Composition}'s Definition \ref{appdef:component_prof} restricted to preference profiles is in fact \emph{equivalent} to \citeauthor{Tideman87:Independence}'s definition of a set of clones (Definition \ref{def:clones})! For instance, in the profile from Figure \ref{fig:appeg_prof}, $\{a,c\}$ is \textit{not} a component according to Definition \ref{appdef:component_prof}, since it is not a component of Voter 3's vote. 

Since most of the rest of the paper by \citet{Laffond96:Composition} focuses on tournaments rather than profiles, later work have largely focused on Definition \ref{appdef:component_tour} for tournaments rather than Definition \ref{appdef:component_prof} for profiles, leading to some papers arguing \citeauthor{Laffond96:Composition}'s definition for components is a ``more liberal notion'' than \citeauthor{Tideman87:Independence}'s clones~\citep{Conitzer24:Position,Holliday24:Simple}, even though the former's definitions for components in preference profiles is identical to that of Tideman's definition for clones. In the rest of the paper, we will be sticking to the term ``clone sets'' for consistency. 

Since their independent introduction by \citeauthor{Tideman87:Independence} and by \citeauthor{Laffond96:Composition}, clone sets in preference profiles have been studied at length in the computational social choice literature. Notably, \citet{Elkind10:Clone} have axiomatized the  structure of clone sets in preference profiles, and introduced a compact representation of clone sets using a data structure called PQ-trees, which we will later introduce in detail.

\subsection{Social choice functions and axioms}\label{appsubsec:axioms}
Of course, identifying ``similar'' candidates in a voting profile is useless unless one can say something meaningful about their impact on the election result. This impact depends on the voting rule we are using to compute the winners. More formally, say $ \mathcal{P}(\cand)$ is the power set of $A$ (set of all subsets). Then, a \textit{social choice function (SCF)} is a function $f: \mathcal{L}(\cand)^n \rightarrow \mathcal{P}(\cand)$ that maps each preference profile $\profile$ to a subset of $A$, which are termed the winner(s) of $\profile$ under $f$. In an election without ties, the output of an SCF contains a single candidate.

Before introducing axioms for robustness against strategic nomination, it is worth noting that for rules that are \emph{not} robust, the exact influence of addition of similar candidates can vary: for example, introducing clones of a candidate can hurt that candidate for an SCF like plurality  voting (which simply picks the candidates ranked first by the most voters) by splitting the vote (as was the case from the Oregon governor race from the introduction of the main body of the paper), making plurality what we call \textit{clone-negative}. On the other hand, having clones helps a candidate win if the SCF being used is Borda's rule, which gives a point for each candidate for every other candidate it beats in each voter's ranking, and picks the candidates with the most points as the winners.

\begin{example}\label{ex:borda}
    Consider the following voting profile for candidates $a$ and $b$:

\begin{center} 
 \begin{tabular}{|c|c|}
  \hline
  62 voters & 38 voters\\ \hline 
  \cellcolor{yellow!25} $a$ & \cellcolor{blue!25} $b$  \\
  \hline
   \cellcolor{blue!25} $b$ & \cellcolor{yellow!25} $a$ \\
  \hline
\end{tabular}
\end{center}

In this case, candidate $a$ receives 62 Borda points, whereas candidate $b$ receives 38 Borda points. Thus, candidate $a$ wins the election. Next we introduce a clone of $b$, obtaining the following voting profile:

\begin{center} 
 \begin{tabular}{|c|c|}
  \hline
  62 voters & 38 voters\\ \hline 
  \cellcolor{yellow!25} $a$ & \cellcolor{blue!25} $b$  \\
  \hline
   \cellcolor{blue!25} $b$ & \cellcolor{blue!35} $b_2$ \\
  \hline
  \cellcolor{blue!35} $b_2$ & \cellcolor{yellow!25} a \\
  \hline
\end{tabular}
\end{center}

Now candidate $a$ receives 124 Borda points, $b$ receives 138 Borda points, and clone $b_2$ receives 38 Borda points. Hence, candidate $b$ now becomes the winner. 
\end{example}

\Cref{ex:borda} shows that unlike with plurality voting, having a clone can positively impact a candidate under Borda's rule, thus making Borda's rule \textit{clone-positive} for this specific profile (in order profiles, having clones can in fact hurt your Borda score). Either of these impacts are undesirable, considering they incentivize strategic nomination, either of candidates similar to one's opponents, or of candidates similar to one's self, either of which can be arbitrarily easy. As such, we would like to find \textit{axioms} such that if an SCF satisfies them, then they are in some way robust to this type of strategic nomination.  

Along with their (equivalent) definitions for similar candidates in preference profiles, \citet{Tideman87:Independence} and \citet{Laffond96:Composition} each introduce their own axiom for identifying SCFs that behave ``desirably'' in response to addition/removal of such candidates. Since we deal with preference profiles (say $\profile$) with some candidates (say $\cand'\subset \cand$) removed, it will be useful to use $\boldsymbol{\sigma} \setminus \cand'$ to denote the profile obtained by removing the elements of $\cand'$ from each voter's ranking in $\boldsymbol{\sigma}$ and preserving the order of all other candidates.

We begin with the axiom by \citet{Tideman87:Independence}, who explicitly identified the goal of achieving robustness against strategic nomination. \citet{Zavist89:Complete} later presented the definition with more precise language, which is the version we use for clarity.

\begin{definition}[\citealt{Zavist89:Complete}]\label{appdef:ioc}
A voting rule $f$ is \emph{independent of clones (IoC)} if the following two conditions are met for all profiles $\boldsymbol{\sigma}$ and for all non-trivial clone sets $\clone \subset A$ with respect to $\boldsymbol{\sigma}$:
\begin{enumerate}
    \item For all $a \in \clone$, we have:
    \begin{align*}
        \clone \cap f(\boldsymbol{\sigma}) \neq \emptyset \Leftrightarrow \clone \setminus \{a\} \cap f(\boldsymbol{\sigma} \setminus \{a\}) \neq \emptyset.  
    \end{align*}
    \item For all $a \in \clone$ and all $b \in A \setminus \clone$ we have:
    \begin{align*}
       b \in f(\boldsymbol{\sigma}) \Leftrightarrow b \in f(\boldsymbol{\sigma} \setminus \{a\}).
    \end{align*}
\end{enumerate}
\end{definition}

Intuitively, IoC dictates that deleting one of the clones must not alter the winning status of the set of clones as a whole, or of any candidate not in the set of clones. This is a desirable property in SCFs, since it imposes that the winner must not change due to the addition of a non-winning candidate who is similar to a candidate already present. This prevents candidates from influencing the election by nominating new copy-cat candidates. 

\begin{example}
    Consider running plurality voting ($PV$) on the profile $\profile$ in Figure \ref{fig:bg_eg} (left). We have $f_{PV}(\profile)=\{b\}$, as it is the top choice of 4 voters, more than any other candidate. Moreover, $\{a_1,a_2\}$ is a clone set with respect to $\profile$. However, $f_{PV}(\profile \setminus \{a_2\}) = \{a_1\}$, since with $a_2$ gone, $a_1$ now has 5 voters ranking it top, beating $b$ (Figure \ref{fig:bg_eg}, right). This violates both conditions 1 and 2 from Definition \ref{appdef:ioc}, as the removal of a clone ($a_2$ from clone set $\{a_1,a_2\}$) results in another clone $(a_1)$ winning, while previously none did, and eliminates a previous winner outside the clone set ($b$).

    Instead, consider running Single Transferable Vote ($STV$) on $\profile$, which is an SCF that iteratively removes the candidates with the least plurality votes from the ballot, returning the last remaining candidate. We have $STV(\profile)=\{a_1\}$ as $a_2$ is removed with 2 plurality votes (causing $a_1$ to now have 5 plurality votes), followed by the $c$ with 3 plurality votes, and finally $b$ with 4 plurality votes. Similarly, $STV(\boldsymbol{\sigma} \setminus\{a_2\})=\{a_1\}$, since $a_2$ was going to be the first candidate to be eliminates anyway. Hence, in both cases, a member of the clone set $\{a_1,a_2\}$ wins. This is in line with the fact that $STV$ is IoC~\citep{Tideman87:Independence}. 
\end{example}

Like Tideman's presentation of IoC, \citet{Laffond96:Composition} are also concerned with the manipulability of elections through cloning. However, the core of the presentation of their axiom, aptly named \textit{composition consistency}, focuses on the consistency between applying a rule directly, or applying it through a two-stage mechanism. In order to formalize this mechanism, we introduce a few concepts:
  \begin{definition}\label{appdef:clone_grouping}
Given a preference profile $\boldsymbol{\sigma}$ over candidates $\cand$, a set of sets $\decomp=\{\clone_1,\clone_2,\ldots,\clone_\ell\}$ where $\clone_i \subseteq A$ for all $i\in [\ell]$ is a \emph{clone decomposition} with respect to $\boldsymbol{\sigma}$ if:
\begin{enumerate}
    \item $\decomp$ is a disjoint partitioning of $\cand$, \emph{i.e.}: $\cand=\bigsqcup_{i \in [\ell]} K_i$ and $\clone_i \cap \clone_j = \emptyset$ for $i \neq j$, and
    \item each $\clone_i$ is a non-empty clone set with respect to $\boldsymbol{\sigma}$.
\end{enumerate}
\end{definition}
A given profile $\boldsymbol{\sigma}$ can have multiple distinct clone decompositions. Indeed, every profile has at least two decompositions: the \textit{null} decomposition $\decomp_{null}=\{A\}$ and the \textit{trivial} decomposition $\decomp_{triv}=\{ \{a\}\}_{a \in A}$. Given a clone decomposition $\decomp$ with respect to $\boldsymbol{\sigma}$, for each $i \in N$, say $\sigma^\decomp_i$ is voter $i$'s ranking over the clone sets in $\decomp$ (which is well defined, since each clone set appears as an interval in $\sigma_i$). We call $\boldsymbol{\sigma}^\decomp = \{\sigma^\decomp_i\}_{i \in [n]}$ the \textit{summary} of $\boldsymbol{\sigma}$ with respect to decomposition $\decomp$, which is a preference profile treating the elements of $\decomp$ as the set of candidates. Lastly, for each $K \in \decomp$, say $\boldsymbol{\sigma}^{K}$ is $\boldsymbol{\sigma}$ with $A\setminus K$ removed (\emph{i.e.}, $\boldsymbol{\sigma}^{K} \equiv \boldsymbol{\sigma} \setminus (A \setminus K)$). We are now ready to introduce \textit{composition products}:

\begin{definition}[Composition product]\label{appdef:gloc} 
    Given any SCF $f$, the \emph{composition product} function of $f$ is a function $\gloc_f$ that takes as input a profile $\boldsymbol{\sigma}$ and a clone decomposition $\decomp$ with respect to $\boldsymbol{\sigma}$ and outputs $\gloc_f(\boldsymbol{\sigma}, \decomp) \equiv \bigcup_{K \in f\left(\boldsymbol{\sigma}^\decomp\right)  }f(\boldsymbol{\sigma}^K)$.
\end{definition}

Intuitively, $\gloc_f$ first runs the input voting rule $f$ on the summary (as specified by $\decomp$), ``packing'' the candidates in each set to treat it as a meta-candidate $\clone_i$. It then ``unpacks'' the clones of each winner clone set, and runs $f$ once again on each. We demonstrate this in the following example.
\begin{example}\label{appeg:GLOC}

Once again consider $\boldsymbol{\sigma}$ from Figure \ref{fig:bg_eg}. Notice $\decomp=\{K_a,K_b,K_c\}$ with $K_a=\{a_1,a_2\}$, $K_b=\{b\}$ and $K_c=\{c\}$ is a valid clone decomposition with respect to $\boldsymbol{\sigma}$. Figure \ref{fig:bg_eg_2} shows $\profile^\decomp$ and $\profile^{\clone_a}$. Notice that we have $STV(\boldsymbol{\sigma}^\decomp)=\clone_a$ ($K_c$ gets eliminated first followed by $K_b$) and $STV(K_a)=\{a_2\}$, implying $\gloc_{STV}(\boldsymbol{\sigma}, \decomp) =\{a_2\}$.

\end{example}

Example \ref{appeg:GLOC} demonstrates that $STV(\boldsymbol{\sigma}) \neq \gloc_{STV}(\boldsymbol{\sigma}, \decomp)$ for this specific $\boldsymbol{\sigma}$ and $\decomp$, showing $STV$ is not \textit{consistent} with respect to this decomposition, even though the winners in two cases are from the same clone set. It is also easy to see that for all rules $f$ and all $\boldsymbol{\sigma}$, we have $f(\boldsymbol{\sigma})=\gloc_f(\boldsymbol{\sigma}, \decomp_{null})=\gloc_f(\boldsymbol{\sigma}, \decomp_{triv})$. To satisfy composition consistency, a rule must satisfy this equality for all non-trivial decompositions too: 
  \begin{definition}[{\citealt[Def. 11]{Laffond96:Composition}}]\label{appdef:oioc}
      An SCF $f$ is \emph{composition-consistent (CC)} if for all preference profiles $\boldsymbol{\sigma}$ and all clone decompositions $\decomp$ w.r.t. $\boldsymbol{\sigma}$, we have $f(\boldsymbol{\sigma}) =\gloc_f(\boldsymbol{\sigma}, \decomp)$.
\end{definition}

Intuitively, an SCF is CC if it chooses the ``best'' candidates from the ``best'' clone sets~\citep{Laffond96:Composition}. Notice that while any other member of the clone set winning after the removal of a winner clone is sufficient by IoC, CC also specifies which exact clones should be winning. We formalize this hierarchy in Proposition \ref{prop:cctoioc} by showing CC implies IoC. Example \ref{appeg:GLOC} already demonstrates that the other direction is untrue, as it proves that $STV$, which is IoC, is not CC. In later sections, we will be analyzing other IoC rules to show whether they are CC.  

While the hierarchical relationship between CC and IoC may seem clear to the reader when presented this way, with few exceptions (more on which below), many papers that mention both axioms do not explicitly identify CC to be a strictly stronger property than IoC. It appears that part of the this unclarity can once again be attributed to SCFs taking a relatively small space in \citet{Laffond96:Composition}, which is mostly dedicated to tournaments. Accordingly, subsequent papers have also studied CC primarily in the context of tournaments~\citep{Cornaz13:Kemeny,Brandt11:Fixed,Brandt18:Extending}, even describing components/CC to be the ``analogue'' of clones/IoC for tournaments~\citep{Elkind11:Cloning,Dellis13:Multiple,Karpov22:Symmetric}. Other works, while identifying a link between IoC and CC, have not been precise about their relationship~\citep{Brandt09:Some,Öztürk20:Consistency,Koray07:Self,Laslier12:Loser,Lederer24:Bivariate,Saitoh22:Characterization,Elkind17:What,Camps12:Continuous,Laslier16:Strategic}, describing them as ``similar'' notions~\citep{Laslier97:Tournament,Heitzig10:Some} or simply ``related''~\citep{Brandt11:Fixed}.

There are papers that come much closer in identifying CC as a stronger axiom than IoC: working in a more general setting where voters' preferences are neither required to be asymmetric (ties are allowed)  nor transitive, \citet{Laslier00:Aggregation} introduces the notion of \textit{cloning consistency}, which he explains is weaker than CC and is the ``same idea'' as \citeauthor{Tideman87:Independence}'s IoC in voting theory. However, there are significant differences between \citeauthor{Laslier00:Aggregation}'s definition and IoC: first, his ``clone set'' definition requires every voter being \textit{indifferent} between any two alternatives in the set (as opposed to having the same relationship to all other candidates), and his cloning consistency dictates that if one clone wins, then so must every other member of the same clone set. Another property for tournament solutions named \emph{weak composition consistency} (this time in fact analogous to IoC) is discussed by \citet{Brandt18:Extending,Kruger18:Permutation} and \citet{Laslier97:Tournament}, although none of them points out the connection to IoC. Perhaps the work that does the most justice to the relationship between CC and IoC is  by \citet{Brandl16:Consistent}, who explicitly state that \citeauthor{Tideman87:Independence}'s IoC (which they refer to as cloning consistency) is weaker than \citeauthor{Laffond96:Composition}'s CC. They work with probabilistic social choice functions (PSCF), which return a set of lotteries over $\cand$ rather than a subset of candidates. As they note, deterministic SCFs can be viewed as PSCFs that return \emph{all} lotteries over a subset of candidates, therefore their observation that CC is stronger than IoC applies to our setting, although the adaptation is not obvious; thus, we formalize this statement in \Cref{prop:cctoioc}. The PSCFs analyzed by \citeauthor{Brandl16:Consistent} are all non-deterministic; hence, to the best of our knowledge, no previous work has studied whether IoC SCFs also satisfy CC, which we do in the main body of the paper.

\section{On \Cref{sec:ioc_rules} (\nameref{sec:ioc_rules})}

In this section, we provide the proofs omitted from \Cref{sec:ioc_rules} of the main body.
\subsection{Proof of \Cref{prop:cctoioc}}

We first prove the relationship between CC and IoC:
\cctoioc*
\begin{proof}
For a CC rule $f$, take any profile $\profile$ over candidates $\cand$, non-trivial clone set $\clone \subset \cand$, and candidate $a\in \cand$.
Consider the clone decomposition $\decomp = \{\clone\} \cup \{\{b\}\}_{b\in \cand \setminus \clone}$ for $\profile$ and the clone decomposition $\decomp' = \{ \clone\setminus \{a\} \} \cup \{\{b\}\}_{b\in \cand \setminus \clone}$ for $\profile \setminus \{a\}$ (i.e., the decomposition which groups all existing members of $\clone$ together, and everyone else is a singleton). Notice that $\profile^\decomp$ and  $(\profile\setminus\{a\})^\mathcal{K'}$ are identical except the meta-candidate for $\clone$ in the former is replaced with the meta-candidate for $\clone \setminus \{a\}$ in the latter. Since $f$ is neutral by \Cref{def:oioc}, this implies:
\begin{align}
    \clone \in f\left(\profile^\decomp\right) &\iff  \clone \setminus\{a\} \in f\left((\profile\setminus\{a\})^\mathcal{K'}\right) \label{eq:cond1s}\\
    \forall b \in \cand \setminus C: \quad \quad& \nonumber \\\{b\} \in f\left(\profile^\decomp\right) &\iff  \{b\} \in f\left((\profile\setminus\{a\})^\mathcal{K'}\right) \label{eq:cond2s}
\end{align}
    
    By Definition \ref{def:gloc}, it is easy to see that for any $K \in \decomp$, we have $K \cap \gloc_f(\profile, \decomp) \neq \emptyset \iff K \in f\left(\profile^\decomp\right)$. Based on this, (\ref{eq:cond1s}) and  (\ref{eq:cond2s}) respectively imply:
    \begin{align}
    \clone \cap \gloc_f(\profile, \decomp) \neq \emptyset &\iff  \clone \setminus\{a\}\cap \gloc_f(\profile \setminus \{a\}, \decomp') \neq \emptyset\label{eq:cond1f}\\
  \forall b \in \cand \setminus \clone:\quad \quad& \nonumber \\  b \in \gloc_f(\profile, \decomp) &\iff  b \in \gloc_f(\profile \setminus \{a\}, \decomp')\label{eq:cond2f}
\end{align}
Since $f$ is CC, we have $\gloc_f(\profile, \decomp) = f(\profile)$ and $\gloc_f(\profile\setminus\{a\}, \decomp') = f(\profile\setminus\{a\})$ so (\ref{eq:cond1f}) and (\ref{eq:cond2f}) respectively imply conditions 1 and 2 in Definition \ref{def:ioc}, proving that $f$ is IoC.
\end{proof}

\subsection{(Extended) Proof of \Cref{thm:cc_fails}}\label{appsec:extended_ioc}

The proof of \Cref{thm:cc_fails} is given in the main body of the paper, but the winners under each SCF are stated without detailed calculations. Here, we give a more extensive proof that walks through the implementation of each SCF.

\ccfails*
\begin{proof}
Since CC implies $f=\gloc_f(\profile, \decomp)$ for \textit{all} profiles $\profile$ and clone decomposition $\decomp$, a single counterexample is sufficient to show a rule failing CC.

For $\STV$ and $\AS$, we will be using $\profile$ over $\cand = \{a_1,a_2,b,c\}$ from \Cref{fig:bg_eg}. Consider $\decomp=\{K_a,K_b,K_c\}$, with $K_a=\{a_1,a_2\}$, $K_b=\{b\}$, and $K_c=\{c\}$. Figure \ref{fig:bg_eg_2} shows $\profile^\decomp$ and $\profile|_{K_a}$. The procedure for $\STV$ is detailed in \Cref{eg:ioc,eg:GLOC}. To run $\AS$ on $\profile$, we will alternate between eliminating all non-Smith candidates and eliminating the candidate with the least plurality score:
\begin{itemize}
    \item First, we have $\Sm(\profile)=A$ due to the cyclicity of the profile, hence no one gets eliminated.
    \item Then, we eliminate $a_2$, the candidate with the least plurality score.
    \item Once again, $\Sm(\profile \setminus\{a_2\})=\cand \setminus\{a_2\}$, so no candidate is eliminated.
    \item $c$ gets eliminated as the next candidate with least plurality scores.
    \item $\Sm(\profile\setminus\{b,a_2\})=\{a_1\}$ since $a_1$ pairwise defeats $b$.
\end{itemize}
Therefore $\AS(\profile)=\{a_1\}$. Running $\AS$ on $\profile^\decomp$, on the other hand, we get:
\begin{itemize}
    \item We have $\Sm(\profile^\decomp)=\decomp$ due to the cyclicity of the profile, so no (meta-)candidate is eliminated.
    \item Then, we eliminate $K_c$, the candidate with the least plurality score.
    \item $\Sm(\profile^\decomp \setminus\{K_c\})=\{K_a\}$ since $K_a$ pairwise defeats $K_b$.
\end{itemize}
Therefore $\AS(\profile^\decomp)=\{K_a\}$. Further, $\AS(\profile|_{K_a})=\{a_2\}$ since $a_1$ is pairwise defeated by $a_2$ and therefore eliminated in the first step. Thus, we have $\AS(\profile)=\{a_1\} \neq \{a_2\} = \gloc_{\AS}(\profile,\decomp)$, proving $\AS$ is not CC. For both $\STV$ and $\AS$, the main idea is that $a_2$ gets eliminated first even though it is a majority winner over $a_1$, since the few voters that prefer $a_1$ over $a_2$ happens to put them to the top of their ballot, giving $a_1$ more plurality votes than $a_2$.

For $BP$ and $SC$, consider the following profile $\profile'$ (the same as the one from \Cref{fig:eg_prof}, with relabeled candidates):

\begin{center}
\begin{tabular}{|c|c|c|c|}
    \hline
    6 voters & 5 voters & 2 voters & 2 voters \\ \hline
    \cellcolor{yellow!25} $a_1$ & \cellcolor{blue!25} $c$ & \cellcolor{green!25} $b$ & \cellcolor{green!25} $b$ \\ \hline
    \cellcolor{red!25} $a_2$ & \cellcolor{red!25} $a_2$ & \cellcolor{blue!25} $c$ & \cellcolor{blue!25} $c$ \\ \hline
    \cellcolor{green!25} $b$ & \cellcolor{yellow!25} $a_1$ & \cellcolor{yellow!25} $a_1$ & \cellcolor{red!25} $a_2$ \\ \hline
    \cellcolor{blue!25} $c$ & \cellcolor{green!25} $b$ & \cellcolor{red!25} $a_2$ & \cellcolor{yellow!25} $a_1$ \\
    \hline
\end{tabular}
\end{center}

To find $BP(\profile')$ and $SC(\profile')$, we construct the majority matrix $M_{\profile'}$ below:

\begin{center}
\begin{tabular}{|c|c|c|c|c|}
    \hline
    $M_{\profile'}$ & \cellcolor{yellow!25} $a_1$ & \cellcolor{red!25} $a_2$ & \cellcolor{green!25} $b$ & \cellcolor{blue!25} $c$ \\ \hline
    \cellcolor{yellow!25} $a_1$ & $0$ & $1$ & $7$ & $-3$ \\ \hline
    \cellcolor{red!25} $a_2$ & $-1$ & $0$ & $7$ & $-3$ \\ \hline
    \cellcolor{green!25} $b$ & $-7$ & $-7$ & $0$ & $5$ \\ \hline
    \cellcolor{blue!25} $c$ & $3$ & $3$ & $-5$ & $0$ \\ 
    \hline
\end{tabular}
\end{center}

Notice that there are 3 simple cycles in $M_{\profile'}$: $(a_1,b,c)$, $(a_2,b,c)$, and $(a_1,a_2,b,c)$, with smallest margins $[c,a_1]$, $[c,a_2]$, and $[a_1,a_2]$, respectively. Removing these three edges form the graph leaves $a_1$ and $a_2$ without any incoming edges, indicating $SC(\profile')=\{a_1,a_2\}$. 

Similarly, $M_{\profile'}$ induces the following strength matrix $S_{\profile'}$, which shows that $BP(\profile')=\{a_1,a_2\}$, since $S[x,y] \geq S[y,x]$ for $x\in \{a_1,a_2\}$ and all $y\in \cand$. 
\begin{center}
\begin{tabular}{|c|c|c|c|c|}
    \hline
    $S_{\profile'}$ & \cellcolor{yellow!25} $a_1$ & \cellcolor{red!25} $a_2$ & \cellcolor{green!25} $b$ & \cellcolor{blue!25} $c$ \\ \hline
    \cellcolor{yellow!25} $a_1$ & $0$ & $3$ & $7$ & $5$ \\ \hline
    \cellcolor{red!25} $a_2$ & $3$ & $0$ & $7$ & $5$ \\ \hline
    \cellcolor{green!25} $b$ & $3$ & $3$ & $0$ & $5$ \\ \hline
    \cellcolor{blue!25} $c$ & $3$ & $3$ & $3$ & $0$ \\ 
    \hline
\end{tabular}
\end{center}

However, using clone decomposition $\decomp$ from above (which is also a valid decomposition with respect to $\profile'$), the graph for $M_{\profile'^\decomp}$ is composed of a single simple cycle, with $M[K_a,K_b]=7$, $M[K_b,K_c]=5$ and $M[K_c,K_a]=3$. Clearly, we have $SC(\profile'^\decomp)=BP(\profile^\decomp)=\{K_a\}$. However, $a_1$ is the majoritarian winner against $a_2$ in $\profile'|_{K_a}$, without any cycles. Hence $\prod_{SC}(\profile', \decomp)=\prod_{BP}(\profile', \decomp)=\{a_1\}$, showing both rules fail CC. Intuitively, both $BP$ and $SC$, while picking their winners for $\profile'$, `discard' the relationship between $a_1$ and $a_2$, $BP$ since neither the strongest path from $a_1$ to $a_2$ nor vice versa go through the $(a_1,a_2)$ edge, and $SC$ since the $(a_1,a_2)$ edge forms the weakest margin in a 4-candidate cycle. As a result, both rules pick both candidates as winner, even though they both agree $a_1$ wins over $a_2$ when applied to $\profile'|_{K_a}$ alone.
\end{proof}

\subsection{Proof of \Cref{thm:rp}}\label{appsec:rpi}
We now prove our main positive characterization result from \Cref{sec:ioc_rules}.

\rpcc*

Without loss of generality, fix $1\in N$. We will show that $RP_1$ satsifes CC. The same proof follows for $RP_i$ for any $i \in N$. We write $\{a,b\} \succ_{\Sigma_1} \{c,d\}$ if $\Sigma_1$ ranks $\{a,b\}$ before $\{c,d\}$. \citeauthor{Zavist89:Complete} show that $\Sigma_1$ is \textit{impartial}; that is, for all $a,b,c,d\in A$, if $\{a,c\}\succ_{\Sigma_1}\{b,c\}$ then  $\{a,d\}\succ_{\Sigma_1}\{b,d\}$. 
Using $\Sigma_1$, we construct a complete \emph{priority order} $\mathcal{L}$ over ordered pairs: pairs are ordered (in decreasing order) according to $M$, and ties are broken by $\Sigma_1$ (and according to $\sigma_1$ when $M[a,b]=0$). Formally, given distinct ordered pairs $(a,b)$ and $(c,d)$ such that $(c,d) \neq (b,a)$, we have:
\begin{align*}
    (a,b) \succ_{\mathcal{L}} (c,d) & \text{ iff:} \begin{cases} M[a,b]> M[c,d] \text{ or }\\  M[a,b]= M[c,d], \{a,b\} \succ_{\Sigma_1} \{c,d\},\end{cases}
\end{align*}
and if $(c,d)=(b,a)$ we have:
\begin{align*}
    (a,b) \succ_{\mathcal{L}} (b,a) & \text{ iff:}\begin{cases} M[a,b]> M[b,a] \text{ or }\\ M[a,b]= M[b,a]=0, a  \succ_{\sigma_1} b. \end{cases}
\end{align*}
Then, \textit{the Ranked Pairs method using voter 1 as a tie-breaker} (hereon referred to as $RP_1$) add edges from $M$ to a digraph according to $\mathcal{L}$, skipping those that create a cycle. 

\citet{Zavist89:Complete} show that $RP_1$ is indeed IoC. We now strengthen this result:

\begin{proof}
\citet{Zavist89:Complete} show that the original RP rule (without tie-breaking) has an equivalent definition using ``stacks''. We introduce an analogous notion and equivalency with respect to a specific $\mathcal{L}$.
\begin{definition} Given a complete ranking $R$ over candidates $\cand$ and a priority order over ordered pairs $\mathcal{L}$, we say $x$ \emph{attains} $y$ through $R$ and with respect to $\mathcal{L}$ if there exists a sequence of candidates $a_1,a_2,\ldots,a_j$ such that $a_1=x$, $a_j=y$ and for all $i \in [j-1]$, we have $a_i \succ_R a_{i+1}$ and $(a_i,a_{i+1}) \succ_{\mathcal{L}} (a_j , a_1)$. We say $R$ is a \emph{stack} with respect to $\mathcal{L}$ if $x \succ_R y$ implies $x$ attains $y$ through $R$ with respect to $\mathcal{L}$.
\end{definition}
\begin{lemma}\label{lemma:stack_winner}
    $RP_1$ with $\Sigma_1$ as a tie-breaker will pick candidate $a$ as a winner if and only if there exists a stack with respect to $\mathcal{L}$ that ranks $a$ first, where $\mathcal{L}$ is the priority order over ordered pairs constructed using $\Sigma_1$ as a tie-breaker.
\end{lemma}
\begin{proof}
    $(\Rightarrow):$ Say $a$ is the $RP_1$ winner with $\mathcal{L}$ as the priority order over ordered pairs (constructed from $\Sigma_1$). Notice that the final graph from the RP procedure will be a DAG. Say $R$ is the topological ordering of this DAG and $a$ is the source node (hence ranked first by $R$), which we call the \textit{winning ranking}. By definition, the rule will pick $a$ as the winner. For any $x,y\in A$ such that $x \succ_R y$, the edge $(y,x)$ was skipped in the RP procedure, implying it would have created a cycle. Hence, there exists candidates $a_1, \ldots a_j$ such that $a_1=x$ and $a_j=y$, and each $(a_i,a_{i+1})$ was locked in the RP graph before $(y,x)$ was considered, implying  $(a_i,a_{i+1}) \succ_\mathcal{L} (y,x)=(a_j,a_1)$. Moreover, since each $(a_i,a_{i+1})$ was locked, we must have $a_i \succ_R a_{i+1}$ in the final ranking. This implies $R$ is indeed a stack with respect to $\mathcal{L}$, with $a$ ranked first.\\
    \noindent $(\Leftarrow):$ Say $R$ is a stack with respect to $\mathcal{L}$, with $a$ ranked first. We argue this is the final ranking produced by running $RP_1$ with $\mathcal{L}$ as priority order (constructed using $\Sigma_1$). Assume instead that RP outputs final ranking $R^*$ with $R^* \neq R$. Then there exists at least one pair $x,y$ such that $x\succ_Ry$ but $y\succ_{R'}x$, so $(y,x)$ was locked by the RP procedure. Of all such pairs, say $x^*,y^*$ is the one where $(y^*,x^*)$ was locked by the RP procedure first. Since $x^*\succ_Ry^*$ and since $R$ is a stack with respect to $\mathcal{L}$, there exists a series of candidates $a_1,\ldots a_j$ such that $a_1=x^*$, $a_j=y^*$, and for all $i \in [j-1]$, we have $a_i \succ_R a_{i+1}$ and $(a_i,a_{i+1}) \succ_{\mathcal{L}} (a_j , a_1)=(y^*,x^*)$. Since $(a_i,a_{i+1}) \succ_{\mathcal{L}} (y^*,x^*)$ for all $i$, all such edges were considered by the RP procedure before $(y^*,x^*)$. At least one of these edges must have been skipped, otherwise locking $(y^*,x^*)$ would have caused a cycle. Say $(a_k,a_{k+1})$ was the first edge that was skipped. This implies locking this edge would have caused a cycle with the already-locked edges; however, since $a_k \succ_R a_{k+1}$, this cycle must contain an edge $(z,\ell)$ such that $\ell \succ_R z$. However, this implies $z\succ_{R'} \ell$ in the final ranking and that $(z,\ell) \succ_{\mathcal{L}}(a_k,a_{k+1})\succ_{\mathcal{L}}(y^*,x^*)$. Since $(y^*,x^*)$ was assumed to be the first such edge to be considered, this is a contradiction. 
\end{proof}
Note that since $RP_1$ results in a single unique ranking over candidates (as a single tie-breaker is fixed), the proof of Lemma \ref{lemma:stack_winner} also shows that there is a unique stack with respect to $\mathcal{L}$. We also use an existing lemma by \citet{Zavist89:Complete}.
\begin{lemma}[{\citealt[\S VII]{Zavist89:Complete}}]\label{lemma:impartial} Say $C$ is a clone set with respect to profile $\profile$. The winning ranking $R$ resulting from running $RP_1$ on $\profile$ with an impartial tie-breaker $\Sigma_1$ based on a ranking $\sigma_1$ will have no element of $A \setminus C$ appear between two elements of $C$ in $R$. 
\end{lemma}

We will now prove that $RP_1$ is composition consistent.
Given $\profile$, say $\decomp=\{K_1,K_2,\ldots,K_k\}$ is a clone decomposition. Say $\profile^\decomp$ is the summary of $\profile$ with respect to $\decomp$ (where the clone sets in each $\sigma_i$ is replaced by the meta candidates $\{K_i\}_{i \in [k]}$) and $\profile|_{K_i}$ is $\profile$ restricted to the candidates in $K_i$. We would like to show that $RP_1(\profile)= \bigcup_{K \in RP_1(\profile^\decomp)} RP_1(\profile|_K)$. Since RP with a specific tie-breaking order always produces a single unique winner, showing containment in a single direction is sufficient. 

Say $RP_1(\profile)=\{a\}$, implying $a$ comes first in the winning ranking $R$. By the proof of the forward direction of Lemma \ref{lemma:stack_winner}, $R$ is a stack with respect to $\mathcal{L}$ (the order that the RP procedure follows, which uses tie-breaking order $\Sigma_1$ based on vote $\sigma_1$). By Lemma \ref{lemma:impartial}, each $K_i \in \decomp$ appears as an interval in $R$, hence we can define a corresponding ranking $R^{\decomp}$ over clone sets in $\decomp$. We would like to show that $R^\decomp$ is a stack with respect to $\mathcal{L}^\decomp$, which is the order of ordered pairs in $\decomp$ according to decreasing order of $M^\decomp$ (the majority matrix of $\profile^\decomp$), using $\sigma_1^\decomp$ (voter 1's vote in the summary) as a tie-breaker. 

We can relabel the clone sets in $\decomp$ such that $R^\decomp=(K_1\succ K_2 \succ \ldots \succ K_k)$. Since $a$ is the ranked first in $R$, we have $a \in K_1$. Notice that if $k=1$, then $R^\decomp$ vacously. Otherwise, take any $K_x, K_y$ such that $K_x \succ_{R^\decomp} K_y$. Say $x$ is the element of $K_x$ that appears last in $R$ and $y$ is the element of $K_y$ that appears first in $R$. Since $R$ is a stack with respect to $\mathcal{L}$, and since $x \succ_R y$, there exists a sequence of candidates $a_1,\ldots a_j$ and for all $i \in [j-1]$, we have $a_i \succ_R a_{i+1}$ and $(a_i,a_{i+1}) \succ_{\mathcal{L}} (a_j , a_1)$. Since the $a_i$ in this sequence appear according to their order in $R$, by Lemma \ref{lemma:impartial}, consecutive candidates in the sequence $a_1,\ldots,a_j$ can be grouped up to form a sequence $K'_1, \ldots K'_{j'}$ such that $K'_{i'} \in \decomp$ for each $i'\in [j']$, $K'_1 = K_x$, $K'_{j'}=K_y$, and $K'_{i'} \succ_{R^\decomp}K'_{i'+1}$ for each $i'\in [j'-1]$. Notice that since $x$ and $y$ are in different clone sets, $j'>1$. Take any $i' \in [j'-1]$ and consider the last element $K'_{i'}$ and the first element of $K'_{i'+1}$ to appear in $(a_1,a_2,\ldots, a_j)$. By construction, these two elements appear consecutively in $(a_1,a_2,\ldots, a_j)$, so they are $a_i$ and $a_{i+1}$, respectively, for some $i \in [j]$. Since $(a_i,a_{i+1})\succ_\mathcal{L}(a_j,a_1)=(y,x)$, based on the way $\mathcal{L}$ was constructed, there are two possible cases:
\begin{enumerate}
    \item $M[a_i,a_{i+1}] > M[y,x]$, in which case we must have $M^\decomp[K'_{i'}, K'_{i'+1}]= M[a_i,a_{i+1}] > M[y,x]= M^\decomp[K_y,K_x]$ by definition of clones, and hence $(K'_{i'},K'_{i'+1})\succ_{\mathcal{L}^\decomp}(K_y,K_x) = (K'_{j'}, K'_{1})$.
    \item $M[a_i,a_{i+1}] = M[y,x]$. In this case, we also have $M^\decomp[K'_{i'}, K'_{i'+1}]= M^\decomp[K_y,K_x]$ by definition of clones. However, since $(a_i,a_{i+1})\succ_\mathcal{L}(y,x)$,  there are four options: 
    \begin{enumerate}
        \item $i'=1$ and $i'+1=j'$, so $a_i=x$ and $a_{i+1}=y$. In this case,  $(a_i,a_{i+1})\succ_\mathcal{L}(y,x)$ implies $x \succ _{\sigma_{1}} y$ and hence $K_x \succ_{\sigma^\decomp_1} K_y$ by definition of clone sets, and hence: 
        $(K'_{i'},K'_{i'+1})=(K_{x},K_{y})\succ_{\mathcal{L}^\decomp}(K_y,K_x)=(K'_{j'},K'_1)$
        \item $i'=1$ and $i'+1 \neq j'$, so $a_i=x$ and $a_{i+1} \neq y$. In this case,  $(a_i,a_{i+1})\succ_\mathcal{L}(y,x)$ implies $a_{i+1}\succ_{\sigma_{1}} y$ and hence $K'_{i'+1} \succ_{\sigma^\decomp_1} K_y$ by definition of clone sets, and hence: 
        $(K'_{i'},K'_{i'+1})=(K_{x},K'_{i'+1})\succ_{\mathcal{L}^\decomp}(K_y,K_x)=(K'_{j'},K'_1)$.
        \item $i' \neq 1$ and $i'+1 = j'$, so $a_i \neq x$ and $a_{i+1} = y$. In this case,  $(a_i,a_{i+1})\succ_\mathcal{L}(y,x)$ implies $a_{i}\succ_{\sigma_{1}} x$ and hence $K'_{i'} \succ_{\sigma^\decomp_1} K_x$ by definition of clone sets, and hence: 
        $(K'_{i'},K'_{i'+1})=(K'_{i'},K_{y})\succ_{\mathcal{L}^\decomp}(K_y,K_x)=(K'_{j'},K'_1)$.
        \item $i' \neq 1$ and $i'+1 \neq j'$, so $a_i \neq x$ and $a_{i+1} \neq y$. 
        
        In this case,  $(a_i,a_{i+1})\succ_\mathcal{L}(y,x)$ implies for some $\alpha \in \{0,1\}$, we have $a_{i+\alpha}\succ_{\sigma_{1}} z$ for each $z \in \{x,y,a_{i+1-\alpha}\}$, and hence $K'_{i'+\alpha}\succ_{\sigma_1^\decomp} Z$ for each $Z \in \{K_x,K_y,K'_{i'+1-\alpha}\}$ by definition of clone sets, and hence: 
        $(K'_{i'},K'_{i'+1})\succ_{\mathcal{L}^\decomp}(K_y,K_x)=(K'_{j'},K'_1)$.

    \end{enumerate}
\end{enumerate}
    In each case, we end up having $(K'_{i'},K'_{i'+1})\succ_{\mathcal{L}^\decomp}(K'_{j'},K'_1)$, which proves that $K_x$ attains $K_y$ through $R^\decomp$ with respect to $\mathcal{L}^\decomp$, and hence that $R^{\decomp}$ is a stack with respect to $\mathcal{L}^\decomp$. By Lemma \ref{lemma:stack_winner}, this implies $RP_1(\profile^\decomp)=\{K_1\}$, as $K_1$ comes first in $R^\decomp$. 

    Since $a\in K_1$, $a$ will be a competing candidate in $\profile|_{K_1}$. Again by Lemma \ref{lemma:impartial}, we know that all elements of $K_1$ appears as a block in the start of $R$. Say $R|_{K_1}$ is this section of $R$. We would like to show that $R|_{K_1}$ is a stack with respect to $\mathcal{L}^{K_1}$, which is the priority order of ordered pairs in $\decomp$ according to decreasing order of $M^{K_1}$ (the majority matrix of $\profile|_{K_1}$), using $\sigma_1|_{K_1}$ (voter 1's vote restricted to $K_1$) as a tie-breaker. Note that for any $a,b,c,d \in K_1$, $(a,b) \succ_\mathcal{L}  (c,d)$ implies $(a,b) \succ_{\mathcal{L}^{K_1}}  (c,d)$, since $\mathcal{L}$ is entirely based on pairwise comparisons and the relative ranking of candidates in $\sigma_1$, neither of which is affected by the deletion of candidates in $A \setminus K_1$ and hence is directly carried to $\mathcal{L}^{K_1}$. Now take any $x,y \in K_1$ such that $x\succ_{R|_{K_1}}y$. Since $R|_{K_1}$ is just an interval of $R$, we must have $x\succ_R y$. Since $R$ is a stack, this implies there exists a sequence of candidates $a_1,a_2, \ldots, a_j$ such that $a_1=x$, $a_j=y$ and for all $i \in [j-1]$, we have $a_i \succ_R a_{i+1}$ and $(a_i,a_{i+1}) \succ_{\mathcal{L}} (a_j , a_1)$. Since all elements of $K_1$ appear as an interval in $R$ by Lemma \ref{lemma:impartial}, $x=a_1 \succ_R a_2 \succ_R \ldots \succ_R a_j = y$ and $x,y\in K_1$ implies $a_i \in K_1$ for all $i \in [j]$. This implies  $a_i \succ_{R|_{K_1}} a_{i+1}$ and $(a_i,a_{i+1}) \succ_{\mathcal{L}^{K_1}} (a_j , a_1)$, implying $R|_{K_1}$ is a stack with respect to $\mathcal{L}^{K_1}$. Since $a$ is first in $R|_{K_1}$, by Lemma \ref{lemma:stack_winner}, this implies $RP_1(\profile|_{K_1})=\{a\}$. Since $ =RP_1(\profile^\decomp)=\{K_1\}$, we have $\bigcup_{K \in RP_1(\profile^\decomp)} f(\profile|_K)=\{a\}=RP_1(\profile)$, completing the proof.
\end{proof}

\subsection{Proof of \Cref{prop:ucg}}

Next, we prove that unlike $\UCf$ (which is not even IoC~\citep{Holliday23:Split}), $\UCg$ does indeed satisfy CC as an SCF.

\ucg*
\begin{proof}
    Recall from \Cref{tab:scfs} that given $\profile$ and $a,b \in \cand$, we say that $a$ \emph{left-covers} $b$ in $\profile$ if any $c \in \cand$ that pairwise defeats $a$ also pairwise defeats $b$. Then $\UCg$ is defined as
    \begin{align*}
        \UCg=\{a \in \cand: \nexists b \in \cand\text{ such that }b\text{ left-covers AND pairwise defeats }a\}.
    \end{align*}

    Fix any profile $\profile$ and clone decomposition $\decomp$ with respect to $\profile$. We will show that $\UCg(\profile)= \gloc_\UCg(\profile, \decomp)$. Equivalently, for any $a \in \cand$, we will show that $a \notin \UCg(\profile) \Leftrightarrow a \notin \gloc_\UCg(\profile, \decomp)$. 
    
    \noindent $(\Rightarrow):$ Say $a \notin \UCg(\profile)$, then $\exists b \in A$ such that $b$ left-covers and pairwise defeats $a$ in $\profile$. Say $K_a \in \decomp$ is the clone set that contains $a$. We will consider two cases:
    \begin{enumerate}
        \item $b \in K_a$. For each $c \in K_a$ that pairwise defeats $b$ in $\profile|_{K_a}$, we must have that $c$ pairwise defeats $a$ in $\profile|_{K_a}$, since $b$ left-covers $a$ in $\profile$ and deletions of other candidates do not affect pairwise victories of the remaining candidates. Hence, $b$ left-covers and pairwise defeats $a$ in $\profile|_{K_a}$. This implies $a \notin \UCg(\profile|_{K_a})$. 
        \item $b \notin K_a$. Say $K_b \in \decomp \setminus \{K_a\}$ is the clone set containing $b$. Since $b$ pairwise defeats $a$ in $\profile$, $K_b$ pairwise defats $K_a$ in $\profile^{\decomp}$ by the clone set definition. Take any $K \in \decomp$ that pairwise defeats $K_b$ in $\profile^\decomp$. This implies there exists some $c \in K$ that pairwise defeats $b$ in $\profile$. Since $b$ left-covers $a$ in $\profile$, this implies $c$ pairwise defeats $a$ in $\profile$ and thus $K$ pairwise defeats $K_a$ in $\profile^\decomp$. Hence, $K_b$ left-covers and pairwise defeats $K_a$ in $\profile^\decomp$, implying $K_a \notin \UCg(\profile^\decomp)$.
    \end{enumerate}
    This implies we either have $a \notin \UCg(\profile|_{K_a})$ or $K_a \notin \UCg(\profile^\decomp)$. By $\Cref{def:gloc}$, this implies $a \notin \gloc_{\UCg}(\profile, \decomp)$.

    \noindent $(\Leftarrow):$ Say $a \notin \gloc_{\UCg}(\profile, \decomp)$ and $K_a \in \decomp$ is the clone set that contains $a$. This implies at least one of the two following two cases must be true:
    \begin{enumerate}
        \item $a \notin \UCg(\profile|_{K_a})$. Then there exists $b \in K_a$ that left-covers and pairwise defeats $a$ in $\profile|_{K_a}$. Since pairwise defeats are not affected by the addition of other candidates, $b$ also pairwise defeats $a$ in $\profile$. Take any $c \in \cand$ that pairwise defeats $b$. If $c \in K_a$, then $c$ must pairwise defeat $a$ because $b$ left covers $\profile|_{K_a}$. If $c \notin K_a$, then $c$ must pairwise defeat $a$ by the clone set definition, since $a,b \in K_a$. Thus, $b$ left-covers and pairwise defats $a$ in $\profile$, implying $a \notin \UCg(\profile)$.
        \item $K_a \notin \UCg(\profile^\decomp)$. Then there exists $K \in \decomp$ that left-covers and pairwise defeats $K_a$ in $\profile^\decomp$. Since $K_a$ cannot pairwise defeat itself, this implies $K \neq K_a$. Take any $b \in K$. Since $K$ pairwise defeat $K_a$ in $\profile^\decomp$, this implies $b$ pairwise defeats $a$ in $\profile$. Take any $c \in \cand$ that pairwise defeats $b$ in $\profile$. We cannot have $c \in K_a$ since $K$ pairwise defeats $K_a$. If $c \in K$, then $c$ must pairwise defeat $a$ since $K$ pairwise defeats $K_a$. If $c \notin K_a$, say $K_c \decomp \setminus \{K_a, K_b\}$ is the clone set that contains $c$. Since $c$ pairwise defeats $b$ in $\profile$, $K_c$ pairwise defeast $K$ in $\profile^\decomp$. Since $K$ left-covers $K_a$ in $\profile^\decomp$, this implies $K_c$ pairwise defeats $K_a$, and therefore $c$ pairwise defeats $a$ in $\profile$. Hence, $b$ left-covers and pairwise defeats $a$, implying $a\notin \UCg(\profile).$
    \end{enumerate}
    Hence, $a \notin \gloc_{\UCg}(\profile, \decomp)$ implies $a \notin \UCg(\profile)$, completing the proof.
\end{proof}
\section{On \Cref{sec:cc_transform} (\nameref{sec:cc_transform})}

In this section, we provide the proofs omitted from \Cref{sec:cc_transform} of the main body, as well as an extended discussion of PQ-trees and clone-aware axioms. 

\subsection{Extended discussion of clone structures and PQ-trees}\label{subsec:extended_pqtrees}
Here, we expand on our discussion of the PQ-trees, first defined by \citet{Booth76:Testing} and later used by \citet{Elkind10:Clone} for representing clone sets. For the full set of formal definitions, see \citet{Elkind10:Clone}. 

Given $\profile$ we can use a PQ-tree to represent its \emph{clone structure} $\family(\profile) \subseteq \mathcal{P}(\cand)$, which is the collection of \textit{all} clone sets on $\profile$. For example, if $\profile$ is the profile from \Cref{fig:eg_prof}, then $\family(\profile)=\{\{a\},\{b\}, \{c\}, \{d\}, \{b,c\}, \{a,b,c,d\}\}$.
Given a set of candidates $\cand=\{a_i\}_{i \in [m]}$, a PQ-tree $T$ over $\cand$ is an ordered tree where the leaves of the tree correspond to a particular permutation of the elements of $\cand$.
Each internal node is either of ``type P'' or of ``type Q''. 
If a node is of type P, then its children can be permuted arbitrarily. 
If a node is of type Q, then the only allowable operation is the reversal of its children's order.

\citet{Elkind10:Clone} begin by defining two special types of clone structures: a \emph{maximal} clone structure (which they also call a \emph{string of sausages}) and a \emph{minimal} clone structure (also called a \emph{fat sausage}). 
A string of sausages corresponds to the clone structure that arises when all rankings in the profile $\profile$ consist of a single linear order (WLOG, $\sigma_1 : a_1 \succ a_2 \succ \cdots \succ a_m$) or its reversal. Then $\family(\profile) = \{ \{a_k\}_{i \leq j}: i \leq j\}$, meaning that the clone structure contains all intervals of candidates in $\sigma_1$. The \emph{majority ranking} of the Q-node is $\sigma_1$ or its reverse, depending on which one appears more in $\profile$. The ``opposite'' scenario (a fat sausage) is when $\family(\profile') = \{A\} \cup \{\{a_i\}\}_{i \in [m]}$, meaning we only have the trivial clone sets in our structure. This arises, for example, when $\profile'$ corresponds to a cyclic profile on $A$, \emph{i.e.}, $\profile' = (\sigma_1', \ldots, \sigma'_m)$, and the preferences of the $i$-th voter are given by $\sigma'_i : a_i \succ_{\sigma_i} a_{i+1} \succ_{\sigma_i} \cdots \succ_{\sigma_i} a_m \succ_{\sigma_i} a_1 \succ_{\sigma_i} \cdots \succ_{\sigma{i}} a_{i-1}$.

We need a few more definitions before describing the construction of the PQ-tree.
Let $\mathcal{F}$ be a family of subsets on a finite set $F$, and likewise $\mathcal{E}$ for $E$, where $F \cap E = \emptyset$.
Then, we can \emph{embed} $\mathcal{F}$ into $\mathcal{E}$ as follows: given $e \in E$, we replace each set $X$ containing $e$ with $(X \setminus \{e\}) \cup F$. 
The resulting family of subsets is denoted by $\mathcal{E}(e \rightarrow \mathcal{F})$.
The inverse operation of embedding is called \emph{collapsing}; note that for a family of subsets $\family$ on $A$ to be collapsible, it should contain a set that does not intersect non-trivially (\emph{i.e.}, not as a sub/superset) with any other set in $\family$), which motivates the definition of a \emph{proper subfamily} of $\mathcal{F}$:

\begin{definition}\label{def:irreducible}
Let $\mathcal{F}$ be a family of subsets on a finite set $F$. A subset $\mathcal{E} \subseteq \mathcal{F}$ is called a \emph{proper subfamily} of $\mathcal{F}$ if there is a set $E \in \mathcal{F}$ such that (i) $\mathcal{E} = \{F \in \mathcal{F} \mid F \subseteq E\}$; (ii) for any $X \in \mathcal{F} \setminus \mathcal{E}$, either $E \subseteq X$ or $X \cap E = \varnothing$, (iii) $E$ is a proper subset of $F$. 
A family of subsets with no proper subfamily is called \emph{irreducible.}
\end{definition}

The key result that we require in the construction of a PQ-tree is that \emph{any irreducible clone structure is either a fat sausage or a string of sausages}   \cite[Thm. 3.10]{Elkind10:Clone}.
Given a clone structure $\family \subseteq \mathcal{P}(\cand)$, we construct its corresponding PQ-tree $T(\family)$ iteratively:
\begin{enumerate}
    \item Pick some non-singleton, irreducible minimal set of clones  $\mathcal{E}_1 \subseteq \family$. By \Cref{def:irreducible}, there exists $C_1 \in \family$ such that $\mathcal{E}_1 = \{F \in \mathcal{F} \mid F \subseteq C_1\}$. 
    \item Update $\family$ to $\family(\mathcal{E}_1 \rightarrow C_1)$, \emph{i.e.}, substitute all appearances of the members of $C_1$ in $\family$ by a meta-candidate $C_1$, and remove the sets in $\family$ that correspond to subsets of $C_1$. Since $C_1$ is either a subset of a superset of each $K \in \family$ is overlaps with, this transformation is well-defined.
    \item Build the subtree for $C_1$. By Theorem 3.10 in \citet{Elkind10:Clone}, $C_1$ is either a fat sausage or a string of sausages.
    If it is a fat sausage, then $C_1$ is set to be of type P- and label it as the $\odot$-product of the candidates in $C_1$. If $|C_1|=2$, then it is both a fat sausage and a string of sausages. 
    In this case, we treat it as a string of sausages sausage (following the convention by \citet{Elkind10:Clone}).
    The candidates in $C_1$ are placed as the children leaves in the subtree.
    If it is a string of sausages, then $C_1$ is of type Q-, and we label it as the $\oplus$-product of the candidates in $C_1$.
    The candidates in $C_1$ are similarly placed as the children leaves in the subtree, following the order dictated by $C_1$.
    \item We repeat the previous three steps for $C_i$, with $i=2,\ldots$,  until we cannot find any non-singleton, irreducible, minimal set. For any child node of $C_i$ that corresponds to a previously-collapsed subset $C_j$ with $j<i$, the node is replaced with the subtree of $C_j$, already constructed by assumption. For child nodes of $C_i$ that correspond to original candidates from $A$, the node is a leaf.
    \item Eventually, no proper irreducible subfamilies are left, and all of the remaining candidates form either a string of sausages or a fat sausage, so we place them as children of the root of $T(\family)$, similarly labeling it as type P or Q. 
\end{enumerate}

The order in which we choose $C_i$ does not impact the final construction, as the irreducible proper subsets of a clone structure $\family$ is non-overlapping \cite[Proposition 4.2.]{Elkind10:Clone}, implying a unique decomposition of candidates into irreducible proper subsets at each step. This ensures that the PQ-tree of a preference profile is unique~\citep{Karpov19:Group}.

\subsection{Discussion of PQ-tree algorithms}\label{sec:app:pq-trees}

The original PQ-tree algorithm is due to Booth and Luecker, who introduced it as a way to represent a family of permutations on a set of elements~\citep{Booth76:Testing}.
Later, \citet{Elkind10:Clone} showed its use in the context of computational social choice.
Cornaz, Galand, and Spanjaard carefully analyze the Booth and Luecker algorithm in the context of voting rules and establish the runtime of $O(nm^3)$ that we use in Lemma~\ref{lemma:pq-poly}~\citep{Cornaz13:Kemeny}.
In this section, we provide some more context on the general relationship between PQ-tree constructions (and tournament decomposition tree constructions) with the graph theoretic literature on modular tree decomposition (particularly with the modular tree decomposition algorithm by \citet{Capelle02:Graph}, following the observations made by \citet{Brandt11:Fixed}.

\citet{Brandt11:Fixed} study CC tournament solutions, following the definition of composition consistency for tournaments first given by~\citep{Laffond96:Composition}.
They provide a \emph{decomposition tree of a tournament} $T$ meant for efficiently implementing CC tournament solutions.
In their analysis, they use what they call the \emph{decomposition degree} of a tournament, which is a parameter that reflects its decomposability (the lower the degree, the better well-behaved its decomposition).
Their decomposition tree is the tournament version of the PQ-tree construction of \citet{Elkind10:Clone}: both use trees with two different types of internal nodes as a suitable way of representing clone structures.
The correspondence between the two constructions is the following:
\begin{enumerate}
    \item \citet{Elkind10:Clone} use a PQ-tree to represent clone structures given a profile $\profile$, while \citet{Brandt11:Fixed} use a decomposition tree to represent components (Definition~\ref{appdef:component_tour}) given a tournament $T$.
    \item \citet{Elkind10:Clone} divide the internal nodes into types P and Q, whereas \citet{Brandt18:Extending} calls them \emph{irreducible} and \emph{reducible}, respectively (but the definition is the same one).
\end{enumerate}

Besides the naming of the internal nodes and the difference between having a profile $\sigma$ versus a tournament $T$ as input, given the equivalence between components in $T$ and clones in $\sigma$, \citet{Brandt18:Extending}'s definition of a decomposition tree of $T$ is equivalent to \citet{Elkind10:Clone}'s definition of a PQ-tree for $\sigma$.
In particular, claims relating to the running time required to compute a decomposition tree of a tournament can be transferred to the running time required to compute PQ-trees.

\citet{Brandt18:Extending} make the following two observations about the running time required to compute the decomposition tree of a tournament $T$:
\begin{enumerate}
    \item First, we compute a \emph{factorization permutation} of $T$, which is a permutation of the alternatives in $A$ such that each component of $T$ is a contiguous interval in the permutation.
    \citet{McConnell05:Linear} provide a linear time algorithm for computing a factorizing permutation of a tournament in linear time
    \item Second, given $T$ and a factorization permutation of $T$, we can use the graph theoretic algorithm by \citet{Capelle02:Graph} to obtain the decomposition tree of $T$.
\end{enumerate}

In our settings of profiles, we can adapt the running time argument from~\citep{Brandt18:Extending} as follows:
\begin{enumerate}
    \item In the case of profiles $\profile$, we do not need to do more work to compute the factorization permutation of $T$; we can read it directly from $\profile$ (from any one voter's ranking).
    By definition of a clone set, \emph{every} voter ranks the members of a clone set continguously in their ranking. 
    Therefore, every single voter's ranking is a factorization permutation of $\profile$.
    Thus, computing the factorization permutation requires $O(|A|)$ running time.
    \item The graph theoretic algorithm \citet{Capelle02:Graph} that \citet{Brandt18:Extending} use for computing decomposition tree tournaments is not directly related to tournaments (or to computational social choice).
    Rather, \citet{Capelle02:Graph} deal with a broad definition of a \emph{modular decomposition} of a directed graph.
    Zooming out from tournaments, for a given graph $G=(V, E)$, a decomposition tree $T_G$ is such that the vertices of $G$ are in one-to-one correspondence with the leaves of $T_G$, and the internal nodes correspond to subsets of $V$.
    They call the nodes of
    a decomposition tree (and the sets of vertices they induce) \emph{decomposition sets}.
    They study the general case where the decomposition sets correspond to \emph{modules}: 
    \begin{definition}
        A \emph{module} in a graph $G = (V, E)$ is a set $X$ of vertices such that 1) if $y \in V \setminus X$, then $y$ has either directed edges to all members of $X$ or to none of them, and 2) all members of $X$ have either directed edges to $y$, or none of them do.
    \end{definition}
    Intuitively, a set of vertices in a graph forms a module if every vertex in $V \setminus X$ has a ``uniform'' relationship to all members of $X$~\citep{McConnell05:Linear}.
    Note that the definition of a module imposes no requirements on whether the vertices in $X$ should be connected or not.
    Observe also that connected components are a particular case of modules.
    A module is \emph{strong} if it does not overlap with any other module.
    Then, \citet{Capelle02:Graph} call the decomposition tree of a graph $G$ into its strong modules the \emph{modular decomposition tree} of $G$, and they provide a (complicated) linear time algorithm for computing it (which we can treat as a black-box algorithm).
    
    The literature on modular decomposition graphs is extensive and a popular topic in graph theory. 
    However, as noted in \citet{Brandt18:Extending}, the literature on composition-consistency (and in social choice more broadly) and on modular decompositions in graph theory is not well-connected.
    In this section, we help clarify part of this connection by detailing how we can use the modular decomposition algorithm by \citet{Capelle02:Graph} to compute the PQ-tree.
    Given that the notion of a module is the graph-theoretic generalization of clone sets in profiles and components in tournaments, we hope that there can be further interesting connections between the two fields.
\end{enumerate}

As observed by \citet{Brandt18:Extending},
to compute the decomposition tree of a tournament, we can simply input the graph induced by the tournament (\emph{i.e.}, we draw an edge from $a$ to $b$ if $a$ beats $b$) to the modular decomposition tree algorithm of \citet{Capelle02:Graph}.
In our case, for computing the PQ-tree using the algorithm of~\citep{Capelle02:Graph}, we need to input a graph $G$ built from $\profile$ such that the modules of $G$ are in bijection with the clone sets of $\profile$. 

\subsection{Clone-aware axioms}\label{appsec:ca-axioms}
First, we introduce three axioms show that they are not necessarily satisfied by $f^{CC}$ (\Cref{def:cc-transform}), even if $f$ satisfies them, implying our CC transformation does not preserve them. We will then introduce \emph{clone-aware} relaxations of these axioms, which are in fact preserved by the CC transformation (see \Cref{thm:cc_transform}). 

\begin{definition}\label{def:mono}
    An SCF $f$ satisfies \emph{monotonicity} if $a \in f(\profile)$ implies $a \in f(\profile')$ if for all $i \in N$ and $b,c \in A \setminus\{a\}$, we have $a \succ_{\sigma_i} b \Rightarrow a \succ_{\sigma'_i} b$ and $b \succ_{\sigma_i} c \Rightarrow b \succ_{\sigma'_i} c$.
\end{definition}

Inuitively, monotonicity dictates that promoting a winner in a profile while keeping all else constant should not cause them to lose. We see that monotonicity is not necessarily preserved by our CC transformation.

\begin{example}\label{ex:counter-mono}
    Consider Plurality Voting $(PV)$, which is monotonic, and the profile $\profile$ from \Cref{fig:counter_mono}. Notice $\{a_1,a_2,a_3\}$ is a fat sausage and is grouped up by the PQ tree. $\{a_1,a_2,a_3\}$ wins against $b$ in the root, and the $a_1$ wins against $a_2$ and $a_3$ with 5 plurality votes, hence $PV^{CC}(\profile)=\{a_1\}$. However, say \emph{one} of the rightmost voters move $a_1$ up, submitting $a_1 \succ b \succ a_2 \succ a_3$ instead. Then there are no longer any nontrivial clone sets, and $a_3$ wins with 4 plurality votes ($a_1$ only has 3). Hence, with this new profile (call it $\profile$'), we have $PV^{CC}(\profile')=PV(\profile')=\{a_3\}$, showing that $PV^{CC}$ is \emph{not} monotone.
\end{example}
\begin{figure}
    \centering
    \begin{subfigure}{.49\textwidth}
        \centering
        \begin{tabular}{|c|c|c|c|}
            \hline
            2 voters & 4 voters & 2 voters & 3 voters \\
            \hline
            \cellcolor{yellow!25}$a_1$ & \cellcolor{green!25}$a_3$ & \cellcolor{orange!25}$a_2$ & \cellcolor{blue!25}$b$ \\ \hline
            \cellcolor{orange!25}$a_2$ & \cellcolor{yellow!25}$a_1$ & \cellcolor{green!25}$a_3$ & \cellcolor{yellow!25}$a_1$ \\ \hline
            \cellcolor{green!25}$a_3$ & \cellcolor{orange!25}$a_2$ & \cellcolor{yellow!25}$a_1$ & \cellcolor{orange!25}$a_2$ \\ \hline
            \cellcolor{blue!25}$b$ & \cellcolor{blue!25}$b$ & \cellcolor{blue!25}$b$ & \cellcolor{green!25}$a_3$ \\
            \hline
        \end{tabular} \caption{Example profile $\boldsymbol{\sigma}$}\label{fig:counter_mono}
    \end{subfigure}
    \begin{subfigure}{.49\textwidth}
        \centering
        \begin{tabular}{|c|c|c|c|}
            \hline
            6 voters & 5 voters & 2 voters & 2 voters \\ \hline
            \cellcolor{yellow!25} $a_1$ & \cellcolor{blue!25} $c$ & \cellcolor{green!25} $b$ & \cellcolor{green!25} $b$ \\ \hline
            \cellcolor{red!25} $a_2$ & \cellcolor{red!25} $a_2$ & \cellcolor{blue!25} $c$ & \cellcolor{blue!25} $c$ \\ \hline
            \cellcolor{green!25} $b$ & \cellcolor{yellow!25} $a_1$ & \cellcolor{yellow!25} $a_1$ & \cellcolor{red!25} $a_2$ \\ \hline
            \cellcolor{blue!25} $c$ & \cellcolor{green!25} $b$ & \cellcolor{orange!25} $z$ & \cellcolor{yellow!25} $a_1$ \\ \hline
            \cellcolor{orange!25} $z$ & \cellcolor{orange!25} $z$& \cellcolor{red!25} $a_2$ & \cellcolor{orange!25} $z$ \\
            \hline
            \end{tabular}
        \caption{Example profile $\profile$}\label{fig:isda-counter}
    \end{subfigure}
\caption{Two example profiles}
\end{figure}
Given a preference profile $\profile=(\sigma_1,\sigma_2,\ldots, \sigma_n) \in \mathcal{L}(A)^n$ over voters $N=[n]$ and a new $(n+1)$th voter with ranking $\sigma_{n+1}$ over $A$, we denote by $\profile+\sigma_{n+1}$ the profile $(\sigma_1,\sigma_2,\ldots, \sigma_n, \sigma_{n+1}) \in \mathcal{L}(A)^{n+1}$. Also, given any voter $i \in N \cup \{n+1\}$ with ranking $\sigma_i$ over $A$ and a non-empty subset $B \subseteq A$, we denote by $\max_{i}(B)$ the candidate in $B$ that is ranked highest by $\sigma_i$. For example, if $\sigma_i = a \succ b \succ c \succ d$, and $B=\{b,c,d\}$, then $\max_i(B)=b$.

\begin{definition}[\citealt{Brandt17:Optimal}]\label{appdef:participation}
    An SCF $f$ satisfies \emph{(optimistic) participation} if given any profile $\profile \in \mathcal{L}(\cand)^n$ and any ranking $\sigma_{n+1} \in\mathcal{L}(\cand) $, we have $\max_{n+1}(f(\profile)) \succeq_{n+1} \max_{n+1}(f(\profile+\sigma_{n+1})).$

\end{definition}
 Participation dictates that a new voter cannot hurt themselves (in terms of their most preferred winner\footnote{One can alternatively use a pessimistic definition focusing on the new voter's lowest ranked candidate in the winner set.}) by participating in the election.

\begin{example}\label{ex:counter-part}
    Once again, Plurality Voting $(PV)$, which satisfies participation, and the profile $\profile$ from \Cref{fig:counter_mono}. As explained in \Cref{ex:counter-mono}, we have $PV^{CC}(\profile)=\{a_1\}$. Consider $\sigma_{n+1}: a_1 \succ b \succ a_2 \succ a_3$ and $\profile'=\profile+\sigma_{n+1}$. Since there are no non-trivial clone sets in $\profile'$, we have $PV^{CC}(\profile')=PV(\profile')= \{a_3\}$. Since $a_1 \succ_{n+1} a_3$, this shows $PV^{CC}$ violates participation, and that someone ranking $\sigma_{n+1}$ is better off staying away from this election. 
\end{example}
    
    Given a set $E$ and a family of its subsets $\mathcal{E} \subseteq 2^E$, for any $a \in E$, we denote $\mathcal{E}-\{a\} = \{K \setminus \{a\}: K \in \mathcal{E}\}$ 

\begin{definition}\label{appdef:isda}
    An SCF $f$ satisfies \emph{independence of Smith-dominated alternatives (ISDA)} if given any profile $\profile \in \mathcal{L}(\cand)^n$ over candidates $A$ and any candidate $a\in A$ such that $a \notin \textit{Sm}(\profile)$ and $\family(\profile \setminus\{a\}) = \family(\profile) - \{a\}$, we have $f(\profile)=f(\profile \setminus \{a\})$
\end{definition}

In words, the winner(s) under any rule satisfying ISDA is not affected by the addition of a non-Smith candidate.

\begin{example}\label{ex:counter-isda}
 Consider Beatpath ($BP$), which satisfies ISDA~\citep{Schulze10:New} and the profile $\profile$ from \Cref{fig:isda-counter}, which is a minor modification from the counterexample for $BP$ in the proof of \Cref{thm:cc_fails}. With $z$ in the ballot, there are no non-trivial clone sets, so $BP^{CC}(\profile)=BP(\profile)=\{a_1,a_2\}$. Notice also that $Sm(\profile)=A \setminus\{z\}$, so $z$ is indeed a non-Smith candidate. With $z$ gone however, $\{a_1,a_2\}$ is a clone set again, and hence $BP^{CC}$ first groups them up, picks $\{a_1,a_2\}$, and then picks $a_1$ in the restriction. Hence, $BP^{CC}(\profile \setminus\{z\}) =\{a_1\}$, showing that the CC-transformation does not necessarily preserve ISDA.
\end{example}

The common thread in \Cref{ex:counter-mono,ex:counter-part,ex:counter-isda} is that the changes in profile (whether promoting a winner on a ranking or the addition/removal of a voter/candidate) significantly alters the clone structure of the profile, causing the behavioral of any $f^{CC}$ to significantly change. Instead, we can relax each of these axioms by limiting the changes they consider to those that leave the clone structure unaffected. We present these relaxations (called the clone-aware version of each axiom) below. These new axioms implicitly assume that the clone structures are \textit{inherent}, based on the candidates' location is some perceptual space (which is in fact the interpretation put forward by \citet{Tideman87:Independence}), so any ``realistic'' change we will do to the profile will not alter the clone sets.

\camono*

\begin{definition}\label{appdef:clone_participation}
    An SCF $f$ satisfies \emph{clone-aware (optimistic) participation (participation$\ca$)} if given any profile $\profile \in \mathcal{L}(\cand)^n$ and any ranking $\sigma_{n+1} \in\mathcal{L}(\cand) $ such that $\family(\profile)=\family(\profile+\sigma_{n+1})$, we have $\max_{n+1}(f(\profile)) \succeq_{n+1} \max_{n+1}(f(\profile+\sigma_{n+1})).$

\end{definition}

\begin{definition}\label{appdef:clone_isda}
    An SCF $f$ satisfies \emph{clone-aware ISDA (ISDA$\ca$)} if given any profile $\profile \in \mathcal{L}(\cand)^n$ over candidates $A$ and any candidate $a\in A$ such that $a \notin \textit{Sm}(\profile)$ and $\family(\profile \setminus\{a\}) = \family(\profile) - \{a\}$, we have $f(\profile)=f(\profile \setminus \{a\})$
\end{definition}

\subsection{Proof of \Cref{thm:cc_transform}}
In this section, we prove the theoretical guarantees of our CC-transform for SCFs.

\cctransform* 

\begin{proof}
We prove each condition one by one.
    \paragraph{Condition 1.} We first prove an intermediary lemma.
    \begin{lemma}\label{lemma:triv decomp}
        Given neutral SCF $f$ and profile $\profile$ over candidates $A$, we have $f(\profile)=\gloc_{f}(\profile,\decomp_{triv})$, where $\decomp_{triv}=\{\{a\}\}_{a \in A}$.
    \end{lemma}
    \begin{proof}
        Note that $\profile^{\decomp_{triv}}$ is isomorphic to $\profile$, with each $a\in A$ replaced with $\{a\}$. By neutrality, we must have $f(\profile^{\decomp_{triv}}) =\{\{a\}\}_{a \in f(\profile)}$. Moreover, since an SCF always returns a non-empty subset, $f(\profile|_{\{a\}})=\{a\}$ for any $a \in A$. This gives us
        \begin{align*}
            \gloc_f(\profile, \decomp_{triv})= \bigcup_{K \in f(\profile^{\decomp_{triv}})} f(\profile|_{K}) =   \bigcup_{a \in f(\profile)} f(\profile|_{\{a\}})= \bigcup_{a \in f(\profile)} \{a\}= f(\profile).
        \end{align*}
    \end{proof}

    If $\profile$ has no non-trivial clone sets, then $\family(\profile)$ is a fat sausage, so the PQ tree of $\profile$ (say $T$) is simply a single P-node (say $\node$) with all of the candidates in $A$ as its children leaf nodes. Since decomp$(\node,T)=\{\{a\}\}_{a \in A}=\decomp_{triv}$,  \Cref{alg:cc-transform} simply outputs $f^{CC}(\profile)=\gloc_{f}(\profile,\decomp_{triv})$. By \Cref{lemma:triv decomp}, this implies $f^{CC}(\profile)=f(\profile)$.

    \paragraph{Condition 2.} The fact that $f^{CC}$ is neutral follows from the neutrality of $f$ and that $\Cref{alg:cc-transform}$ is robust to relabeling of candidates. To prove $f^{CC}$ satisfies $CC$, we first prove an important lemma.

    \begin{lemma}\label{lemma:fcc_intermediary}
        Given neutral SCF $f$ and profile $\profile$, say $\decomp, \decomp'$ are two clone decomposition with respect to $\profile$, such that $\decomp= \{K_1,K_2,\ldots, K_z\} \cup \{\{a\}\}_{a \in A \setminus \left( \bigcup_{i \in [z]}K_i \right)}$ for some $z \in \mathbb{Z}_{\geq 0}$, and there exists some $K \subseteq A \setminus \left( \bigcup_{i \in [z]}K_i \right)$ with $|K|>1$ that satisfies $\decomp' = \decomp \setminus \calD \cup \{K\}$, where $\calD=\{\{a\}\}_{a \in K}$. In words, $\decomp'$ is the same decomposition as $\decomp$, except a group of singleton clone sets in $\decomp$ is now combined into a single new clone set $K$. Then $\gloc_{f^{CC}}(\profile, \decomp)=\gloc_{f^{CC}}(\profile, \decomp')$.
    \end{lemma}
    The proof of \Cref{lemma:fcc_intermediary} relies on the observation that the PQ trees for $\profile^\decomp$ and  $\profile^{\decomp'}$ are identical, except the subtree(s) corresponding to $\calD$ in the former (by \Cref{lemma:pq_clones}) is replaced by a single leaf node $K$ in the latter. Hence, we first show that \Cref{alg:cc-transform} proceeds identically on inputs $\profile^\decomp$ and  $\profile^{\decomp'}$, picking the same set of leaves from $\decomp \setminus \calD$ in both cases and returning \emph{some} descendants of $\calD$ in the former case if it returns $K$ in the latter. We then show that \emph{if} some descendants of $\calD$ are returned by the algorithm on input $\profile^\decomp$, these are exactly the same as the output of the algorihtm when run on input $\profile|_K$. Combining these gives us the lemma statement.
    \begin{proof}[Proof of \Cref{lemma:fcc_intermediary}]
        Say $\decomp, \decomp'$ satisfies the conditions in the lemma statement. Say $T$ and $T'$ are the PQ trees of $\profile^\decomp$ and $\profile^{\decomp'}$, respectively.\footnote{It is worth making a notational point here: when dealing with PQ trees of a profile $\profile$ over a candidates $A$, and each leaf node corresponded to a candidate in $a \in A$ and each internal node could be represented as a subset $\node \subseteq A$. Since $T$ (resp., $T'$) is the PQ trees of a summary $\profile^\decomp$ (resp., $\profile^{\decomp'}$), each leaf node now corresponds to a clone set $K \in \decomp$ (resp., $K' \in \decomp'$) and each internal node can be represented as a subset $\calB \subseteq \decomp$ (resp., $\calB' \subseteq \decomp'$).} Given an interval node $\calB \subseteq \decomp$ (resp., $\calB' \subseteq \decomp')$ in $T$ (resp., $T'$), we denote by $T(\calB$) (resp., $T'(\calB')$) the subtree of $T$ (resp., $T'$) rooteed at $\calB$ (resp., $\calB'$). We will be comparing the structure of $T$ and $T'$, using the fact that PQ trees are built by iteratively collapsing irreducible subfamilies (see \Cref{subsec:extended_pqtrees} above, and also \citet{Elkind10:Clone}). Since $\mathcal{D}=\{\{a\}\}_{a \in K}$ is a clone set with respect to $\profile^\decomp$ (which follows from the assumption that $K$ is a clone set with respect to $\profile$), by \Cref{lemma:pq_clones}, there are two options:
        \begin{itemize}
            \item $\mathcal{D}$ is a node of $T$, in which case its members are leaves of a subtree ($T(\calD)$). In this case, the tree $T'$ is identical to $T$, except $T(\calD)$ is replaced by a single leaf node $K$. The PQ tree for $\profile|_{K}$, on the other hand, is exactly $T(\calD)$ (except the leaf for each singleton $\{a\}$ is replaced with the leaf for $a$).
            \item $\mathcal{D}$ union of an interval of nodes ($\{B_k(\calB,T)\}_{i \leq k \leq j}$ for some $i < j$) that are adjacent children of the same Q-node $\calB \subseteq \decomp$, in which case its members are leaves of the same interval of subtrees ($\{T(B_k(\calB,T))\}_{i \leq k \leq j}$ ) . In this case, the tree $T'$ is identical to $T$, except the children of $\calB$ corresponding to $\calD$ ($\{T(B_k(\calB,T))\}_{i \leq k \leq j}$) are now replaced by a single leaf node $K$, placed in appropriate place in the majority ranking of $\calB$ (in this case, $i$th position), which is well-defined, since the replaced children formed an interval. The PQ tree for $\profile|_{K}$, on the other hand, is exactly $\{T(B_k(\calB,T))\}_{i \leq k \leq j}$, united by a single Q-node that is the root of the tree (except the leaf for each singleton $\{a\}$ is replaced with the leaf for $a$).
        \end{itemize}
        Now take any internal node $\calB \subseteq \decomp$ of the tree $T$ such that either $\calD \subsetneq \calB$ or  $\calD \cap \calB =\emptyset$ (in words, $T(\calB)$ either strictly contains $\calD$, or is not overlapping with it $\calD$ at all). In both  cases, there is (by the analysis above) a corresponding node $\calB'$ in the tree $T'$: if $\calD \subseteq \calB$, then $\calB'=\calB \setminus \calD \cup \{K\}$, and if $\calD \cap \calB =\emptyset$ then $\calB'=\calB$. We would like to compare the children node that are enqueued by \Cref{alg:cc-transform} if (\textbf{case (a)}) it enqueues $\calB$ when run on input $\profile^\decomp$ versus if (\textbf{case (b)}) it enqueues $\calB'$ when run on input $\profile^{\decomp'}$. We consider each possible scenario:
        \begin{enumerate}
            \item[1.] If $\calD \cap \calB =\emptyset$. In this case, $\calB'=\calB$ so the algorithm proceeds the same way in both cases \casea and \caseb, enqueing the same children regardless of whether $\calB$ is a Q-node or a P-node.
            \item[2.] If $\calD \subsetneq \calB$, and $\calB$ has a child node $\mathcal{E}$ such that $\calD \subseteq \mathcal{E}$. In words, $\calB$ is either a non-immediate ancestor of the subtree(s) corresponding to $\calD$ or the parent node of a single subtree $T(\calD)$. In this case, decomp($\calB',T'$)=  decomp($\calB,T)\setminus \{\mathcal{E}\} \cup \{\mathcal{E}'\}$, where $\mathcal{E}'=\mathcal{E} \setminus \mathcal{D} \cup \{K\}$. Then, $\profile^{\text{decomp}(\calB,T)}$ and $\profile^{\text{decomp}(\calB',T')}$ are isomorphic (with $\mathcal{E}$ relabeled as $\mathcal{E}'$). Hence, the algorithm proceeds the same way in both cases \casea and \caseb, enqueing the same children regardless of whether $\calB$ is a Q-node or a P-node, since $f$ is neutral. In other words, any child node $\mathcal{F} \neq \mathcal{E}$ will be enqueued in case \casea iff it is enqueued in case \caseb; $\mathcal{E}$ will be enqueued in case \casea iff $\mathcal{E}'$ is enqueued in case \caseb.
            \item[3.] If $\calD \subsetneq \calB$ and no child node of $\calB$ entirely contains $\calD$. By \Cref{lemma:pq_clones}, this implies that $\calB$ is a Q-node and $\exists i,j: 0 \leq i < j \leq |\text{decomp}(\calB,T)|$ such that $\calD=\bigcup_{k: i \leq k \leq j} B_k(\calB, T)$. In words, $\calD$ corresponds to the leaves of multiple (specifically, $j-i+1$) subtrees ($\{T(B_k(\calB,T))\}_{i \leq k \leq j}$) whose roots ($\{B_k(\calB,T)\}_{i \leq k \leq j}$) are an interval of children nodes of $\calB$. We cannot have $j-i+1 =|\text{decomp}(\calB,T)|$, since this would imply $\calB= \calD$, even though we assumed $\calD$ is a strict subset of $\calB$. Hence, $\calB'$ (the node in $T'$ corresponding to $\calB$) is a Q-node with $|\text{decomp}(\calB,T)|-(j-i)$ children nodes,\footnote{If $|\text{decomp}(\calB,T)|-(j-i)=2$, then $\calB'$ is technically both a Q- and a P- node, which does not affect our analysis since \Cref{alg:cc-transform} treats these cases identically.} with $\{T(B_k(\calB,T))\}_{i \leq k \leq j}$ replaced by a single leaf node $K$ (respecting the rest of the order). Say $\ell = |\text{decomp}(\calB,T)|$ and $\ell'=|\text{decomp}(\calB',T')|=|\text{decomp}(\calB,T)|-(j-i)$. By the clone set definition, we have that $\profile^\decomp|_{\{B_1(\calB,T),B_2(\calB,T)\}}$ and $\profile^{\decomp'}|_{\{B_1(\calB',T'),B_2(\calB',T')\}}$ are isomorphic, and since $f$ is neutral, we have  $$B_i(\calB,T)\in f(\profile^\decomp|_{\{B_1(\calB,T),B_2(\calB,T)\}}) \Leftrightarrow B_i(\calB',T')\in f(\profile^{\decomp'}|_{\{B_1(\calB',T'),B_2(\calB',T')\}})$$ for $i \in \{1,2\}$. Then we consider the three possible cases separately:
            \begin{itemize}
                \item[3a.] If $f(\profile^{\decomp}|_{\{B_1(\calB,T),B_2(\calB,T)\}})=\{B_1(\calB,T)\}$, then $B_1(\calB,T)$ gets enqueued in case \casea, and $B_1(\calB',T')$ gets enqueued in case \caseb. If $i>1$ (\emph{i.e.}, $B_1(\calB,T) \cap \calD =\emptyset$)  then $B_1(\calB,T)=B_1(\calB',T')$, so the same node gets enqueued in both cases. If $i=1$, then $B_1(\calB',T')=\{K\}$, so $B_1(\calB,T)  \subseteq \calD$ gets enqueued in \casea and $\{K\}$ gets enqueued in \caseb.
                \item[3b.] If $f(\profile^{\decomp}|_{\{B_1(\calB,T),B_2(\calB,T)\}})=\{B_2(\calB,T)\}$, then $B_\ell(\calB,T)$ gets enqueued in case \casea and $B_{\ell'}(\calB',T')$ gets enqueued in case \caseb. If $j<\ell$ (\emph{i.e.}, $B_\ell(\calB,T) \cap \calD =\emptyset$)  then $B_{\ell}(\calB,T)=B_{\ell'}(\calB',T')$, so the same node gets enqueued in both cases. If $j=\ell$, then $B_{\ell'}(\calB',T')=\{K\}$, so $B_\ell(\calB,T)  \subseteq \calD$ gets enqueued in \casea and $\{K\}$ gets enqueued in \caseb.
                \item[3c. ] $f(\profile^{\decomp}|_{\{B_1(\calB,T),B_2(\calB,T)\}})=\{B_1(\calB,T),B_2(\calB,T)\}$, then all the children nodes get enqueued in both cases
            \end{itemize}
        \end{enumerate}
    Together, the cases above imply that starting from corresponding nodes $\calB$ and $\calB'$ in $T$ and $T'$ (respectively) that either contain $\calD$ and $\{K\}$ (respectively) or do not overlap with them,
    \begin{itemize}
        \item \Cref{alg:cc-transform} enqueues any child node of $\calB$ that either contains $\calD$ or do not overlap with it in \casea if and only if it enqueues corresponding childnode of $\calB'$ in \caseb
        \item \Cref{alg:cc-transform} enqueues some subtree(s) corresponding to $\calD$ in \casea if and only if it outputs $|K|$ as one of the winners in \caseb.
    \end{itemize}
    Since \Cref{alg:cc-transform} run on both input $\profile^\decomp$ or input $\profile^{\decomp'}$ start at their root nodes (which indeed contain $\calD$ and $\{K\}$, respectively), inductively applying this argument implies that for all $K' \in \decomp \setminus \calD$, we have:
    \begin{align}
        K' \in f^{CC}(\profile^\decomp) &\iff  K' \in f^{CC}(\profile^{\decomp'})\label{eq:non_d}\\
        \calD \cap f^{CC}(\profile^\decomp) \neq \emptyset &\iff  K \in f^{CC}(\profile^{\decomp'})\label{eq:yes_d}
    \end{align}
    
    What remains to be shown is if $ \calD \cap f^{CC}(\profile^\decomp) \neq \emptyset$, then $\calD \cap f^{CC}(\profile^\decomp)=\{\{a\}\}_{a \in f^{CC}(\profile|_{K})}$. In words, we must show that  if \Cref{alg:cc-transform} outputs any descendants of $\calD$ in case \casea, then these decedents are the same as those that are output by the algorithm on input $\profile|_K$. Consider the three cases, assuming $ \calD \cap f^{CC}(\profile^\decomp) \neq \emptyset$: 
        \begin{itemize}
            \item $\calD$ corresponds to a single subtree $T(\calD)$. Then by case (2.) above, we have that $\calD$ will be enqueued by \Cref{alg:cc-transform} when running on input $\profile^\decomp$ annd $\{K\}$ will be enqueued by the algorithm when run on input $\profile^{\decomp'}$. Since the PQ-tree for $\decomp|_K$ is identical to $T(\calD)$, the descendants of $\calD$ that will be output by \Cref{alg:cc-transform} when running on input $\profile^\decomp$ (after dequeuing $\calD$) are the same as the ones the algorithm would output on input $\profile|_K$.
            \item $\calD$ corresponds to an interval of children nodes ($\{B_k(\calB,T)\}_{i \leq k \leq j}$ ) under a Q-node ($\calB$) in $T$ and $f(\profile^{\decomp}|_{{\{B_1(\calB,T),B_2(\calB,T)\}})=\{B_1(\calB,T),B_2(\calB,T)\}}$. Then by case (3c.) above, \Cref{alg:cc-transform} will enqueue all of these children nodes when running on input $\profile^\decomp$ and will enqueue $\{K\}$ when running on input $\profile^{\decomp'}$. The root of the PQ tree of $\profile|_K$ (say $T_K$) is a Q-node (denoted $K$) connecting these subtrees ($\{T(B_k(\calB,T))\}_{i \leq k \leq j}$ ). Since $f$ is neutral, we have $f(\profile^{\decomp}|_{\{B_1(K,T_K),B_2(K,T_K)\}})=\{B_1(\calB,T_K),B_2(\calB,T_K)\}$ as, by definition of Q-nodes, any voter $i \in \voter$ will have $B_1(\calB,T) \succ_{i} B_2(\calB,T)$ if and only if $B_1(K,T_K) \succ_{i} B_2(K,T_K)$. Hence, once again all of these subtrees will be enqueued on the first step of \Cref{alg:cc-transform} when it is run on input $\profile|_K$. The rest of the algorithm will follow identically in both cases.
            \item $\calD$ corresponds to an interval of children nodes ($\{B_k(\calB,T)\}_{i \leq k \leq j}$ ) under a Q-node ($\calB$) in $T$ with $f(\profile^{\decomp}|_{{\{B_1(\calB,T),B_2(\calB,T)\}})=\{B_1(\calB,T)\}}$ (resp., $f(\profile^{\decomp}|_{{\{B_1(\calB,T),B_2(\calB,T)\}})=\{B_2(\calB,T)\}}$). Then by case (3a.) (resp., (3b.)) above, \Cref{alg:cc-transform} will only enqueue $\mathcal{E}\equiv B_1(\calB,T)\subsetneq \calD$ (resp., $\mathcal{E}\equiv B_j(\calB,T)\subsetneq \calD$). Then, the root of the PQ tree of $\profile|_K$ (say $T_k$) is a Q-node (denoted $K$) with $B_1(K,T_K)$ (resp., $B_{j-i+1}(K,T_K)$) corresponding to $\mathcal{E}$. By neutrality of $f$, we have $f(\profile^{\decomp}|_{\{B_1(K,T_K),B_2(K,T_K)\}})=\{B_1(\calB,T_K)\}$ (resp., $f(\profile^{\decomp}|_{\{B_1(K,T_K),B_2(K,T_K)\}})=\{B_2(\calB,T_K)\}$). Therefore, the first step of \Cref{alg:cc-transform} when it is run on input $\profile|_K$ will pick the subtree corresponding to $\mathcal{E}$. The rest of the algorithm will follow identically in both cases.
    \end{itemize}
    Together, these cases show that the if any descendent of $\calD$ will be output when \Cref{alg:cc-transform} is run on $\profile^\decomp$, then they are the same as those output when its run on $\profile|_K$. In other words, if $ \calD \cap f^{CC}(\profile^\decomp) \neq \emptyset$, then $\calD \cap f^{CC}(\profile^\decomp)=\{\{a\}\}_{a \in f^{CC}(\profile|_{K})}$. Combined with (\ref{eq:non_d}) and (\ref{eq:yes_d}), this gives us
    
    \begin{align*}
        \calD \cap f^{CC}(\profile^\decomp) = \emptyset \Rightarrow \gloc_{f^{CC}}(\profile, \decomp) &=\bigcup_{K' \in f^{CC}(\profile^\decomp)} f^{CC}(\profile|_{K'})\\&=\bigcup_{K' \in f^{CC}(\profile^{\decomp'})} f^{CC}(\profile|_{K'})= \gloc_{f^{CC}}(\profile, \decomp')\text{, and}
    \end{align*}
    \begin{align*}
        \calD \cap f^{CC}(\profile^\decomp) \neq \emptyset \Rightarrow \gloc_{f^{CC}}(\profile, \decomp) &=\left(\bigcup_{K' \in f^{CC}(\profile^\decomp) \setminus \calD} f^{CC}(\profile|_{K'})\right) \cup \left(\bigcup_{K' \in f^{CC}(\profile^\decomp) \cap \calD} f^{CC}(\profile|_{K'})\right) \\
        &=\left(\bigcup_{K' \in f^{CC}(\profile^{\decomp'}) \setminus \{K\}} f^{CC}(\profile|_{K'})\right) \cup \left(\bigcup_{a \in f^{CC}(\profile|_{K})} f^{CC}(\profile|_{\{a\}})\right)\\
        &=\left(\bigcup_{K' \in f^{CC}(\profile^{\decomp'}) \setminus \{K\}} f^{CC}(\profile|_{K'})\right) \cup f^{CC}(\profile|_{K})\\
        &=\bigcup_{K' \in f^{CC}(\profile^{\decomp'})} f^{CC}(\profile|_{K'})=\gloc_{f^{CC}}(\profile, \decomp').
    \end{align*}
    In both cases, we have $\gloc_{f^{CC}}(\profile, \decomp)=\gloc_{f^{CC}}(\profile, \decomp')$, completing the proof of the lemma.
    \end{proof}
We now turn to proving $f^{CC}$ satisfies CC for any neutral $f$. Now, given any neutral SCF $f$, profile $\profile$ over $A$, and decomposition $\decomp=\{K_1,K_2,\ldots, K_\ell\}$ with respect to $\profile$, define $\decomp_i= \{K_1,K_2,\ldots, K_i\} \cup \{\{a\}\}_{a \in A \setminus \left( \bigcup_{j \in [i]}K_j \right)}$ for each $i \in [\ell]\cup\{0\}$. In words, $\decomp_i$ is the decomposition with the first $i$ clone sets in $\decomp$, and the remaining candidates are left as singletons. We have $\decomp_0=\{\{a\}\}_{a \in A}=\decomp_{triv}$ and $\decomp_\ell=\decomp$. Then
\begin{align*}
    f^{CC}(\profile)= \gloc_{f^{CC}}(\profile, \decomp_0) = \gloc_{f^{CC}}(\profile, \decomp_1) = \ldots =  \gloc_{f^{CC}}(\profile, \decomp_{\ell-1}) =  \gloc_{f^{CC}}(\profile, \decomp_\ell)=\gloc_{f^{CC}}(\profile, \decomp), 
\end{align*}
where first equality follows from \Cref{lemma:triv decomp} and subsequent inequalities follow from \Cref{lemma:fcc_intermediary}. Since $\decomp$ was arbitrarily chosen, this proves that $f^{CC}$ satisfies CC.

\paragraph{Condition 3.} Say $f$ is a composition-consistent SCF. By \Cref{def:oioc}, $f$ must be neutral. We will prove $f(\profile)=f^{CC}(\profile)$ by inducting on the depth of the PQ-tree of $\profile$ (say $T$). As a base case, say $T$ has depth one (\emph{i.e.}, it is a single leaf node). This implies there is a single candidate in $\profile$, so both $f$ and $f^{CC}$ will return that candidate. Now, assume $f$ and $f^{CC}$ agree on all profiles with PQ-trees of depth 1,2,\ldots,i-1, and say $\profile$ has a PQ-tree (say $T$) of depth $i$. Say $\cand$ is the root node of $T$ and consider two cases:
\begin{enumerate}
    \item $\cand$ is a P-node. Say $\decomp =$decomp$(K,T)$. By the recursive construction of \Cref{alg:cc-transform}, we will have $f^{CC}(\profile)= \bigcup_{\node \in f(\profile^\decomp)}f^{CC}(\profile|_B)$. Since $f$ is CC, we must also have $f(\profile)=\gloc_{f}(\profile,\decomp)= \bigcup_{\node \in f(\profile^\decomp)}f(\profile|_B)$. For each $B \in \decomp$, $\profile|_B$ can have a PQ-tree of depth at most $i-1$. Thus, by the inductive hypothesis we have $f^{CC}(\profile|_B)=f(\profile|_B)$, implying $f^{CC}(\profile)=f(\profile)$, as desired.
    \item $\cand$ is a Q-node. Say $\decomp =$decomp$(K,T)$ and $\ell =|\decomp|$. For each $i,j \in [\ell]$, say $\node_{i} = \node_i(\cand, T)$ and $B_{i,j}= \cup_{[k\in [i,j]}B_k$. Consider three cases:
    \begin{enumerate}
        \item[(2a)] $f(\profile^\decomp|_{\{B_1,B_2\}})=\{B_1\}$. By \Cref{alg:cc-transform}, we have $f^{CC}(\profile)=f^{CC}(\profile|_{B_1})$. Consider the decomposition $\decomp_1 = \{B_1, B_{2,\ell}\}$ (this is indeed a valid decomposition by the definition of a Q-node). Since $f$ is CC and neutral, we must have $f(\profile)=\gloc_f(\profile,\decomp_1)=\bigcup_{B \in f(\profile^{\decomp_1})}f(\profile|_B)=f(\profile|_{B_1})$. Since $\profile|_{B_1}$ must have a PQ-tree of depth at most $i-1$, by the inductive hypothesis we must have $f^{CC}(\profile|_{B_1})= f(\profile|_{B_1})$, implying $f^{CC}(\profile)=f(\profile)$, as desired. 
        \item[(2b)] $f(\profile^\decomp|_{\{B_1,B_2\}})=\{B_2\}$. By \Cref{alg:cc-transform}, we have $f^{CC}(\profile)=f^{CC}(\profile|_{B_\ell})$. Consider the decomposition $\decomp_2 = \{B_{1,\ell-1}, B_{\ell}\}$ (this is indeed a valid decomposition by the definition of a Q-node). Since $f$ is CC and neutral, we must have $f(\profile)=\gloc_f(\profile,\decomp_2)=\bigcup_{B \in f(\profile^{\decomp_2})}f(\profile|_B)=f(\profile|_{B_\ell})$. Since $\profile|_{B_\ell}$ must have a PQ-tree of depth at most $i-1$, by the inductive hypothesis we must have $f^{CC}(\profile|_{B_\ell})= f(\profile|_{B_\ell})$, implying $f^{CC}(\profile)=f(\profile)$, as desired. 
        \item[(2c)] $f(\profile^\decomp|_{\{B_1,B_2\}})=\{B_1, B_2\}$. By \Cref{alg:cc-transform}, we have $f^{CC}(\profile)=\bigcup_{i \in [\ell]}f^{CC}(\profile|_{B_i})$. For each $i \in [\ell-1]$, define $\decomp_i= \{B_i, B_{i+1,\ell}\}$ as a decomposition of $\profile|_{B_{i,\ell}}$. By successively using the fact that $f$ is CC and neutral, we get
        \begin{align*}
            f(\profile)&=\gloc_f(\profile,\decomp_1)= \bigcup_{\node \in f(\profile^\decomp_{1})} f(\profile|_B)= f(\profile|_{B_1}) \cup f(\profile|_{B_{2,\ell}}) =  f(\profile|_{B_1}) \cup \gloc_f(\profile|_{B_{2,\ell}}, \decomp_2)\\
            &= f(\profile|_{B_1}) \cup  f(\profile|_{B_2}) \cup  f(\profile|_{B_{3,\ell}}) = f(\profile|_{B_1}) \cup  f(\profile|_{B_2}) \cup  \gloc_f(\profile|_{B_{3,\ell}}, \decomp_3)=\ldots\\
            &= \bigcup_{i \in [\ell]} f(\profile|_{B_i}).
        \end{align*}
        For each $i\in [\ell]$, $\profile|_{B_i}$ must have a PQ-tree of depth at most $i-1$. Thus by the inductive hypothesis we must have $f^{CC}(\profile|_{B_i})= f(\profile|_{B_i})$, implying $f^{CC}(\profile)=f(\profile)$, as desired. 
    \end{enumerate}
\end{enumerate}
The inductive proof above shows that in all cases, we have $f(\profile)=f^{CC}(\profile)$ for all $\profile$, as long as $f$ is CC.

\paragraph{Condition 4.} The statement that $f$ being anonymous implies $f^{CC}$ being anonymous follows from the fact that \Cref{alg:cc-transform} is robust to relabeling of voters. For each of the remaining properties, we will prove that it is preserved as a separate lemma. We start with Condorcet consistency. Recall that $f$ is Condorcet-consistent if it returns $Sm(\profile)$ whenever  $|Sm(\profile)|=1$, where $Sm$ is defined in \Cref{tab:scfs}. In words, if there is a candidate $a \in A$ that pairwise defeats every other candidate in $\profile$ (\emph{i.e.}, $a$ is the \emph{Condorcet winner}), then we must have $f(\profile)=\{a\}$.
\begin{lemma}\label{lemma:cc_condorcet}
If (neutral) $f$ is Condorcet-consistent, then $f^{CC}$ is Condorcet-consistent.
\end{lemma}
\begin{proof}

Consider running \Cref{alg:cc-transform} on input SCF $f$, which is Condorcet-consistent, and profile $\profile$, where $a \in A$ is the Condorcet winner.

Assume the algorithm dequeues node $B \subseteq A$ whose subtree contains $a$. If $|B|=1$, then $\node$ is the leaf corresponding to $\{a\}$, and $a$ is added to the winner list $W$. If $|B|>1$, then say $\decomp =$decomp($B,T$). Since $|B|$ is an internal node, we have $|\decomp|>1$ (all internal nodes in the PQ-tree has at least two children---see \Cref{subsec:extended_pqtrees} above). We would like to the show that \emph{only} the child node that contains $a$ (say $K_a \in \decomp$) will be enqueued by the algorithm. Given any $K' \in \decomp \setminus\{K_a\}$ and $b \in K'$, we have $M[a,b] =M^\decomp[K_a,K']$ by the clone definition (\emph{i.e.}, the majority relationship between $a$ and $b$ is the same as the majority relationship between their clone sets). Since $a$ pairwise defeats all $b \in A \setminus\{a\}$, this implies $K_a$ pairwise defeats all $K' \in \decomp \setminus\{K_a\}$, \emph{i.e.}, $K_a$ is the Condorcet winner of $\profile^\decomp$. Then, there are two options:

\begin{enumerate}
    \item If $B$ is a P-node: since $\clone_a$ is the Condorcet winner of $\profile^\decomp$ and $f$ is Condorcet-consistent we have that $f(\profile^\decomp) = \{\clone_a\}$. Hence, only $K_a$ is enqueued by the algorithm among the children nodes of $K_a$.
    \item If $B$ is a Q-node: since $\clone_a$ is a Condorcet winner of $\profile^\decomp$, we must have $K_a= B_1(B,T)$. Moreover, since $f$ is Condorcet-consistent, we must have $f(\profile^\decomp|_{\{B_1(B,T),B_2(B,T)\}})=\{B_1(B,T)\}$ (as $B_1(B,T)=K_a$ pairwise defeats $B_2(B,T)$). Hence, only $K_a$ is enqueued by the algorithm among the children nodes of $B$.
\end{enumerate}

This implies that starting from a node whose subtree contains $a$, \Cref{alg:cc-transform} will iteratively pick only the children node containing $a$, until arriving at $a$'s leaf node and adding it to the winner list. Since the queue $\queue$ initially has only the root node (denoted $A$), whose subtree $(T)$ indeed contains $a$, this implies only $a$ will be added to $W$ by the algorithm. Hence, $f^{CC}(\profile) = \{a\}$, \emph{i.e.}, $f^{CC}$ is Condorcet-consistent. 
\end{proof}

Before proving the preservation of the stronger axiom of Smith consistency, which dictates $f(\profile) \subseteq Sm(\profile)$ for \emph{all} profiles $\profile$, we first prove a useful intermediary lemma.

\begin{lemma} \label{lemma:smith_intersect}
    Given any profile $\profile$ and clone set $\clone$ with respect to $\profile$, it must be that $\clone$ and $Sm(\profile)$ cannot intersect nontrivially. That is, it must be that either $Sm(\profile) \subseteq \clone$, $\clone \subseteq Sm(\profile)$, or $Sm(\profile) \cap \clone = \emptyset$. 
    
    \begin{proof}
Suppose $K$ and $Sm(\profile)$ intersects nontrivially. Take any $a \in \clone \setminus Sm(\profile)$,  $b \in \clone \cap Sm(\profile)$, and $c \in Sm(\profile) \setminus K$. We must have that $c$ pairwise defeats $a$, since $a \notin  Sm(\profile) $ and $c \in  Sm(\profile)$. By the clone definition this implies $c$ also pairwise defeats $b$. Thus each candidate in $Sm(\profile) \setminus \clone$ pairwise defeats any candidate out of it, and is strictly smaller than $Sm(\profile)$. This contradicts the definition of $Sm(\profile)$. 
    \end{proof}
    
\end{lemma}
\begin{corollary}\label{cor:smith_clone}
If $\decomp$ is a clone decomposition with respect to $\profile$, either there exists $K \in \decomp$ such that $Sm(\profile) \subseteq K$, or there exists $\decomp' \subseteq \decomp$ such that $Sm(\profile) = \bigsqcup_{K \in \decomp'}K$.
\end{corollary}

\begin{lemma}\label{lemma:cc_smith}
    If (neutral) $f$ is Smith-consistent, then $f^{CC}$ is Smith-consistent. 
\end{lemma}
\begin{proof}
    We first show that when run on SCF $f$ (which is Smith-consistent) and profile $\profile$ (with PQ tree $T$), for any node $B \subseteq A$ whose subtree contains the entirety of $Sm(\profile)$, \Cref{alg:cc-transform} either only enqueues a single child node that is also a superset of $Sm(\profile)$, or only enqueues (possibly multiple) children nodes that are subsets of $Sm(\profile)$. By \Cref{cor:smith_clone}, when the algorithm is at node $B$ that is a superset of $Sm(\profile)$, the corresponding decomposition $\decomp=$decomp$(B,T)$ will satisfy one of two cases:

    \begin{enumerate}
        \item One child node ($\clone_S \in \decomp$) will contain all candidates in the Smith set. By the clone definition, this implies $K_S$ pairwise defeats every other clone set in $\profile^\decomp$. This means that $\{\clone_S\}$ is the Smith set of $\profile^\decomp$. Consider the cases for $B$: 
        \begin{enumerate}
            \item[(1a)] If $B$ is P-node, then because $f$ is Smith-consistent, we have that $f(\profile^\decomp) \subseteq Sm(\profile^\decomp) = \{\clone_S\}$. Hence, again, only $K_S$ gets enqueued.
            \item[(1b)] If $B$ is a Q-node, then we must have $B_1(B,T)=K_S$, which is the only child that gets enqueued. Moreover, since $f$ is Smith-consistent, we must have $f(\profile^\decomp|_{\{B_1(B,T),B_2(B,T)\}})=\{B_1(B,T)\}$ (as $B_1(B,T)=K_a$ pairwise defeats $B_2(B,T)$, so $\Sm(\profile^\decomp|_{\{B_1(B,T),B_2(B,T)\}})=\{B_1(B,T)\}$). Hence, only $K_S$ is enqueued by the algorithm among the children nodes of $B$.
        \end{enumerate} 
        \item There exists some $\decomp' \subseteq \decomp$ such that $Sm(\profile)=\bigsqcup_{K \in \decomp'} K$. In this case, $\decomp'$ is the Smith set of $\profile^\decomp$ (since $Sm(\profile^\decomp) \subseteq \decomp'$ by the clone definition, and $\decomp' \subseteq Sm(\profile^\decomp)$ by the minimality of $Sm(\profile)$). Consider the cases for $B$: \begin{enumerate}
            \item[(2a)] $B$ is a P-node. Since $f$ is Smith-consistent, it must be that $f(\profile^\decomp) \subseteq Sm(\profile^\decomp)=\decomp'$. Therefore, only child nodes that are subsets of $Sm(\profile)$ will be enqueued. 
            \item[(2b)]  $B$ is a Q-node and $\decomp \setminus \decomp'=\emptyset$. Then we have $B=Sm(\profile^\decomp)$, so any children of $\node$ that is enqueued is a subset of $Sm(\profile^\decomp)$ by definition.
            \item[(2c)]  $B$ is a Q-node and $\decomp \setminus \decomp' \neq \emptyset$. Since any $K \in \decomp'$ must pairwise any $K'$, we must have $\decomp' = \{B_i(\node, T)\}_{i=1}^j$ for some $j < |\decomp|$. Further, we must $j=1$, as otherwise $B_1(B,T)$ also pairwise defeats the remaining members of $\decomp'= \Sm(\profile^\decomp)$, which contradicts the minimality of $\Sm$. Therefore this case is identical to that of (1b), and only the Smith set gets enqueued. 
        \end{enumerate} 
    \end{enumerate}

Starting from a node $B$ that is a superset of $Sm(\profile)$, there can only $|B\setminus Sm(\profile)|$ number of subsequent nodes that falls into case (1) above, since each time this happens at least \emph{some} non-Smith candidates are dropped by the algorithm. Hence, starting from a node $B$ that is a superset of $Sm(\profile)$, the algorithm will eventually come to a node that fulfills case (2) above, in which case only the child nodes that are entirely subsets of $Sm(\profile)$ are enqueued, after which it is impossible for any $B\setminus Sm(\profile)$ to win. Since the root node of the tree ($\cand$), where the algorithm starts, is by definition a superset of $Sm(\profile)$, this implies that $f^{CC} (\profile) \subseteq \Sm(\profile)$, \emph{i.e.}, $f^{CC}$ satisfies Smith-consistency. 
\end{proof}

Recall that unlike the other axioms, we have so far only defined decisiveness on a specific profile $\profile$, \emph{i.e.} $|f(\profile)|=1$ (see \Cref{sec:bg}). Having fixed the voters $N$ and candidates $\cand$, we say $f$ is (overall) decisive if it is decisive on all $\profile\in \mathcal{L}(\cand)^n$. 
\begin{lemma}\label{lemma:cc_decisive}
If (neutral) $f$ is decisive, then $f^{CC}$ is decisive. 
\end{lemma}
\begin{proof}
When run on $f$ and any profile $\profile$ (with PQ-tree $T$), for each node $\node$ that is dequeued, \Cref{alg:cc-transform} will always enqueue a single child node of  $\node$: if $\node$ is a P-node, this is $f(\profile^\decomp)$ (where $\decomp =$decomp$(\node,T)$), which has cardinality 1 since $f$ is decisive; if $\node$ is a Q-node, this is $B_1(\node,T)$ or $B_{|\decomp|}(\node,T)$ (we cannot have $f(\profile^\decomp|_{\{B_1(\node,T),B_2(\node,T)\}}=\{B_1(\node,T),B_2(\node,T)\}$ since $f$ is decisive). This implies that \Cref{alg:cc-transform} will start from the root node of $T$ and go down one child node at a time, until reaching a leaf node, which will be the single winner added to $W$. Hence, $|f^{CC}(\profile)|=1$ for all $\profile$, \emph{i.e.} $f^{CC}$ is decisive.
\end{proof}
We now move to the clone-aware axioms, formally defined in the preceding section.

\begin{lemma}\label{lemma:cc_mono}
If (neutral) $f$ satisfies monotonicity$\ca$  (Def.~\ref{def:weak-mono}), then $f^{CC}$ satisfies monotonicity$\ca$. 
\end{lemma}

\begin{proof}
    Fix a profile $\profile$, a candidate $a \in f^{CC}(\profile)$, and a second profile $\profile'$ with (1) $\family(\profile)=\family(\profile')$ and (2) for all $i \in N$ and $b,c \in A \setminus\{a\}$, we have $a \succ_{\sigma_i} b \Rightarrow a \succ_{\sigma'_i} b$ and $b \succ_{\sigma_i} c \Rightarrow b \succ_{\sigma'_i} c$. We would like to show that $a \in f^{CC}(\profile').$ Since, $\family(\profile)=\family(\profile')$ the node structure of the PQ-trees of the two profiles (say $T$ and $T'$, respectively) are identical, but the number of each vote in $\profile^\decomp$ and $\profile'^\decomp$ might be different for a given $\decomp$. Hence, we only need to show that at each node $B \subseteq A$ that contains $a$, the child node containing $a$ (and possibly others) will be enqueued. Fix an internal node $\node$ cotaining $a$ in the PQ-tree, and say $\decomp=$ decomp($B,T$)=decomp($B,T'$) and $K_a$ is the clone set in $\decomp$ containing $a$ (\emph{i.e.}, $a \in K_a \in \decomp$). Consider two options:

    \begin{enumerate}
        \item $\node$ is a P-node. Since $a \in f^{CC}(\profile)$, we must have $K_a \in f(\profile^\decomp)$. Furthermore, for any two clone sets $K_b, K_c \in \decomp \setminus \{K_a\}$, it must be that $K_a \succ_{\sigma_i^\decomp} K_b \implies K_a \succ_{{\sigma_i^\decomp}'} K_b$ and 
        $K_b \succ_{\sigma_i^\decomp} K_c \implies K_b \succ_{{\sigma_i^\decomp}'} K_c$ for all $i \in N$ by the clone set definition, since the only difference between $\profile$ and $\profile'$ is $a$ moving up in some rankings. As $f$ satisfies clone-aware monotonicity, $K_a \in f(\profile^\decomp) \implies K_a \in f({\profile'^\decomp})$, implying $K_a$ is enqueued in both cases.
        \item $\node$ is a Q-node. This implies everyone in $\profile^\decomp$ has either ranked $B_1(\node, T) \succ B_2(\node, T) \succ \ldots \succ B_{|\decomp|}(\node,T)$ or $B_{|\decomp|}(\node, T) \succ B_{|\decomp|-1}(\node, T) \succ \ldots \succ B_{1}(\node,T)$. Say $K_a = B_{k}(\node,T)$ for some $k \in [|\decomp|]$. Consider two cases:
        \begin{itemize}
            \item[(2a)] $|\decomp|>2$. For any $i \in \voter$, we will show that  $\sigma_i^\decomp= \sigma_i'^{\decomp}$. By construction, the order in which all $B_{i}(\node, T)$ for $i \in [|\decomp|]\setminus \{k\}$ are the same in the two rankings, as only $K_a$ can move up. Assume for the sake of contradiction $\sigma_{i}'^{\decomp}$ ranks $K_a$ in the $j$th position for some $j<k$. If $j>1$, this implies $\{B_{j-1}(\node,T), B_{j}(\node,T)\}$ is a clone set in $\profile^\decomp$ (by definition of a Q-node) but not a clone set in $\profile'^{\decomp}$ (as they are interrupted by $K_a$ in $\sigma_i'^{\decomp}$), and therefore $ B_{j-1}(\node,T) \cup B_{j}(\node,T) \in \family(\profile) \setminus \family(\profile')$, which contradicts the assumption that $\family(\profile)=\family(\profile')$. Similarly, if $k < |\decomp|$, then $ B_{k}(\node,T) \cup B_{k+1}(\node,T) \in \family(\profile) \setminus  \family(\profile')$, once again leading to a contradiction. Lastly, if $j=1$ and $k=|\decomp|$, we have  $ B_{k-1}(\node,T) \cup B_{k}(\node,T) \in \family(\profile) \setminus  \family(\profile')$, as they are now interrupted by $B_1(\node, T)$ (we have $k-1>1$ since $k>2$). In all cases, assuming $j<k$ leads to a contradiction. Hence,   $\sigma_{i}'^{\decomp}$ ranks $K_a$ in the $k$th position, implying $\sigma_i^\decomp= \sigma_i'^{\decomp}$ and therefore $\profile^\decomp= \profile'^{\decomp}$. Thus, \Cref{alg:cc-transform} enqueues the same children nodes (and by assumption $K_a$) in both cases. 
            \item[(2b)] $|\decomp|=2$, in this case, $\node$ is a P-node and a Q-node at the same time (Q-nodes with two children are treated identically to P-nodes by \Cref{alg:cc-transform}), therefore we have this case is identical to case (1) above.
        \end{itemize}
    \end{enumerate}

 Therefore, at every step in the PQ-tree, since the clone set that contains $a$ is enqueued when running \Cref{alg:cc-transform} on $\profile$, it will also be enqueued when running \Cref{alg:cc-transform} on $\profile'$. Hence, $a \in f^{CC}(\profile')$, as desired.
\end{proof}

\begin{lemma}\label{lemma:cc_isda}
If  (neutral) $f$ satisfies ISDA$\ca$ (Def.~\ref{appdef:clone_isda}), then $f^{CC}$ satisfies ISDA$\ca$.
\end{lemma}
\begin{proof}
    Assume $f$ satisfies ISDA$\ca$, and take any profile $\profile \in \mathcal{L}(\cand)^n$ over candidates $A$ and any candidate $a\in A$ such that $a \notin \textit{Sm}(\profile)$ and $\family(\profile \setminus\{a\}) = \family(\profile) - \{a\}$. Denote $\profile'=\profile \setminus\{a\}$. Say $T$ is the PQ-tree of $\profile$ and $\decomp=\{K_1,K_2, \ldots , K_\ell\}=$decomp($A,T$) are children nodes of the root node of $T$. WLOG, say $a \in K_\ell$. Since $\family(\profile') = \family(\profile) - \{a\}$, we have that $\decomp' = \{\clone_1, ... , \clone_\ell \setminus \{a\} \}$ is a clone decomposition with respect to $\profile'$. We will argue $f^{CC}(\profile) = f^{CC}(\profile\setminus \{a\})$ by induction on the depth of the PQ-tree of $\profile$ (say $T$). 

    \noindent\textbf{Base case:} Say $T$ has depth 2 (depth 1 is impossible, since $a \notin \Sm(\profile)$ implies $\profile$ is over at least 2 candidates). In this case, $|K_i|=1$ for each $i \in [\ell]$. If the root is a P-node, this implies $\profile$ has no non-trivial clone sets. Since $\family(\profile') = \family(\profile) - \{a\}$, this implies $\profile'$ also  has no non-trivial clone sets. Then:
    \begin{align*}
        f^{CC}(\profile) =   f(\profile) =f(\profile')=f^{CC}(\profile')  
    \end{align*}
    where the first and last inequality follows from Condition 1 of \Cref{thm:cc_transform} proven above, and the second inequality follows from the assumption that $f$ satisfies ISDA$\ca$. If the root of $T$ is a Q-node (with majority ranking $\sigma$) on the other hand, it must be an untied Q-node (\emph{i.e.}, strictly more voters rank $\sigma$ than 
    its reverrse), otherwise we would have had $Sm(\profile)=A$, which contradicts the assumption that $a\notin Sm(\profile)$. Since $f$ satisfies ISDA$\ca$, we must have $f(\profile^\decomp|_{\{B_1(\node,T),B_2(\node,T)\}} = \{B_1(\node,T)\}$, otherwise removing $B_2(\node,T)\}$, which is not in the Smith set of $\profile^\decomp|_{\{B_1(\node,T),B_2(\node,T)\}}$ would change the election result. Therefore, $f^{CC}(\profile) =  B_1(A,T)$. Since $a$ is not in the Smith set, this implies $K_\ell=\{a\} \neq B_1(B,T)$. Since the removal of $a$ does not change the clone structure, the PQ tree of $\profile'$ (say $T'$) is either a single leaf node corresponding to $B_1(\node, T)$ (if $\ell=2$) or is also a single Q-node with $\ell-1$ children nodes that are all leaves and majority matrix $\sigma \setminus\{a\}$ (if $\ell >2$). Since $a$ did not come first in $\sigma$, this implies  $f^{CC}(\profile')=B_1(A \setminus\{a\},T')= B_1(A,T)= f^{CC}(\profile)$. This finishes the base case.

    \noindent \textbf{Inductive:} Assume that $f^{CC}(\profile)= f^{CC}(\profile')$ if the depth of $T$ is $1,2,\ldots k-1$. Fix a profile $\profile$ such that $T$ has depth $k$. Since $f^{CC}$ satisfies CC by Condition 2 proven above, we have $f^{CC}(\profile) = \gloc_{f^{CC}}(\profile, \decomp)$ and $f^{CC}(\profile') = \gloc_{f^{CC}}(\profile', \decomp')$. Hence, it is sufficent to prove that $\gloc_{f^{CC}}(\profile, \decomp)=\gloc_{f^{CC}}(\profile', \decomp')$. Consider two cases:

    \begin{enumerate}
        \item If $|\clone_\ell| = 1$, then $\clone_\ell=\{a\}$ is a Smith-dominated candidate within $\profile^\decomp$. Moreover, $\decomp'= \{K_1,K_2,\ldots,K_{\ell-1}\}$. Since $\decomp$ correspond to the children of the root node of $T$, the PQ-tree of $\profile^\decomp$ (say $T^\decomp$) is either a single P-node or a Q-node. Since $\profile'^{\decomp'}= \profile^{\decomp} \setminus \{K_\ell\}$, it follows by the base case above that $f^{CC}(\profile^{\decomp}) = f^{CC}(\profile'^{\decomp'})$. Then we have:
        \begin{align*}
            \gloc_{f^{CC}}(\profile, \decomp) = \bigcup_{K \in f^{CC}(\profile^{\decomp}) } f(\profile|_{K})=\bigcup_{K \in f^{CC}(\profile'^{\decomp'}) } f(\profile|_{K})= \gloc_{f^{CC}}(\profile', \decomp')
        \end{align*}
        and we are done.
        \item If $|\clone_\ell| > 1$: Then $\profile^\decomp$ and $\profile'^{\decomp'}$ are isomorphic, where the meta-candidate $K_\ell$ in  $\profile^\decomp$ is replaced with the meta-candidate $K'_\ell=K_\ell \setminus
        \{a\}$ in $\profile'^{\decomp'}$. Consider two options: 
        \begin{enumerate}
            \item[(2a)] $K_\ell \notin f^{CC}(\profile^\decomp)$. Then, by neutrality, we have $f^{CC}(\profile'^{\decomp'})=f^{CC}(\profile^\decomp)$, and we have
            \begin{align*}
            \gloc_{f^{CC}}(\profile, \decomp) = \bigcup_{K \in f^{CC}(\profile^{\decomp}) } f(\profile|_{K})=\bigcup_{K \in f^{CC}(\profile'^{\decomp'}) } f(\profile|_{K})= \gloc_{f^{CC}}(\profile', \decomp').
            \end{align*}

            \item[(2b)] $K_\ell \in f^{CC}(\profile^\decomp)$. Then, by neutrality, we have $f^{CC}(\profile'^{\decomp'})=f^{CC}(\profile^\decomp) \setminus \{K_\ell\} \cup \{K'_\ell\}$. We argue that $\Sm(\profile) \cap K_\ell \neq  \emptyset$. Assume for the sake of contradiction that  $\Sm(\profile) \cap K_\ell =  \emptyset$. Then $K_\ell \notin Sm(\profile^\decomp)$. If the root of $T^\decomp$ is a Q-node, this contradicts $K_\ell \in f^{CC}(\profile^\decomp)$, since it cannot not be $B_1(\decomp, T^{\decomp})$, which is the only enqueued child node since $f$ is ISDA$\ca$. If the root of $T^\decomp$ is a P-node, then by Condition 1 of \Cref{thm:cc_transform}, we have $f(\profile^\decomp) = f^{CC}(\profile^\decomp)$, so $K_\ell \in f(\profile^\decomp)$ violates the assumption that $f$ satisfies ISDA$\ca$, since by definition removing $K_\ell$ (which is not in $Sm(\profile^\decomp)$) will change the outcome. Therefore, we must have  $\Sm(\profile) \cap K_\ell \neq  \emptyset$. By \Cref{lemma:smith_intersect}, this implies  $Sm(\profile) \subset K_\ell$, since $a \in K_\ell \setminus Sm(\profile)$. Then, $a \notin Sm(\profile|_{K_\ell})$. Moreover, the PQ-tree of $\profile|_{K_\ell}$ has depth at most $k-1$. By the inductive hypothesis, this implies that $f^{CC}(\profile|_{K_\ell}) = f^{CC}(\profile|_{K_\ell} \setminus\{a\}) =f^{CC}(\profile|_{K'_\ell})$. Therefore
            \begin{align*}
                \gloc_{f^{CC}}(\profile, \decomp) &= \left(\bigcup_{K \in f^{CC}(\profile^{\decomp}) \setminus\{K_\ell\}} f(\profile|_{K})\right) \cup f^{CC}(\profile|_{K_\ell}) \\
                &= \left(\bigcup_{K \in f^{CC}(\profile'^{\decomp'}) \setminus\{K'_\ell\}} f(\profile|_{K})\right) \cup f^{CC}(\profile|_{K'_\ell}) \\
                &=\bigcup_{K \in f^{CC}(\profile'^{\decomp'}) } f(\profile|_{K})= \gloc_{f^{CC}}(\profile', \decomp').
            \end{align*}
        \end{enumerate}
    \end{enumerate}
    In each case, we have shown that $ \gloc_{f^{CC}}(\profile, \decomp)=  \gloc_{f^{CC}}(\profile', \decomp')$, which, by Condition 2, implies $f^{CC}(\profile)=f^{CC}(\profile')$, thus completing the inductive case.
\end{proof}    

\begin{lemma}\label{lemma:cc_participation}

If (neutral) $f$ satisfies participation$\ca$ (Def.~\ref{appdef:clone_participation}), then $f^{CC}$ satisfies participation$\ca$.
\end{lemma}

\begin{proof}

    Fix any profile $\profile \in \mathcal{L}(\cand)^n$ and any ranking $\sigma_{n+1} \in\mathcal{L}(\cand) $ such that $\family(\profile)=\family(\profile+\sigma_{n+1})$, implying that the PQ tree of both (say $T$ and $T'$, respectively) have the same structure. We denote $\profile'=\profile+\sigma_{n+1}$. Fix a node $\node$ in the PQ-tree that was dequeued by \Cref{alg:cc-transform} at some point when run on input $\profile$. Say $\decomp=$ decomp($B,T$)=decomp($B,T'$) and that $\decomp^* \subseteq \decomp$ are the child nodes that were enqueued by the algorithm. We will show that if \Cref{alg:cc-transform} on input $\profile'$ ever dequeues $\node$, then it will either enqueue $\max_{n+1}(\decomp^*)$ or a child node preferred by $\profile^\decomp_{n+1}$. Consider two cases:
    
    \begin{enumerate}
        \item $\node$ is a P-node. In that case, $\decomp^*=f(\profile^\decomp)$ by construction of \Cref{alg:cc-transform}. Similarly, if dequeued when run on input $\profile'$, \Cref{alg:cc-transform} will enqueue $f(\profile'^\decomp)$. Since $f$ satisfies clone-aware participation, we must have  $\max_{n+1}(f(\profile'^\decomp)) \succeq_{n+1}\max_{n+1}(f(\profile^\decomp))$, so the algorithm does indeed enqueue $\max_{n+1}(\decomp^*)$ or a child node preferred by $\profile^\decomp_{n+1}$. 
        \item $\node$ is a Q-node, with majority ranking $\sigma^*$ over $\decomp$. Moreover, since $\family(\profile)=\family(\profile')$, $\sigma^\decomp_{n+1}$ must either be $\sigma^*$ or its reverse. Consider three subcases:
        \begin{enumerate}
            \item[(2a)] $f(\profile^{\decomp}|_{\{B_1(\node, T), B_2(\node,T)\}})=\{B_1(\node,T)\}$. If $\sigma_{n+1}^\decomp=\sigma^*$, then $\profile'^{\decomp}|_{\{B_1(\node, T'), B_2(\node,T')\}}$ is simply $\profile^{\decomp}|_{\{B_1(\node, T), B_2(\node,T)\}}$ with an additional $(B_1(\node, T) \succ B_2(\node, T))$ vote. Since $f$ satisfies participation$\ca$, we must have $f(\profile'^{\decomp}|_{\{B_1(\node, T'), B_2(\node,T')\}})=\{B_1(\node,T')\}=\{B_1(\node,T)\}$. Thus $B_{1}(\node,T)$ get enqueued on input $\profile'$, which is the top ranked candidate in $\sigma^\decomp_{n+1}$. If $\sigma_{n+1}^\decomp$ is the reverse of $\sigma^*$, on the other hand, the child node enqueued at $\node$ on input $\profile$ ($B_1(\node,T)$) is the bottom ranked candidate in $\sigma_{n+1}^\decomp$, so it cannot possibly be ranked above the child node enqueued at $\node$ on input $\profile'$.
            \item[(2b)] $f(\profile^{\decomp}|_{\{B_1(\node, T), B_2(\node,T)\}})=\{B_2(\node,T)\}$. If $\sigma_{n+1}^\decomp=\sigma^*$, the child node enqueued at $\node$ on input $\profile$ ($B_{|\decomp|}(\node,T)$) is the bottom ranked candidate in $\sigma_{n+1}^\decomp$, so it cannot possibly be ranked above the child node enqueued at $\node$ on input $\profile'$. If $\sigma_{n+1}^\decomp$ is the reverse of $\sigma^*$, on the other hand, $\profile'^{\decomp}|_{\{B_1(\node, T'), B_2(\node,T')\}}$ is simply $\profile^{\decomp}|_{\{B_1(\node, T), B_2(\node,T)\}}$ with an additional $(B_2(\node, T) \succ B_1(\node, T))$ vote. By assumption $f$ satisfies participation$\ca$; thus, we must have $f(\profile'^{\decomp}|_{\{B_1(\node, T'), B_2(\node,T')\}})=\{B_2(\node,T')\}=\{B_2(\node,T)\}$. Thus $B_{|\decomp|}(\node,T)$ get on input $\profile'$, which is the top ranked candidate in $\sigma^\decomp_{n+1}$.
            \item[(2c)] $f(\profile^{\decomp}|_{\{B_1(\node, T), B_2(\node,T)\}})=\{B_1(\node,T),B_2(\node,T)\}$. If $\sigma_{n+1}^\decomp=\sigma^*$, $\profile'^{\decomp}|_{\{B_1(\node, T'), B_2(\node,T')\}}$ is simply $\profile^{\decomp}|_{\{B_1(\node, T), B_2(\node,T)\}}$ with an additional $(B_1(\node, T) \succ B_2(\node, T))$ vote. Since $f$ satisfies participation$\ca$, we must have $B_1(\node,T') \in f(\profile'^{\decomp}|_{\{B_1(\node, T'), B_2(\node,T')\}})$. Thus $B_{1}(\node,T)$ is one of the child nodes that get enqueued on input $\profile'$, which is the top ranked candidate in $\sigma^\decomp_{n+1}$. If $\sigma_{n+1}^\decomp$ is the reverse of $\sigma^*$, on the other hand, $\profile'^{\decomp}|_{\{B_1(\node, T'), B_2(\node,T')\}}$ is simply $\profile^{\decomp}|_{\{B_1(\node, T), B_2(\node,T)\}}$ with an additional $(B_2(\node, T) \succ B_1(\node, T))$ vote. Since $f$ satisfies participation$\ca$, we must have $B_2(\node,T)\in f(\profile'^{\decomp}|_{\{B_1(\node, T'), B_2(\node,T')\}})$. Thus $B_{|\decomp|}(\node,T)$ is one of the child nodes that get enqueued on input $\profile'$, which is the top ranked candidate in $\sigma^\decomp_{n+1}$.
        \end{enumerate}
    \end{enumerate}
    
    In each case, we see that if \Cref{alg:cc-transform} is considering $\node$ when run on $\profile'$, it will either enqueue $\max_{n+1}(\decomp^*)$ or a child node strictly preferred by $\profile^\decomp_{n+1}$. 
    
    Say $B$ is the root node $(A)$, which is indeed dequeued by the algorithm  when run on either input. If a $K \succ \max_{n+1}(\decomp^*)$ is enqueued (\emph{i.e.}, a strictly preferred child node) when run on input $\sigma'$, then we are done, since this implies $f^{CC}(\profile') \cap K \neq \emptyset$ and each element of $K$ is preferred to all elements in $f^{CC}(\profile)$ by $\sigma_{n+1}$. Otherwise,  $\max_{n+1}(\decomp^*)$ must have been enqueued, and we can apply the same argument to that node, since it will be considered by the algorithm on both inputs. We repeat following the $\max_{n+1}(\decomp^*)$ on each step until we reach a strictly preferred child node that is enqueued, or we reach a leaf node, in which case $\max_{n+1}(f^{CC}(\profile))= \max_{n+1}(f^{CC}(\profile'))$. In either case, we have $\max_{n+1}(f^{CC}(\profile')) \succeq_{n+1} \max_{n+1}(f^{CC}(\profile'))$, proving $f^{CC}$ satisfies participation$\ca$. 
    \end{proof}    

Together \Cref{lemma:cc_condorcet,lemma:cc_smith,lemma:cc_decisive,lemma:cc_mono,lemma:cc_isda,lemma:cc_participation} prove Condition 4 of \Cref{thm:cc_transform}.

\paragraph{Condition 5.}
We analyze the running time of Algorithm 1: CC transformation for SCF $f$.
First, we construct the PQ-tree $T = PQ(\profile)$ for $\sigma$.
By Lemma~\ref{lemma:pq-poly} (shown by \citet{Cornaz13:Kemeny}), this requires $O(nm^3)$ time.
Next, whenever Algorithm~\ref{alg:cc-transform} encounters a node $B$ that is of type P-, it runs $f$ on $\profile^\mathcal{K}$, where $\mathcal{K} = \mathrm{decomp}(B, T)$. 
By definition of $\delta(T)$, we can upper bound the runtime of running $\profile^\mathcal{K}$ for each node $B$ of type P- by $g(n, \delta(T))$.
Hence, overall, this requires at most $|\mathcal{P}| \cdot g(n, \delta(T))$ runtime, where recall that $\mathcal{P}$ denotes the set of P-nodes in PQ-tree $T$.
On the other hand, whenever Algorithm~\ref{alg:cc-transform} encounters a node $B$ that is of type Q-, it only runs $f$ on the first two child nodes; \emph{i.e.}, it runs $f$ with at most two candidates. 
Again by definition of function $g$, each encounter of a Q-node in Algorithm~\ref{alg:cc-transform} thus adds a running time of at most $g(n, 2)$. 
Overall, this requires $|\mathcal{Q}| \cdot g(n, 2)$ runtime, where $\mathcal{Q}$ denotes the set of Q-nodes in PQ-tree $T$. 

Hence, the total runtime of Algorithm~\ref{alg:cc-transform} is $O(nm^3) + |\mathcal{P}| \cdot g(n, \delta(T)) + |\mathcal{Q}| \cdot g(n, 2)$.
By definition of $\delta(T)$, and excluding the trivial case where $m=1$, it follows that $\delta(T) \geq 2$, and thus $g(n, 2) \leq g(n, \delta(T))$.
Moreover, since all nodes of a PQ-tree are of either type P- or type Q-, it follows that $|\mathcal{P}| + |\mathcal{Q}| \leq m$, as the number of internal nodes in a tree with $m$ leaves is bounded by $m$. 

Therefore, the total running time of Algorithm~\ref{alg:cc-transform} is upper bounded by $O(nm^3) + m \cdot g(n, \delta(T))$. 

\end{proof}

\section{On \Cref{sec:spf} (\nameref{sec:spf})}
In the main body of the paper, we have introduced (to the best of our knowledge) the first extension of the CC definition to social preference functions (SPFs), which returns a set of rankings over $A$ rather than a subset. We recall this definition, along with the for IoC for SPFs by \citet{Freeman14:Axiomatic} below. In the rest of this section, we prove our results about SPFs.

\iocspf*

Intuitively, IoC dictates that removing a member of a clone set should not change the rankings of all non-clones, as well as that of some highest ranked clone.

\ccspf*

In words, CC dictates that all clone sets in the $\decomp$ should appear as an interval in the order(s) specified by $F(\profile^\decomp)$, and the order(s) within each $K \in \decomp$ should be specified by $F(\profile|_K)$.

\subsection{Definitions of social preference functions}\label{appsec:spfdefs}

Here, we give the descriptions of the SPF versions of several SCFs we have discussed in previous sections. Below, the ``RP procedure'' refers to the process of locking in edges from $M$ in non-increasing order, skipping the ones that create a tie. 
\begin{itemize}
    \item $RP$: Return all rankings $r$ that correspond to the topological ordering of the final graph constructed by the RP procedure for \emph{some} tie-breaking order.
    \item  $RP_i$: Return the topological ordering of the final graph constructed by the RP procedure using $\Sigma_i$ as a tie-breaker.
    \item  $RP_N$: Return the union over $RP_i$ for all $i \in N$
    \item $STV$: At each round, eliminate the candidate with the least plurality votes, until only one candidate remains\footnote{Recall that as an SCF, it was sufficient to run $STV$ until a candidate secures majority (see \Cref{tab:scfs}). However, for the SPF version, we need to keep running the elimination process until there is a single candidate left, so we can get a complete order of elimination over the candidates.} Output the reverse order elimination 
    \item $BP$: Construct the strength matrix $S$ as described in \Cref{tab:scfs}. Define relationship $\succ_{BP}$ over candidates as $a \succ_{BP} b$ if $S[a,b]>S[b,a]$ and  $a =_{BP} b$ if $S[a,b]= S[b,a]$. As proven by \citet{Schulze10:New}, $\succeq_{BP}$ satisfies transitivity; hence, it gives a weak ordering over candidates $A$. Return all strict rankings $r$ that are consistent with the weak ordering of $\succeq_{BP}$. For example, for $\profile'$ with the strength matrix given in the extended proof of \Cref{thm:cc_fails} above, we have $a_1=_{BP}a_2 \succ_{BP} b \succ_{BP} c$. This implies $BP^*(\profile') =\{r_1,r_2\}$ where $BP^*$ is the SPF version of $BP$, with $r_1 : a_1 \succ a_2 \succ b\succ  c$ and $r_2 : a_2 \succ a_1 \succ b\succ  c$.
\end{itemize}
\subsection{Proof of \Cref{prop:SPFtoSCF}}
We first prove that our novel definitions of IoC/CC for SPFs are consistent with the ones for SCFs.

\spfscf*

\begin{proof}
    Say $F$ satisfies IoC, and pick any profile $\profile$, non-trivial clone set $K$ with respect to $\profile$, a $a\in K$. Note that for any ranking $r$ over $A$ and $b \in A \setminus K$, we have $b=top(r) \iff b=top(r_{\neg K \rightarrow z})$, since relabeling/adding/removing lower ranked candidates does not change the fact that $b$ is ranked top. Hence, we have:
 \begin{align*}
     b\in f(\profile) \iff   \exists r \in F(\profile)\text{ s.t. } b=top(r) &\iff  \exists r \in F(\profile)_{K \rightarrow z}\text{ s.t. } b=top(r) \\&\overset{(IoC)}{{\iff}}   \exists r \in F(\profile \setminus \{a\})_{(K \setminus \{a\})\rightarrow z}\text{ s.t. } b=top(r)\\
     &\iff \exists r \in F(\profile \setminus \{a\})\text{ s.t. } b=top(r)\\& \iff b \in f(\profile\setminus\{a\})
 \end{align*}
 which give proves $f$ satisfies condition (2) from \Cref{def:ioc}. Next, notice that by definition of the $\neg$ operator, for any ranking $r$ we have $top(r) \in K \iff z=top(r_{\neg K \rightarrow z})$. This implies:
 \begin{align}
     K \cap f(\profile) \neq \emptyset  \iff \exists r \in F(\profile)\text{ s.t. } top(r) \in K &\iff \exists r \in F(\profile)_{\neg K \rightarrow z}\text{ s.t. } z=top(r)
     \\&\overset{(IoC)}{{\iff}}   \exists r \in F(\profile \setminus \{a\})_{(K \setminus \{a\})\rightarrow z}\text{ s.t. } z=top(r)\\
     &\iff \exists r \in F(\profile \setminus \{a\})\text{ s.t. } top(r) \in K\setminus\{a\} \\& \iff  (K\setminus\{a\}) \cap  f(\profile\setminus\{a\}) \neq \emptyset
 \end{align}
 which give proves $f$ satisfies condition (1) from \Cref{def:ioc}. Hence, $f$ is IoC.

 Next, assume $F$ satisfies CC. Take any profile $\profile$ and clone decomposition $\decomp$ with respect to $\profile$. By \Cref{def:cc_spf}, we have:
 \begin{align*}
     F(\boldsymbol{\sigma}) = F(\boldsymbol{\sigma}^\decomp)(K \rightarrow F(\boldsymbol{\sigma}|_\clone)\text{ for }\clone \in \decomp) \Rightarrow f(\profile) = top(F(\profile)) &= \bigcup_{K \in top(F(\profile^\decomp))}top(F(\profile|_{K}))\\&=\bigcup_{K \in f(\profile^\decomp)}f(\profile|_{K})
 \end{align*}
Hence, we have $f(\profile)=\gloc_f(\profile,\decomp)$, so $f$ satisfies CC.
\end{proof}

\subsection{(Nested) nested runoff voting}\label{appsec:nested}

In this section, we briefly discuss how non-anonymous tie-breakers (such as the one introduced by \citet{Zavist89:Complete}) can be used to construct CC SPFs other than $RP_i$. \citet{Freeman14:Axiomatic} introduce an SPF named Nested Runoff ($NR$), which is a modification of $\STV$: at each round, instead of the candidate with the lowest plurality score, the winner of $\STV(\rev(\profile))$ is eliminated, where $\rev(\profile)$ is $\profile$ with every voter's ranking reversed. \citeauthor{Freeman14:Axiomatic} show that $NR$ is IoC as an SPF. By our \Cref{prop:SPFtoSCF}, this implies $NR$ is IoC as an SCF too. However, since it is anonymous, it cannot be CC as an SPF by \Cref{thm:anonspf}. In fact, $NR$ is not CC as an SCF either; to see this, consider the following profile over 4 candidates with 3 voters:
\begin{align}
    \profile= \begin{cases}
        b \succ_1 a_2 \succ_1 a_1 \succ_1 c \\
        a_2 \succ_2 a_1 \succ_2 c \succ_2 b \\
        c \succ_3 b \succ_3 a_1 \succ_3 a_2 \label{eq:nested-counter}
    \end{cases}
\end{align}
Say $\decomp=\{\{a_1,a_2\}, \{b\}, \{c\}\}$. It can be checked that $a_1 \in NR(\profile)$ but $a_1 \notin \gloc_{NR}(\profile, \decomp)$, which violates CC. 

Now, say $STV_i$ is simply the version of $STV$ that uses voter $i$'s vote as a tie-breaker (\emph{i.e.}, if multiple candidates tie for the lowest plurality score at any point, the one ranked lowest by voter $i$ is eliminated). It is straightforward to check that, much like $\STV$, $\STV_i$ is IoC as an SPF and as an SCF, but CC as neither. Unlike $\STV$, however, $\STV_i$ is decisive on all $\profile$. Now, we define $NR_i$ using $\STV_i$ on the reverse profile to decide on the order of elimination. We will show that $NR_i$ is CC as an SCF. Take any profile $\profile$ and any decomposition $\decomp$ with respect to $\profile$. Since we are using a decisive tie-breaker, we will have $|NR_{i}(\profile)|= |\gloc_{NR_i}(\profile, \decomp)|=1$, so it is sufficient to show containment in one direction. Say $\gloc_{NR_i}(\profile, \decomp)=\{a\}$ and $K_a \in \decomp$ is the clone set containing $a$. This implies $NR_i(\profile^{\decomp}) = \{K_a\}$ and $NR_i(\profile|_{K_a})=\{a\}$. Say $K_1,K_2,\ldots,K_{\ell}$ is the order in which clone sets are eliminated when $NR_i$ is run on $\profile^{\decomp}$. This implies $STV_i(\rev(\profile^{\decomp}))=\{K_1\}$. By successive application of the IoC for SCF property, we must have $STV_i(\rev(\profile))=\{b_1\}$ for some $b_1 \in K_1$, implying that the first candidate eliminated by $NR_i$ on input $\profile$ belongs to $K_1$. If $|K_1|>1$, this argument can be repeated again with $STV_i(\rev(\profile \setminus\{b_1\}))$, implying the next eliminated candidate too will belong to $K_1$. Applying this argument repeatedly gives us that $NR_i$ on input $\profile$ will eliminate all elements of $K_1$ before any other candidate. Now, by the assumption on the order of elimination in $NR_i(\profile^{\decomp})$ we have $STV_i(\rev(\profile^{\decomp} \setminus K_1))=\{K_2\}$, and the same IoC argument can be applied to show that $STV_i(\rev(\profile \setminus K_1))=\{b_2\}$ for some $b_2 \in K_2$. Inductively applying this argument gives that $NR_i$ will eliminate all candidates of $K_i$ before any candidate of $K_{i+1}$ for all $i \in [\ell]$, where $K_{\ell+1}=K_a$ (since it never gets eliminated). Hence, at some point in the execution of $NR_i$ on $\profile$, we will have $\profile \setminus \left( \bigcup_{i \in [\ell]} K_i \right)$ left. However, this is precisely $\profile|_{K_a}$, and by assumption we have $NR_i(\profile|_{K_a})=\{a\}$, showing that we must indeed have $NR_i(\profile)=\{a\}$, proving that $NR_i$ is CC as an SCF. 

To see that $NR_i$ is not CC as an SPF, once again consider the profile from \eqref{eq:nested-counter} and $NR^*_{3}$ (\emph{i.e.}, the SPF version of $NP_i$ using $i=3$ as the tie-breaker). It can be checked that $NR^*_3(\profile)= c \succ b \succ a_1 \succ a_2$, but $NR^*_3(\profile|_{\{a_1,a_2\}})=a_2 \succ a_1$, hence violating CC as an SPF. Hence, $NR_i$ serves as a ``natural'' counterexample showing that the reverse of \Cref{prop:SPFtoSCF} does not always hold. 

Finally, let us use $NR_i$ to design a CC SPF. Define the \emph{nested nested runoff rule} using $i$ as a tie-breaker ($NNR_i$) as a modification of $NR_i$ that, instead of $STV_i$, runs $NR_i$ on the reverse profile to decide the next eliminated candidate. Given any profile $\profile$ and decomposition $\decomp$, say $NNR_i(\profile^{\decomp})=K_\ell \succ K_{\ell-1}\succ \ldots \succ K_2 \succ K_1$. This implies $NR_i(\rev(\profile^{\decomp}))=\{K_1\}$. Since $NR_i$ is CC as an SCF, it is IoC as an SCF (by \Cref{prop:cctoioc}). Hence, we must have $NR_i(\rev(\profile))=\{b_1\}$ for some $b_1 \in K_1$. By the CC property of $NR_i$, this implies $NR_i(\rev(\profile)|_{K_1})=\{b_1\}$, implying that the candidate ranked at the bottom of $NNR^*_i(\profile)$ is $\{b_1\}$. If $|K_1|>1$, applying the same argument again gives us $NR_i(\rev(\profile \setminus \{b_1\} ))= NR_i(\rev(\profile|_{K_1} \setminus \{b_1\}))$. Thus, all the candidates in $K_1$ appear in the bottom of $NNR_i(\profile)$, and they appear exactly in the order they do in $NNR_i(\profile)$. Applying this argument inductively to all $K_i$ for $i \in [\ell]$ gives us exactly the CC definition for SPFs (\Cref{def:cc_spf}), completing the proof.

Hence, we have arrived at an interesting hierarchy. $STV_i$ is IoC as an SPF and an SCF, but CC as neither. $NR_i$, which uses $STV_{i}$ to eliminate candidates, is CC as an SCF, but still only IoC as an SPF. Lastly, $NNR_i$, which uses $NR_i$ to eliminate candidates, is CC both as an SPF and an SCF. Based on this observation, we believe studying the axiomatic properties of this type of (nested) nested rules is an interesting future direction. 

\subsection{Proof of \Cref{prop:cctoiocspf}}
We first prove that the CC to IoC relationship extends to the definitions for SCFs we have introduced.

\spfccioc*

\begin{proof}
    Say $F$ satisfies CC. By \Cref{def:cc_spf}, this implies that $F$ is neutral. Pick any profile $\profile$, non-trivial clone set $K$ with respect to $\profile$, a $a\in K$. Consider the clone decomposition $\decomp = \{\clone\} \cup \{\{b\}\}_{b\in \cand \setminus \clone}$ for $\profile$ and the clone decomposition $\decomp' = \{ \clone\setminus \{a\} \} \cup \{\{b\}\}_{b\in \cand \setminus \clone}$ for $\profile \setminus \{a\}$ (i.e., the decomposition which groups all existing members of $\clone$ together, and everyone else is a singleton). Notice that $\profile^\decomp$ and  $(\profile\setminus\{a\})^\mathcal{K'}$ are identical except the meta-candidate for $\clone$ in the former is replaced with the meta-candidate for $\clone \setminus \{a\}$ in the latter. By neutrality, this implies: $F(\profile^\decomp)_{\neg \{K\} \rightarrow K\setminus\{a\}}=F((\profile\setminus\{a\})^\mathcal{K'})$. Each $K'\in \decomp \setminus\{K\}$ is a singleton, and hence $F(\profile|_K)$ is just a single ranking with the only element in $K'$. For any $b \in A$, say $r_b$ is the trivial ranking over $\{b\}$. 
    Using CC, we get
    \begin{align*}
        F(\boldsymbol{\sigma})\neg_{{\clone \to z}} &= \left(F(\boldsymbol{\sigma}^\decomp)(K' \rightarrow F(\boldsymbol{\sigma}|_{\clone'})\text{ for }\clone' \in \decomp)\right)\neg_{{\clone \to z}}\\
        &= \left(F(\boldsymbol{\sigma}^\decomp)(K \rightarrow F(\boldsymbol{\sigma}|_{\clone}); \{b\} \rightarrow r_b \text{ for }b \in A\setminus K)\right)\neg_{{\clone \to z}}\\
        &= F(\boldsymbol{\sigma}^\decomp)(K \rightarrow r_z; \{b\} \rightarrow r_b \text{ for }b \in A\setminus K)\\
        &=F((\profile\setminus\{a\})^\mathcal{K'})( (K\setminus\{a\}) \rightarrow r_z; \{b\} \rightarrow r_b \text{ for }b \in A\setminus K)\\
        &=\left(F((\profile\setminus\{a\})^\mathcal{K'})( (K\setminus\{a\}) \rightarrow F(\profile|_{K \setminus\{a\}}); \{b\} \rightarrow r_b \text{ for }b \in A\setminus K)\right)_{(K \setminus\{a\})\rightarrow z}\\
        &=F(\profile\setminus\{a\})_{(K \setminus\{a\})\rightarrow z}.
    \end{align*}
Hence, $F$ satisfies IoC.
\end{proof}

\subsection{Proof of \Cref{thm:spftaxonomy}}
We now prove that for SCFs for which we described the SPF version above, our results generalize. 

\spftaxonomy*

\begin{proof}
We prove the axioms satisfied by each SPF as a seperate Lemma. 
\begin{lemma}\label{lemma:spfstv}
    The SPF version of STV is IoC but not CC.
\end{lemma}
\begin{proof} 

The fact that SPF version of $\STV$ is IoC is shown by \citet{Freeman14:Axiomatic}. Since SCF version of $STV$ is not CC (\Cref{thm:cc_fails}), then the SPF version of $STV$ is also not CC by \Cref{prop:SPFtoSCF}.
\end{proof}

\begin{lemma}\label{lemma:spfbp}
    The SPF version of $BP$ is IoC but not CC.
\end{lemma}
\begin{proof} 

 Take any profile $\profile$, clone set $K$, and $a\in K$. Say $S$ and $S'$ (resp. $M$ and $M'$) are the strength (resp. majority) matrices that result from running the $BP$ procedure on $\profile$ and $\profile\setminus\{a\}$, respectively. First, notice that for any $b,c \in A \setminus \{a\}$, we have $M[b,c] =M'[b,c]$, since the removal of candidate does not change the pairwise relationship between the remaining candidates. Take any $x \in A\setminus\{a\}$ and $y,z \in A \setminus K$. We would like to show that:
 \begin{align}
     S[x,y]=S'[x,y] \quad   S[y,x]=S'[y,x] \quad S[y,z]=S'[y,z]  \label{eq:strengths}
 \end{align}
 Since $M'$ is simply $M$ with $a$ removed, any path in $M'$ exists in $M$. This gives you the $\geq$ direction of all of the equalities in (\ref{eq:strengths}). For the reverse direction, consider any path $P$ from $x$ to $y$ in $M$. If the path does not contain $a$, then it exists in $M'$. If it does contain $a$, consider the alternative path $P'$ that starts from the last element belonging to $K$ in $P$, but replaces it with $x$ (so $P'$ is also a path from $x$ to $y$). By the clone definition, the first edge in the path is equally strong, and the remaining edges are the same. Since the strength of a path is the minimum weight over the edges in the path, this shows that $P'$ is at least as strong as $P$. The same method can be applied for paths from $y$ to $x$ by replacing the first occurrence of a member of $K$ with $x$. Now take any path $P$ in $M$ from $y$ to $z$. Again, if it does not contain $a$, it still exists in $M'$. If it does contain it, then pick any $b \in K \setminus \{a\}$ (exists sicne $K$ is non-trivial) and construct path $P'$ by replacing the interval in $P$ from the first occurrence of a member of $K$ to the last occurrence of a member of $K$ with $b$. By the clone definition, the incoming and outgoing edge of $b$ will have the same weight as the incoming and outgoing edge to this interval. Since the remaining paths are only subtracted, again $P'$ is at least as strong as $P$. This finishes the  $\leq$ direction of all of the equalities in (\ref{eq:strengths}). 

This implies that for any $b,c \in A \setminus\{a\}$ such that at least one of them is not in $K$, we will have $b \succeq_{BP} c \iff b \succeq'_{BP} c$, where $\succeq_{BP}$ and $\succeq'_{BP}$ are the (weak) linear orderings resulting from running $BP$ on $\profile$ and $\profile \setminus \{a\}$, respectively. This implies that $ BP^*(\profile)\neg_{K \rightarrow z}= BP^*(\profile\setminus\{a\})\neg_{ (K \setminus\{a\}) \rightarrow z}$, proving that $BP^*$ (the SPF version of $BP$) is IoC.

Since SCF version of $BP$ is not CC (\Cref{thm:cc_fails}), then $BP^*$ is also not CC by \Cref{prop:SPFtoSCF}.
\end{proof}

\begin{lemma}\label{lemma:spfrp}
    The SPF version of $RP$ neither IoC nor CC.
\end{lemma}
\begin{proof}
    Since the SCF version of $RP$ is not IoC (\Cref{prop:rpfailioc}) and therefore not CC (by \Cref{prop:cctoioc}), SPF version of $RP$ is neither IoC nor CC \Cref{prop:SPFtoSCF}.
\end{proof}

\begin{lemma}\label{lemma:spfrpi}
    The SPF version of $RP_i$ is both IoC and CC.
\end{lemma}
\begin{proof}
    The proof that $RP_i$ satisfies CC follows easily from the proof of \Cref{thm:rp}. There, (using \Cref{lemma:stack_winner}) we showed that given any decomposition $\decomp$ the RP ranking resulting from running $RP_i$ on $\profile$ has each  clone set in $\decomp$ as an interval, in the order specified by the RP ranking resulting from running $RP_i$ on $\profile^\decomp$. Moreover, we showed that the clone set ranked first in this ranking (say $K_1$) appeared in the order  specified by the RP ranking resulting from running $RP_i$ on $\profile|_{K_1}$. This last proof did not use the fact that $K_1$ was the first clone set to appear in the ranking, but only that it appeared as an interval. Hence, the same proof can be easily applied to all $K \in \decomp$, since each appear as an interval. As a result, we have $RP^*_i(\profile)= RP^*_i(\profile^\decomp)(K \rightarrow RP^*_i(\profile)\text{ for }K\in \decomp)$, where $RP^*_i$ is the SPF version of  $RP_i$.
\end{proof}

\begin{lemma}\label{lemma:spfrpn}
    The SPF version of $RP_N$ is IoC but not CC.
\end{lemma}
\begin{proof}
    Say $RP^*_N$ and $RP^*_i$ are the SPF versions of $RP_N$ and $RP_i$, respectively. Fix any profile $\profile$, non-trivial clone set $K$, and $a \in K$. Since $RP^*_i$ is IoC for each $i \in N$ by \Cref{lemma:spfrpi} and by definition of the $\neg$ operator, we have
    \begin{align*}
        RP^*_N(\boldsymbol{\sigma}) \neg_{\clone \to z}&=
       \left(\bigcup_{i \in N}  RP^*_i(\boldsymbol{\sigma}) \right) \neg_{\clone \to z}
       = \bigcup_{i \in N}  RP^*_i(\boldsymbol{\sigma}) \neg_{\clone \to z} = \bigcup_{i \in N}  RP^*_i(\boldsymbol{\sigma}) \neg_{\clone \to z}\\&=\bigcup_{i \in N}  RP^*_i(\boldsymbol{\sigma} \setminus \{a\})\neg_{(\clone \setminus \{a\})\to z}=\left(\bigcup_{i \in N}  RP^*_i(\boldsymbol{\sigma} \setminus \{a\})\right)\neg_{(\clone \setminus \{a\})\to z}\\&=RP^*_N(\boldsymbol{\sigma} \setminus \{a\}) \neg_{(\clone \setminus \{a\})\to z},
    \end{align*}
    proving $RP^*_N$ satisfies IoC. Since the SCF version of $RP_N$ does not satisfy CC (\Cref{prop:rpn}), this proves that  $RP^*_N$ is not CC by \Cref{prop:SPFtoSCF}.
\end{proof}
\Cref{lemma:spfstv,lemma:spfbp,lemma:spfrp,lemma:spfrpi,lemma:spfrpn}, together with our result from \Cref{thm:cc_fails,thm:rp}, prove the theorem statement.
\end{proof}

\subsection{Proof of \Cref{thm:anonspf}}
We next prove a surprising negative resulting showing the incompatibility of anonymity and composition consistency for SPFs.

\anonspf*

\begin{proof}
    Assume for the sake of contradiction that we have an SPF $F$ that is CC and anonymous. By \Cref{def:cc_spf}, this implies $F$ is also neutral. Consider a profile $\profile$ over $\cand=\{a,b,c\}$ with two votes. Voter 1 ranks $a \succ_1 b \succ_1 c$ and Voter 2 ranks $c \succ_2 b \succ_2 a$. Define $K_1=\{a,b\}$ and $K_2=\{b,c\}$, which are both clone sets with respect to $\profile$. Further, define $\decomp_1=\{K_1,\{c\}\}$ and $\decomp_2=\{
    \{a\},K_2\}$, which are both clone decompositions with respect to $\profile$. Consider $\profile^{\decomp_1}$, which consists of $K_1 \succ_1 \{c\}$ and $\{c\} \succ_2 K_1$. Since $F$ is neutral and anonymous, we must have  $F(\profile^{\decomp_1})=\{K_1\succ \{c\}, \{c\} \succ K_1\}$, otherwise permuting Voter 1 with Voter 2 and $K_1$ with $\{c\}$ gives a contradiction. By the same reasoning, we must have $F(\profile|_{K_1})=\{a\succ b, b\succ a\}$. By composition consistency (\Cref{def:cc_spf}), we must have
    \begin{align}
        F(\boldsymbol{\sigma}) = F(\boldsymbol{\sigma}^{\decomp_1})(K \rightarrow F(\boldsymbol{\sigma}|_\clone)\text{ for }\clone \in \decomp_1)=\left\{\!\begin{aligned}a\succ b \succ c,\\ b\succ a \succ c, \\c\succ a \succ b, \\c\succ b \succ a\end{aligned}\right\}.\label{eq:anonspf1}
    \end{align}
    Similarly, $\profile^{\decomp_2}$, which consists of $\{a\} \succ_1 K_2$ and $K_2 \succ_2 \{a\}$. By neutrality and anonymity, we must have  $F(\profile^{\decomp_2})=\{\{a\} \succ_1 K_2, K_2 \succ_2 \{a\}\}$, otherwise permuting Voter 1 with 2 and $\{a\}$ with $K_2$ gives a contradiction. By the same reasoning, we must have $F(\profile|_{K_2})=\{b\succ c, c\succ b\}$. By CC, we must have
    \begin{align}
        F(\boldsymbol{\sigma}) = F(\boldsymbol{\sigma}^{\decomp_1})(K \rightarrow F(\boldsymbol{\sigma}|_\clone)\text{ for }\clone \in \decomp_1)=\left\{\!\begin{aligned}a\succ b \succ c,\\ a\succ b \succ c, \\b\succ c \succ a, \\c\succ b \succ a\end{aligned}\right\}.\label{eq:anonspf2}
    \end{align}
    Comparing \eqref{eq:anonspf1} with \eqref{eq:anonspf2}, we immediately get a contradiction.
\end{proof}

\section{On \Cref{sec:oioc} (\nameref{sec:oioc})}

In this section, we provide the proofs omitted from \Cref{sec:oioc} of the main body, as well as a formal definition of extensive games and obviously-dominant strategies (in the restricted setting where each agent has a single information set).

\subsection{Proof of \Cref{prop:metric}}
Given $\profile$ over candidates $\cand$ with $|\cand|=m$, consider $d_{\profile}: B \times B \rightarrow [m] \cup \{0\}$ defined for each $a_i,a_j \in \cand$ as:
\begin{align*}
    d_{\profile}(a_i,a_j)= \min_{\substack{{\clone \subseteq \cand: a_i,a_j \in K,}\\{K\text{ is a clone set w.r.t. }\profile}}} |K| -1
\end{align*}

\clonemetric*

\begin{proof} We prove $d_{\profile}$ satisfies all axioms of a metric:
    \begin{itemize}
    \item (Zero distance to self) For each $a \in \cand$, $\{a\}$ is a clone set, so $d(a,a)=|\{a\}|-1=0$.
    \item (Positivity) If $a\neq b$, then any clone set $K$ that contains both of them must have $|K|\geq 2$, so $d(a,b) \geq 2-1=1>0$.
    \item (Symmetry) Clearly, $d(a,b)=d(b,a)$.
    \item (Triangle inequality) Given any $a,b,c \in A$, say $K_1$ is the clone set that includes $a,b$ with $|K_1|=d(a,b)+1$ and $K_2$ is the clone set that includes $b,c$ with $|K_2|=d(b,c)+1$. Since $b \in K_1 \cap K_2$, we have $K_1 \cap K_2 \neq \emptyset$ so by Axiom (A1) by \citet{Elkind10:Clone}, we have that $K_1 \cup K_2$ is a clone set. Notice $|K_1 \cup K_2|= |K_1|+|K_2| - |K_1 \cap K_2| \leq (d(a,b)+1) + (d(b,c)+1) - 1=d(a,b)+d(b,c)+1$. Since $a,c \in K_1 \cup K_2$, we have $d(a,c) \leq |K_1 \cup K_2|-1 \leq d(a,b)+d(b,c)+1-1=d(a,b)+d(b,c)$, satisfying triangle inequality. 
\end{itemize}
\end{proof}
\subsection{Proof of \Cref{prop:ioc_ds}}
Next, we prove that in the strategic candidacy setting where the preferences of candidates are dictates by $d_{\profile}$, IoC rules not only achieve but strengthen candidate stability.
\iocds*

\begin{proof}
Given any $a\in A$, say $u_a(S)$ is the utility of this player in $\Gamma_{\profile}^f$ when exactly the candidates in $S\subseteq A$ play $R$, and all candidates in $A \setminus S$ play $D$, which is a decreasing function of $d_{\profile}(a,f(\profile|_S))$ (and minimized at $u_a(\emptyset)$). Fix any $a \in \cand$ and a pure action for every other candidate. Say $S$ are the candidates among $\cand \setminus \{a\}$ that played $R$. To show that $R$ is a dominant strategy for $a$, we would like to show $u_a(S\cup\{a\}) \geq u_a(S)$. Consider three cases:
\begin{enumerate}
    \item Case 1: If $f(\profile|_{S \cup \{a\}})=\{a\}$, then $u_a(S \cup \{a\})>u_a(S)$ since $a \notin f(\profile|_{S})$, so $d_{\profile}(a,  f(\profile|_{S}))>0$, and we are done.
    \item Case 2: $f(\profile|_{S \cup \{a\}})=f(\profile|_{S})$, then $u_a(S \cup \{a\})=u_a(S)$ and we are done.
    \item Case 3: $S= \emptyset$. Then $u_a(S)$ is the minimizer of $u_a$ and $u_a(S \cup \{a\})=u_a(\{a\})$ is the maximizer, so we are done. 
    \item Case 4: If Cases 1-3 are false, we must have $f(\profile|_{S \cup \{a\}})=\{b\}$ and $f(\profile|_{S})=\{c\}$ for some $b \neq a$ and $c \neq b$. Take any clone set $K \subseteq \cand$ with respect to $A$ containing both $a$ and $c$; we would like to show $b \in K$ (which will automatically apply $d_{\profile}(a,b) \leq d_{\profile}(a,c)$). Say $K' = K \cap (S \cup \{a\})$, which is a clone set in $\profile|_{S \cup \{a\}}$ by \Cref{def:clones}. Since $f$ is IoC, we have
    \begin{align*}
        K' \cap f(\profile|_{S \cup \{a\}}) \neq \emptyset \Leftrightarrow K'\setminus\{a\} \cap f(\profile|_{S}) \neq \emptyset.
    \end{align*}
    Since the right hand side is true (as $c \in K'\setminus\{a\} \cap f(\profile|_{S})$), we must have $ K'\cap f(\profile|_{S \cup \{a\}}) \neq \emptyset$, implying $b \in K' \subseteq K$ and therefore $d_{\profile}(a,b) \leq d_{\profile}(a,c)$ and $u_a(S\cup\{a\}) \geq u_a(S)$.
\end{enumerate} 
\end{proof}

\subsection{Definitions for extensive-form games and obviously dominant strategies}\label{appsec:efg-ods}
In this section, we introduce extensive-form games and obviously dominant strategies, which we use to argue that CC exposes the obviousness of IoC.

\begin{definition}
    We can define an \emph{extensive-form game} $\Gamma$ as follows:
    \begin{enumerate}
        \item  $\Gamma$ is represented by a rooted tree structure. The set of all nodes in this tree is denoted by $\calH$ with each edge of the tree representing a single game \emph{action}. The game begins at the root, and each action traverses down the tree, until the game finishes at a leaf which we call a \emph{terminal node}. The set of terminal nodes is denoted by $\calZ \subset \calH$, and the set of actions available at any nonterminal node $h \in \calH \setminus \calZ$ is denoted by $A_h$.
        \item A finite set of strategic and chance players $|\calN \cup \{c\}| = N+1$ with $N \geq 1$. The set $\calN$ contains the \emph{strategic players}, and $c$ stands for a \emph{chance} ``player'' that models exogenous stochasticity. Each nonterminal node $h$ is assigned to either a strategic player or the chance player, who chooses an action to take from $A_h$. We call the set of nodes assigned to Player $i$ $\calH_i$.
        \item For each chance node $h \in \calH_c$, a probability distribution $\Prob_c(\cdot \mid h)$ on $A_h$ with which chance elects an action at $h$.
        \item For each strategic player $i \in \calN$, a (without loss of generality) nonnegative \emph{utility (payoff)} function $u_i : \calZ \to \RR_{\geq 0}$ which returns what $i$ receives when the game finishes at a terminal node. Player $i$ aims to maximize that utility.
        \item For each strategic player $i \in \calN$, a partition $\calH_i = \sqcup_{I \in \calI_i} I$ of the nodes of $i$ into information sets (\emph{infosets}). Nodes of the same infoset are considered indistinguishable to the player at that infoset. For that, we also require $A_h = A_{h'}$ for $h, h' \in I$. This also makes action set $A_I$ well-defined.
    \end{enumerate}
\end{definition}

\paragraph{Strategies and utilities.}

Players can select a probability distribution---a \emph{randomized action}---over the actions at an infoset. A (behavioral) \emph{strategy} $\pi_i$ of a player $i \in \calN$ specifies a randomized action $\pi_i(\cdot \mid I) \in \Delta(A_I)$ at each infoset $I \in \calI_i$. We say $\pi_i$ is \emph{pure} if it assigns probability 1 to a single action for each infoset. A (strategy) \emph{profile} $\pi = (\pi_i)_{i \in \calN}$ specifies a strategy for each player. We use the common notation $\pi_{-i} = (\pi_1, \dots, \pi_{i-1}, \pi_{i+1}, \dots, \pi_n)$. We denote the strategy set of Player $i$ with $S_i$, and $S = \bigtimes_{i \in \calN} S_i$.

We denote the reach probability of a node $h'$ from another node $h$ under a profile $\pi$ as $\Prob(h' \mid \pi, h)$. It evaluates to $0$ if $h \notin \seq(h')$, and otherwise to the product of probabilities with which the actions on the path from $h$ to $h'$ are taken under $\pi$ and chance. For any infoset, let $I^{\text{1st}}$ refer to the nodes $h\in I$ for which $I$ does not appear in $\obs(h)$. Then the reach probability of $I$ is $\Prob(I\mid \pi, h)\coloneqq \sum_{h' \in I^{\text{1st}}}\Prob(h' \mid \pi, h)$. We denote with $\U_i(\pi \mid h) \coloneqq \sum_{z \in \calZ} \Prob(z \mid \pi, h) \cdot u_i(z)$ the expected utility of Player $i$ given that the game is at node $h$ and the players are following profile $\pi$. Finally, we overload notation for the special case the game starts at root node $h_0$ by defining $\Prob(h \mid \pi) := \Prob(h \mid \pi, h_0)$ and $\U_i(\pi) \coloneqq \U_i(\pi \mid h_0)$.

We now introduce obviously dominant strategies. Since we will focus on games with no exogenous stochasticity (i.e., no chance nodes) and where every player will have a single infoset, our definition is a simplified version of the original definition by \citet{Li17:Obviously}.

\begin{definition}[{\citealt[Obviously Dominant Strategy]{Li17:Obviously}}]
Given an EFG $\Gamma$ with no chance nodes and a single infoset per player (i.e., $\calI_j =\{I_j\}$ for each $j \in \calN$) and a player $i \in \mathcal{N}$, an action $s \in A_{I_i}$ is obviously dominant if:

\[
\forall s' \in I_i: \quad \quad
\sup_{h \in I_i, \pi_{-i}} u_i((\pi_i^{s'}, \pi_{-i}) \mid h) \leq \inf_{h \in I_i, \pi_{-i}} u_i((\pi_i^{s}, \pi_{-i}) \mid h )
\]
where $\pi_i^s$ is the player $i$ strategy that plays action $s$ with probability 1. 
\end{definition}

Inutitively, an action $s$ is obviously dominant  for player $a$ if for any other action $s'$, \emph{starting from when $a$ must take an action}, the best possible outcome from $s'$ is no better than the worst possible outcome from $s$. The $\sup/\inf$ over $h \in I_i$ allows us to compare the best and worst possible for $i$ given what she knows at the point where she must act ($I_i$), and the $\sup/\inf$ over $\pi_{-i}$ allows us to best and worst possible outcomes based on the strategies of all other players (again, given that $I_i$ is reached), including those that have not acted yet. 

For example, $\Gamma^f_{\profile}$ from the main body of the paper can be interpreted as an EFG where players act simultaneously. In this case, even if the $f$ is IoC, running ($R$) is \emph{not} an obviously dominant strategy, due to the uncertainty of the actions of every other candidate: 
\begin{example}
    Consider $\Gamma^{\STV}_{\profile}$, where $\profile$ is from Figure~\ref{fig:eg_prof}. For $b$, the \emph{worst} outcome of running ($R$) is that every other candidate plays $R$ too, making $d$ the winner. The \emph{best} outcome of dropping out ($D$), on the other hand, is for $c$ to play $R$ and $d$ to play $D$, in which case $c$ wins regardless of what $a$ does. Since $2=d_{\profile}(b,c)< d_{\profile}(b,a)=4$, candidate $b$ strictly prefers the latter outcome, showing that $R$ is \emph{not} an obviously-dominant strategy for her, even though it is a dominant strategy by Prop.~\ref{prop:ioc_ds}. 
\end{example}

\subsection{Proof of \Cref{thm:cc_ods}}
Finally, we prove that in the process of implementing a rule $f^{CC}$, $\Lambda_{\profile}^f$ achieves obvious strategy-proofness for candidates.
\ccods*

\begin{proof}[Proof]
    Take any candidate $a\in A$ and consider the point in $\Lambda^f_{\profile}$ where $a$ must decide $R$ or $D$. This happens when Algo.~\ref{alg:cc-transform} is on the parent node of $a$, say $\node$. If $\node$ is a Q-node, the worst possible outcome of playing $R$ is $a$ winning herself, which is her optimal outcome, and hence the best outcome of $D$ cannot be any better. If is a P-node, then $\node$ is the smallest non-trivial clone set that contains $a$ by Lemma~\ref{lemma:pq_clones} (in other words, the members of $\node$ are exactly $a$'s second-favorite candidates after herself). Then the worst possible outcome of $a$ running ($R$) is some other candidate $b \in \node \setminus \{a\}$ winning (since Algo.~\ref{alg:cc-transform} will move out of $\node$ only if all the candidates in $\node$, including $a$, play $D$), whereas the best possible outcome of $a$ dropping out ($D$) is, again, some other candidate $c \in \node \setminus \{a\}$ winning. Since $d_{\profile}(a,b)=d_{\profile}(a,c)$ (both pairs are united by $\node$ as the smallest clone set), the latter outcome is no better than the former, proving  that $R$ is an obviously-dominant strategy for $a$. 
\end{proof}

\end{document}